%% file: paper.tex
\documentclass[longauth]{aa}

\usepackage{pdflscape}
\usepackage{graphicx}
\usepackage{natbib}
\usepackage{scalerel}
\usepackage{tabularx} 

\usepackage{pdflscape}
\usepackage{rotating}
\usepackage{subcaption} 
\usepackage[font=small]{caption} 
\usepackage[labelfont=bf]{caption} 
\usepackage[singlelinecheck=false]{caption} 

\usepackage[table]{xcolor}

\usepackage[nolist,nohyperlinks]{acronym}

\usepackage{txfonts}
\usepackage[pdfencoding=auto,psdextra]{hyperref}
\hypersetup{
    colorlinks=true,
    linkcolor=blue,
    filecolor=magenta,      
    urlcolor=blue,
    citecolor=blue
}
\usepackage{orcidlink} 
\newcommand{\orcid}[1]{\orcidlink{#1}}

\usepackage[nameinlink,capitalise]{cleveref}

\crefname{section}{Sect.}{Sects.}
\Crefname{section}{Section}{Sections}
\crefname{figure}{Fig.}{Figs.}
\Crefname{figure}{Figure}{Figures}
\crefname{equation}{Eq.}{Eqs.}
\Crefname{equation}{Equation}{Equations}

\makeatletter
\renewcommand*\aa@pageof{, page \thepage{} of \pageref*{LastPage}}
\makeatother
\usepackage{lastpage}

\usepackage[utf8]{inputenc}

\usepackage[switch, modulo]{lineno}

\usepackage{float}
\usepackage{euclid}

\begin{document}

\input{acronym}

\title{\Euclid: Early Release Observations of ram-pressure stripping in the Perseus cluster. Detection of parsec-scale star formation within the low surface brightness stripped tails of UGC 2665 and MCG +07-07-070\thanks{This paper is published on behalf of the Euclid Consortium}}

\input{authorsandaffiliations}

\abstract{\textit{Euclid} is delivering optical and near-infrared imaging data over 14\,000\,deg$^2$ on the sky at spatial resolution and surface brightness levels that can be used to understand the morphological transformation of galaxies within groups and clusters. Using the Early Release Observations (ERO) of the Perseus cluster, we demonstrate the capability offered by \textit{Euclid} in studying the nature of perturbations for galaxies in clusters. Filamentary structures are observed along the discs of two spiral galaxies, UGC 2665 and MCG +07-07-070, with no extended diffuse emission expected from tidal interactions at surface brightness levels of $\sim$ $30\,{\rm mag}\,{\rm arcsec}^{-2}$. The detected features exhibit a good correspondence in morphology between optical and near-infrared wavelengths, with a surface brightness of $\sim$ $25\,{\rm mag}\,{\rm arcsec}^{-2}$, and the knots within the features have sizes of $\sim$ 100 pc, as observed through \IE imaging. Using the \textit{Euclid}, CFHT, UVIT, and LOFAR $144\,{\rm MHz}$ radio continuum observations, we conduct a detailed analysis to understand the origin of the detected features. We constructed the \textit{Euclid} $\IE-\YE$, $\YE-\HE$, and CFHT $u - r$, $g - i$ colour-colour plane and showed that these features contain recent star formation events, which are also indicated by their H$\alpha$ and NUV emissions. \textit{Euclid} colours alone are insufficient for studying stellar population ages in unresolved star-forming regions, which require multi-wavelength optical imaging data. There are features with red colours that can be explained by the presence of dust stripped along with the gas in these regions. The morphological shape, orientation, and mean age of the stellar population, combined with the presence of extended radio continuum cometary tails can be consistently explained if these features have been formed during a recent ram-pressure stripping event. This result further confirms the exceptional qualities of \textit{Euclid} in the study of galaxy evolution in dense environments.}


\titlerunning{\Euclid: ERO -- ram-pressure stripping in the Perseus cluster }
\authorrunning{George et al.}
   
\maketitle

\section{\label{sc:Intro}Introduction}

Star-forming galaxies in the Universe follow a tight relation between the star-formation rate and the stellar mass, observed to be present from redshift $z\sim6$ to $0$ \citep{Brinchmann_2004,Salim_2007,Noeske_2007,Elbaz_2007,Daddi_2007,Popesso_2023}. The existence of this relation since early epochs suggests that the process of gas condensation and star formation is well-regulated and closely tied to the galaxy's gravitational potential well. Primarily, star-forming galaxies have spiral morphologies, where gas stabilizes and collapses to form new star-forming regions in the disc, a process regulated by the availability of atomic and molecular hydrogen \citep{Kennicutt_2012}. In dense environments like clusters and groups, external processes such as galaxy mergers, starvation, thermal evaporation, ram-pressure stripping, and tidal interaction with the cluster potential can alter the gas content and star formation process (see for reviews \citealt{Boselli_2006,Boselli_2014,Cortese_2021}). When galaxies fall into galaxy clusters for the first time, they can experience mergers on the outskirts that disrupt their gas and stellar content, potentially triggering a starburst episode that rapidly depletes fuel for regulated star formation \citep{Barnes_1992,Barnes_2004}. The cluster environment can cut off the gas supply to galaxies, leading to a gradual reduction of star formation, known as starvation, which slows down the star formation rate once the galaxy becomes a satellite of a larger halo \citep{Larson_1980}. The cluster potential can create perturbations to galaxies and even disrupt low-mass galaxies falling in radial orbits close to the cluster centre \citep{Valluri_1993,Moore_1996,Mastropietro_2005}. The cold interstellar medium (ISM) of a galaxy can interact with the surrounding hot intracluster medium (ICM) leading to thermal evaporation \citep{Cowie_1977}. Ram-pressure stripping (RPS)  can remove gas from the disk of an infalling gas-rich spiral galaxy, dragging its interstellar medium into the surrounding intergalactic medium \citep{Gunn_1972,Boselli_2002}. This process can be observed in the form of
stripped tails in CO, radio continuum, H\textsc{i}, dust, H$\alpha$, and X-rays \citep{Gavazzi_2001,Vollmer_2004,Kenney_2004,Sun_2006,Chung_2009,Yagi_2010,Merluzzi_2013,Fumagalli_2014,Abramson_2016,Jachym_2017,Moretti_2018,Poggianti_2019,Roberts_2021a,Roberts_2021b,Ignesti_2022}. All these processes can act alone or jointly depending on the trajectory of the infalling galaxy, on its properties, and on that of the cluster \citep{Gullieuszik_2020,Smith_2022}.

Hydrodynamical processes like RPS act only on the diffuse components of the ISM (gas, dust) leaving the stellar component unaffected. They can easily be identified whenever the tails of stripped material do not contain stars. Indeed, several examples exist of stripped tails of cold, ionized, or hot gas without any associated stellar emission \citep{Gavazzi_2001,Boselli_2016,Boissier_2012,Yagi_2007,Yagi_2010,Yagi_2017,Jachym_2017,Laudari_2022,Serra_2023}. However, under specific and still unclear physical conditions \citep{Boselli_2022}, star formation can occur in the tails of stripped gas \citep{Owen_2006,Cortese_2007,Smith_2010,Owers_2012,Ebeling_2014,Fumagalli_2014,Rawle_2014,Poggianti_2016,Poggianti_2019,Bellhouse_2017,Gullieuszik_2017,Boselli_2018,George_2018,George_2023}. Mixing the cold ISM with the hot ICM is expected to warm up the cold gas component, preventing its collapse into giant molecular clouds where star formation occurs. The occurrence of star formation in the stripped tail is intriguing, as it takes place outside the galaxy disc within the very hostile environment of a hot ICM, with a temperature of 
 10$^{7-8}$K \citep{Sarazin_1986}. Within the stripped tails, spectacular trails form and star formation progresses as the galaxy moves within the cluster. The nearly face-on or edge-on orientation of the in-falling galaxy can make the star formation in the stripped tails appear to have different morphology. The smallest scales at which star formation progresses in these regions, as well as their diffuse extent, can set constraints on the nature of gas collapse within the stripped tails \citep{Portegies_2010,Elmegreen_2010,Elmegreen_2014}. To understand the size distribution of star-forming knots in the stripped gas at the faintest levels and the star-forming process in this hostile environment, it is necessary to observe star-forming regions at high spatial resolution and low surface brightness. High-resolution optical imaging of RPS galaxies can resolve knots of star formation in the stripped tails at 50--$100\,{\rm pc}$ scales in nearby galaxy clusters \citep{Abramson_2014,Kenney_2015,Cramer_2019,Boselli_2021,Giunchi_2023a,Gullieuszik_2023,Waldron_2023,Giunchi_2025}.

The recently launched \textit{Euclid} mission with its 
$\approx\,\ang{;;0.16}$ spatial resolution in \IE can resolve these knots at $50\,{\rm pc}$ scales up to distances of $72.5\,{\rm Mpc}$ ($z\sim0.016$). Perseus (Abell 426) is a massive galaxy cluster ($r_{200}$ = $2.2\,{\rm Mpc}$, $M_{200}$ = 1.2 $\times$ 10$^{15} M_\odot$, velocity dispersion = 1040 km s$^{-1}$) located at a redshift $\sim$ 0.0167 \citep{Aguerri_2020,Cuillandre_2024a}. The core region of the Perseus cluster is exceptionally rich in early-type galaxies with a strong deficiency in late-type systems \citep{Kent_1983}. The galaxy cluster is located very close to the Galactic plane (latitude $-$13 degrees). There are four known RPS galaxy candidates in the central regions of the Perseus cluster (MCG +07-07-070, UGC 2654, UGC 2665, and LEDA 2191078). These RPS candidates are identified from the presence of cometary-shaped radio continuum tails at $144\,{\rm MHz}$ and from the peculiar, asymmetric morphology of the stellar disc seen in ground-based optical imaging \citep{Roberts_2022}. The \textit{Euclid} satellite observed the Perseus cluster in optical (\IE) and infrared (\YE,\JE,\HE) bands as part of the \Euclid Early Release Observations \citep[ERO,][]{EROcite} programme. Two of these galaxies (MCG +07-07-070 and UGC 2665) are included in the \textit{Euclid} ERO field with near-simultaneous co-aligned imaging in optical and near-infrared.\\

The goal of this paper is to demonstrate the possibility opened up by \textit{Euclid} in studying galaxy evolution in dense environments. We demonstrate this through a study of star-forming regions in the tails and main bodies of galaxies UGC 2665 and MCG +07-07-070, made possible by the high spatial resolution and low surface brightness regime of \textit{Euclid}'s optical and near-infrared imaging observations. The dominant perturbing mechanism can be first established using \textit{Euclid} imaging data. The features resulting from gravitational perturbations are diffuse and include shells, plumes, and tidal tails \citep{Bilek_2020}. On the other hand, those related to RPS are filamentary and clumpy, with a cometary shape, as expected from star formation in the stripped gas \citep{Boselli_2022}. The sensitivity to low surface brightness features, in addition to the angular resolution offered by \textit{Euclid}, can be used to identify the dominant perturbing mechanism (gravitational or hydrodynamic), resolve the star-forming regions in the stripped material, and reconstruct their star formation history. Finally, our results demonstrate the capabilities of the Euclid Wide Survey (EWS) in detecting low surface brightness features around galaxies across a large region of the sky, enabling the study of galaxy evolution in different environments. The low surface brightness imaging data from EWS, which reaches $29.8\,{\rm mag}\,{\rm arcsec}^{-2}$, will enable surface brightness limits necessary to identify any possible perturbations induced on the stellar component by gravitational perturbations, and thus determine the dominant perturbing mechanism in rich environments on large, statistically significant samples. In identifying galaxies undergoing an RPS event through morphological analysis of broadband imaging data, it is particularly important to note that the lack of any evident perturbation in the stellar distribution is crucial for ruling out gravitational perturbations.

Throughout the paper we adopt a standard flat $\Lambda$CDM cosmology with $\Omega_{\rm m}=0.319$ and $H_0=67$\,km\,s$^{‐1}$Mpc$^{‐1}$ \citep{Aghanim_2020}. Magnitudes are in the AB system, and, in concordance with other Perseus ERO papers, we adopted a distance of 72 $\pm$ $3\,{\rm Mpc}$ to the Perseus cluster, where $\ang{;;1}$ corresponds to $0.338\,{\rm kpc}$ \citep{Cuillandre_2024a}.

\section{Data and analysis}

\citet{Cuillandre_2024b} present details on \textit{Euclid} ERO and data reduction optimized for preserving low surface brightness features. We use the data products generated as part of the Perseus cluster observations, which are described in detail in \citet{Cuillandre_2024a}. The \textit{Euclid} visible imager (VIS) has a broad passband (\IE) that covers the wavelength range 5500--9000\,$\AA$ \citep{Cropper_2014,Cropper_2016,Cropper_2024}. 
The near-infrared spectrometer and photometer (NISP) covers the wavelength range 9200--20\,000\,$\AA$ using the \YE,\JE,\HE passbands \citep{Maciaszek_2014,Maciaszek_2016,Jahnke_2024}. Observations are taken centred on coordinates RA=\ra{03;18;40}, Dec=\ang{41;39;00} with a $\sim$ 0.7\,deg$^2$ field of view. The Perseus imaging from ERO is created by combining four Reference Observation Sequences (ROS), whereas the EWS will consist of one ROS \citep{Scaramella_2022,Mellier_2024}. The final combined image exposure time is 7456\,s in the \IE filter and 1392.2\,s in the \YE,\JE, and \HE filters. \textit{Euclid} VIS and NISP imaging data have pixel scales of 0.1\,arcsec pix$^{-1}$ and 0.3\,arcsec pix$^{-1}$, with angular resolutions of $\approx\,\ang{;;0.16}$ and $\approx\,\ang{;;0.48}$ respectively. This enables us to achieve resolved spatial scales of $\sim$ $54\,{\rm pc}$ in VIS and $135\,{\rm pc}$ in NISP imaging observations of the stripped tails. The limiting surface brightness of the Perseus cluster field is $30.1\,{\rm mag}\,{\rm arcsec}^{-2}$ in \IE, $29.1\,{\rm mag}\,{\rm arcsec}^{-2}$ in \YE, $29.2\,{\rm mag}\,{\rm arcsec}^{-2}$ in \JE, $29.2\,{\rm mag}\,{\rm arcsec}^{-2}$ in \HE for 10$\arcsec$ × 10$\arcsec$ scale at 1$\sigma$ \citep{Cuillandre_2024a}. \\

We used the optical broadband $u, g, r, i, z$ photometry and narrow band H$\alpha$ imaging data of the Perseus cluster field taken with the Canada France Hawaii Telescope (CFHT) before the \textit{Euclid} launch. Details on the CFHT observations, data analysis and image quality for all the bands are given in \citet{Cuillandre_2024a}. The image quality of CFHT band images varies between $u$ $\approx\ \ang{;;1.46}$, $g$  $\approx\,\ang{;;1.23}$, $r$ $\approx\,\ang{;;0.79}$, $i$ $\approx\,\ang{;;0.56}$, $z$  $\approx\,\ang{;;0.60}$, and H$\alpha$
 $\approx\,\ang{;;0.49}$. The narrow-band H$\alpha$ ‘off’ filter (CFHT ID 9604), centred on $\lambda{_c}$ = 6719\,$\AA$, has a width of $\delta\lambda$ = 109\,$\AA$, which corresponds to a heliocentric velocity range of 4660--9600\,km s$^{-1}$ and is used for H$\alpha$ observations. We create the H$\alpha$ stellar continuum-subtracted (pure H$\alpha$+\ion{N}{ii} emission) image by subtracting with the $r$ band image of the field. 

We used the archival data of Perseus cluster observed using the Ultraviolet imaging telescope onboard Astrosat \citep{Agrawal_2006,Tandon_2017}.
Level 2 data is generated using the latest version (7.0.1) of the UVIT pipeline \citep{Joseph_2025}. The observations are in the far ultraviolet [FUV, filter F154W: $\lambda_{\rm mean}$=1541\,{\AA}, $\delta_{\lambda}$=380\,{\AA}. integration time=11099\,s], and the Near Ultraviolet [NUV, filter N245M: $\lambda_{\rm mean}$=2447\,{\AA}, $\delta_{\lambda}$=280\,{\AA}, integration time=10966.4\,s]. The UVIT NUV imaging is performed using a narrow-band filter at $\approx\,\ang{;;1.2}$ compared to the FUV imaging done with a broadband filter at $\approx\,\ang{;;1.4}$ resolution. We used NUV N245M imaging in the analysis owing to a better resolution than FUV \citep{Tandon_2017,Tandon_2020}. \\

We use the LOw Frequency ARray (LOFAR) $144\,{\rm MHz}$ observations of Perseus cluster field for radio continuum emission from the galaxies. These observations are taken as part of LOFAR Two-metre Sky Survey \citep{Shimwell_2017,Shimwell_2019}.
The LOFAR $144\,{\rm MHz}$ image covers the central $\sim$ 2\degree $\times$ 2\degree of the Perseus cluster field, with a resolution of $\ang{;;6}$ and an RMS of \SI{100}{\micro Jy \per beam}. The details of LOFAR data reduction and analysis are presented in \citet{Roberts_2022,vanWeeren_2024}. 

We note that the data from other wavelengths are at different spatial resolution and relative depths. Comparisons of observations from different instruments and wavelengths can be biased by this sensitivity issue, especially when using \textit{Euclid}'s low surface brightness optimized imaging. This makes it difficult to quantitatively compare the details of the galaxies at low surface brightness between different wavelengths. 

We used stellar masses derived from the spectral energy distribution fitting over the $u,g,r,i,z$,\IE,\YE,\JE, and \HE photometry, using the hyperz code \citep{Bolzonella_2000,Bolzonella_2010}. The effective radius is measured from the surface brightness profile using AutoProf/AstroPhot. Details of the derivation of these quantities are given in  \citet{Cuillandre_2024a}.

\begin{figure*}[htbp!]
\centering
\includegraphics[angle=0,width=0.85\hsize]{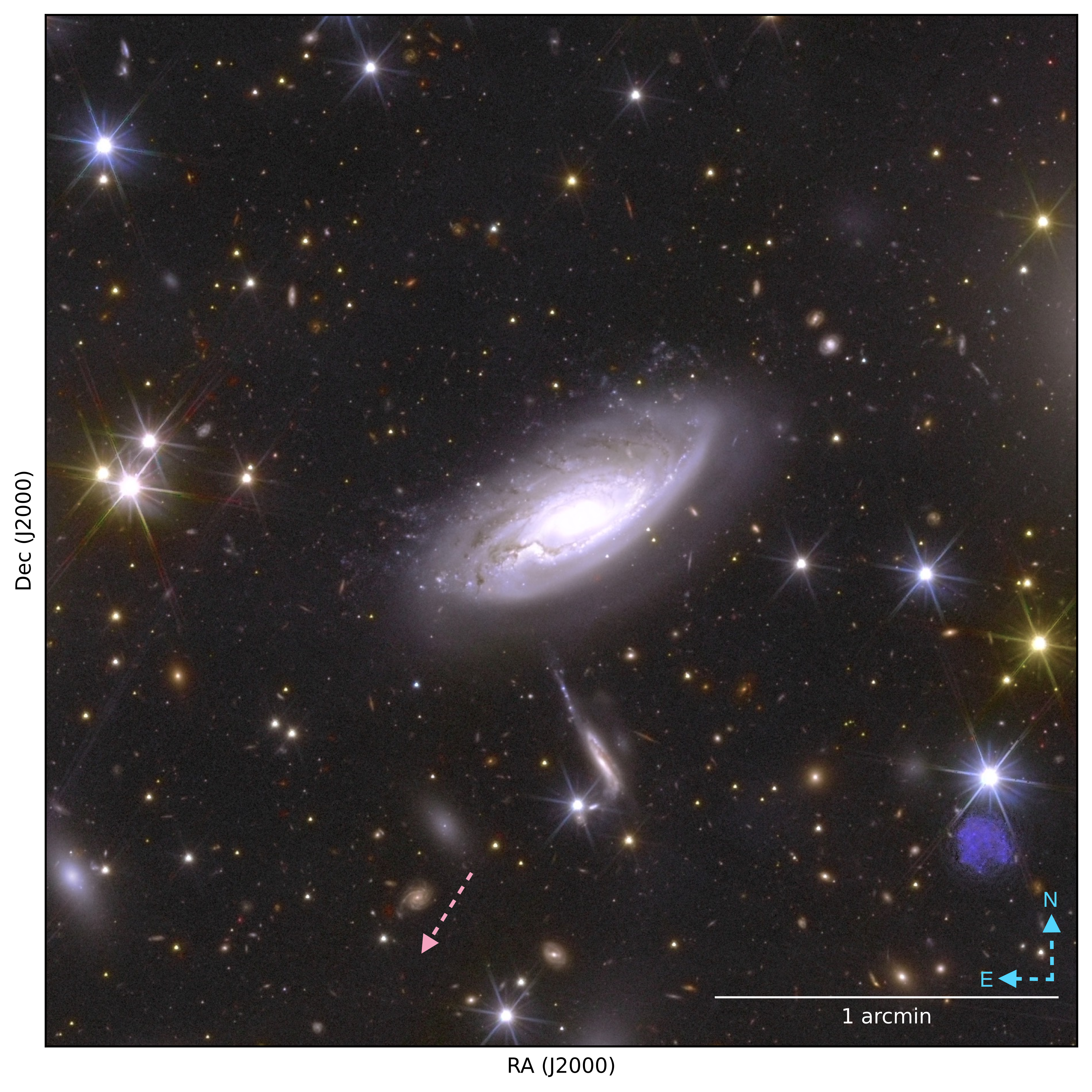}
\caption{Colour-composite image of the UGC 2665 galaxy created by combining and assigning blue, green, and red colours for \IE, \YE, and \HE imaging data. The direction to the cluster centre is shown with a light magenta-coloured arrow. In the bottom right of the image, the blue patch is an artifact caused by dichroic ghost in \IE imaging.}\label{figure:fig1}
\end{figure*}

\begin{figure*}[htbp!]
\centering
\includegraphics[angle=0,width=0.85\hsize]{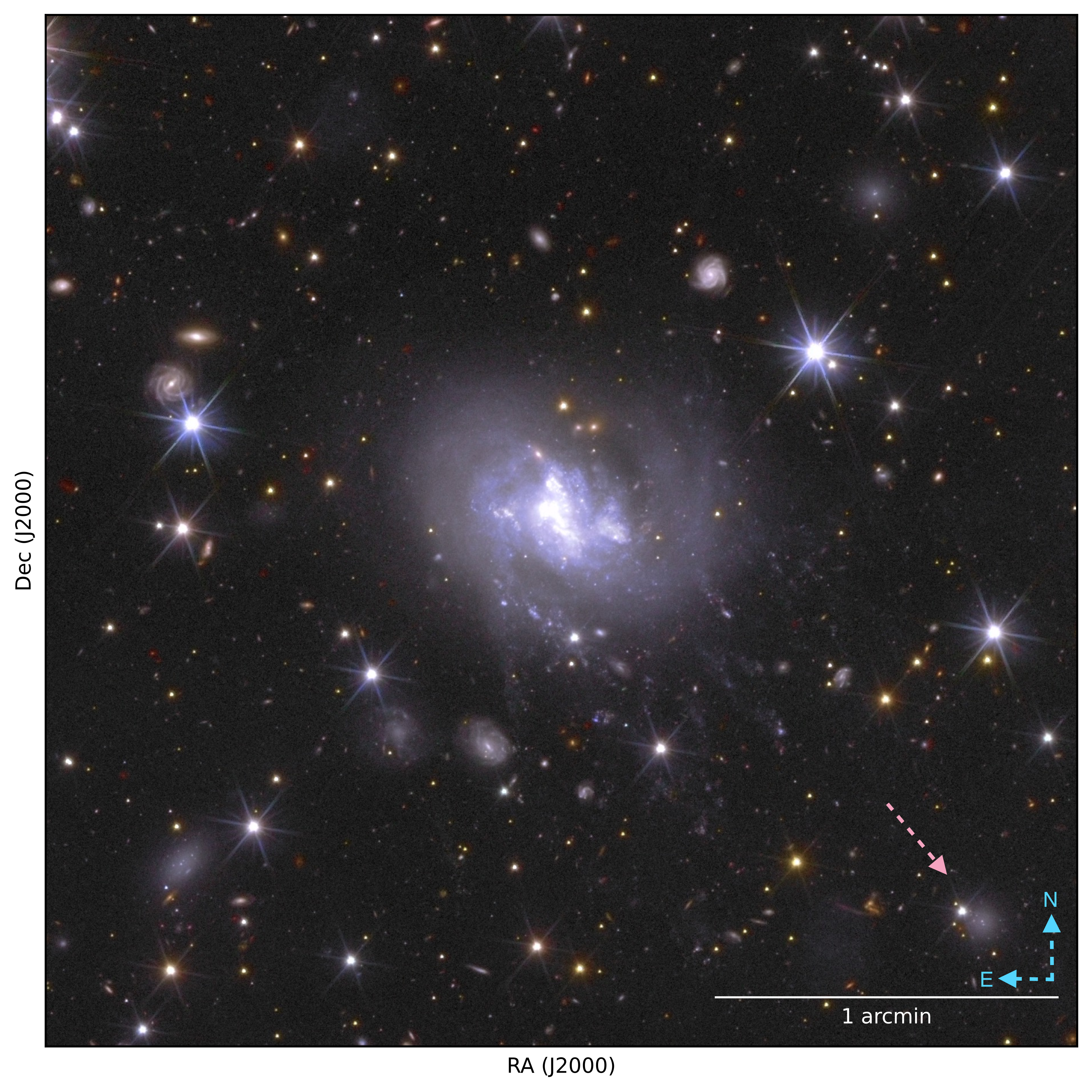}
\caption{Colour composite image of the MCG +07-07-070 galaxy. Details are same as in Fig.~\ref{figure:fig1}.}\label{figure:fig2}
\end{figure*}

A colour composite image of UGC 2665 and MCG +07-07-070 made from \IE, \YE, and \HE band is given in Figs.~\ref{figure:fig1} and ~\ref{figure:fig2}.

\begin{figure*}[htbp!]
\centering
\includegraphics[width=1\textwidth]{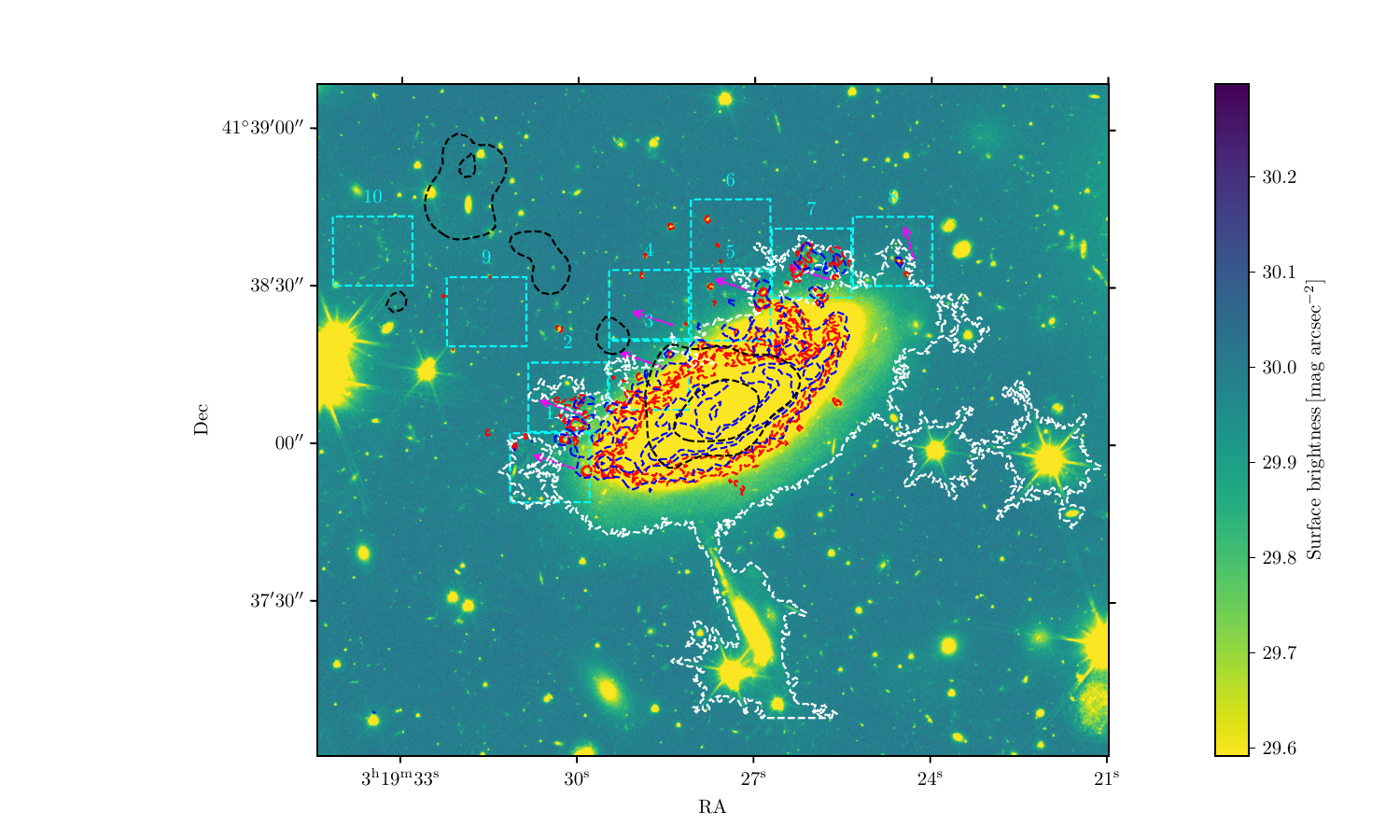}
\caption{Colour scale \IE image of galaxy UGC 2665 with the scaling set to highlight faint stripped features at the galaxy outskirts. The features that are likely part of the stripped tail are marked with cyan boxes. Coloured contours overlaid on the image are for NUV N245M imaging data of the galaxy in blue, H$\alpha$ in red and $144\,{\rm MHz}$ radio continuum in black. The NUV and H$\alpha$ contour levels created for $2\,\sigma$, $4\,\sigma$ and $6\,\sigma$ are shown. The contour levels created for $3\,\sigma$, $6\,\sigma$ and $12\,\sigma$ from LOFAR image are shown. The direction of the stripping feature is marked with magenta-coloured arrows. White contours show isophote generated for surface brightness level of $30\,{\rm mag}\,{\rm arcsec}^{-2}$.}
\label{figure:fig3}
\end{figure*}

\begin{figure*}[htbp!]
\centering
\includegraphics[width=1.0\textwidth]{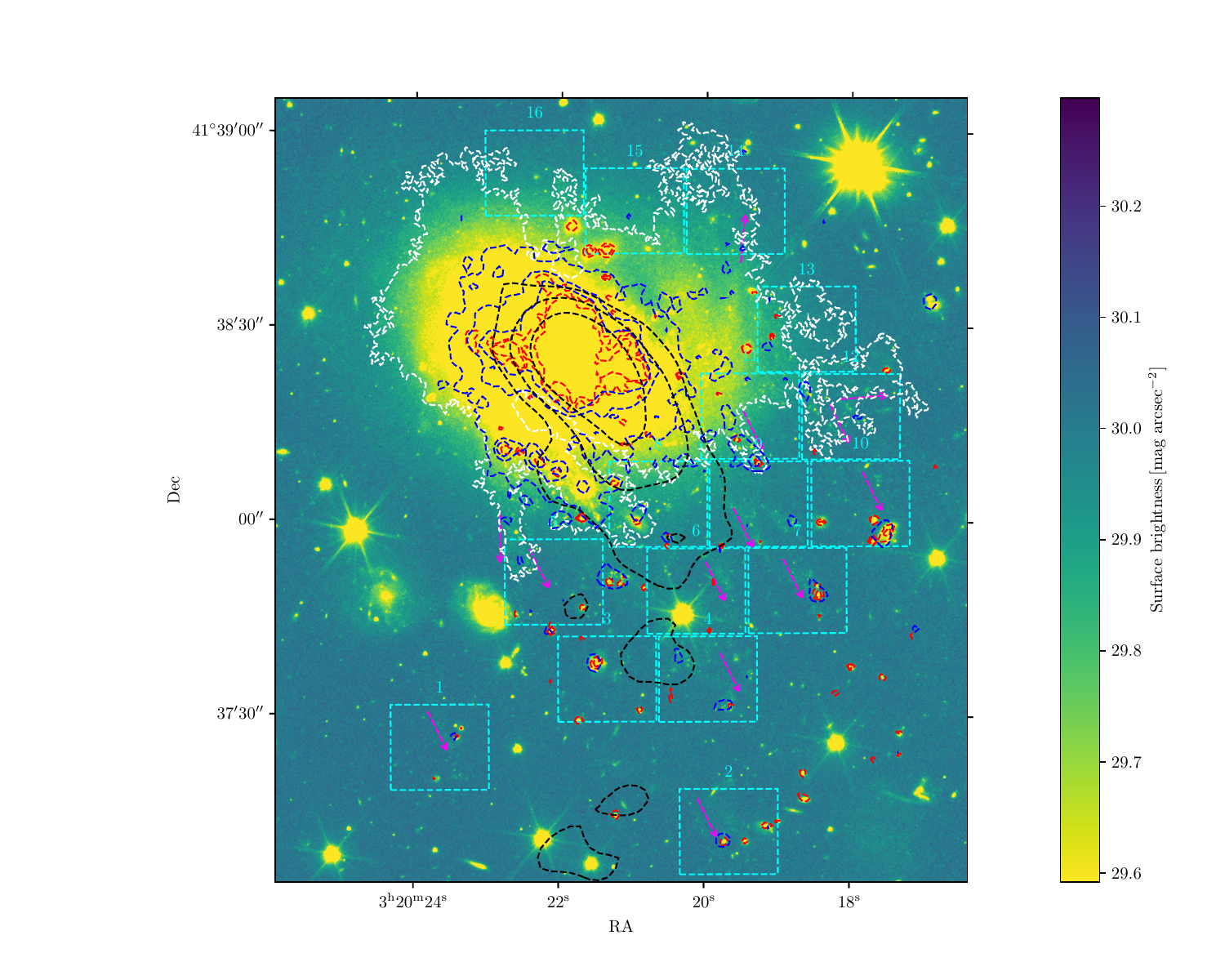}
\caption{Colour scale \IE image of galaxy MCG +07-07-070 with the scaling set to highlight faint stripped features at the galaxy outskirts. Details are same as in Fig.~\ref{figure:fig3}.} 
\label{figure:fig4}
\end{figure*}

\begin{figure*}[htbp!]
\centering
\includegraphics[width=1.01\textwidth]{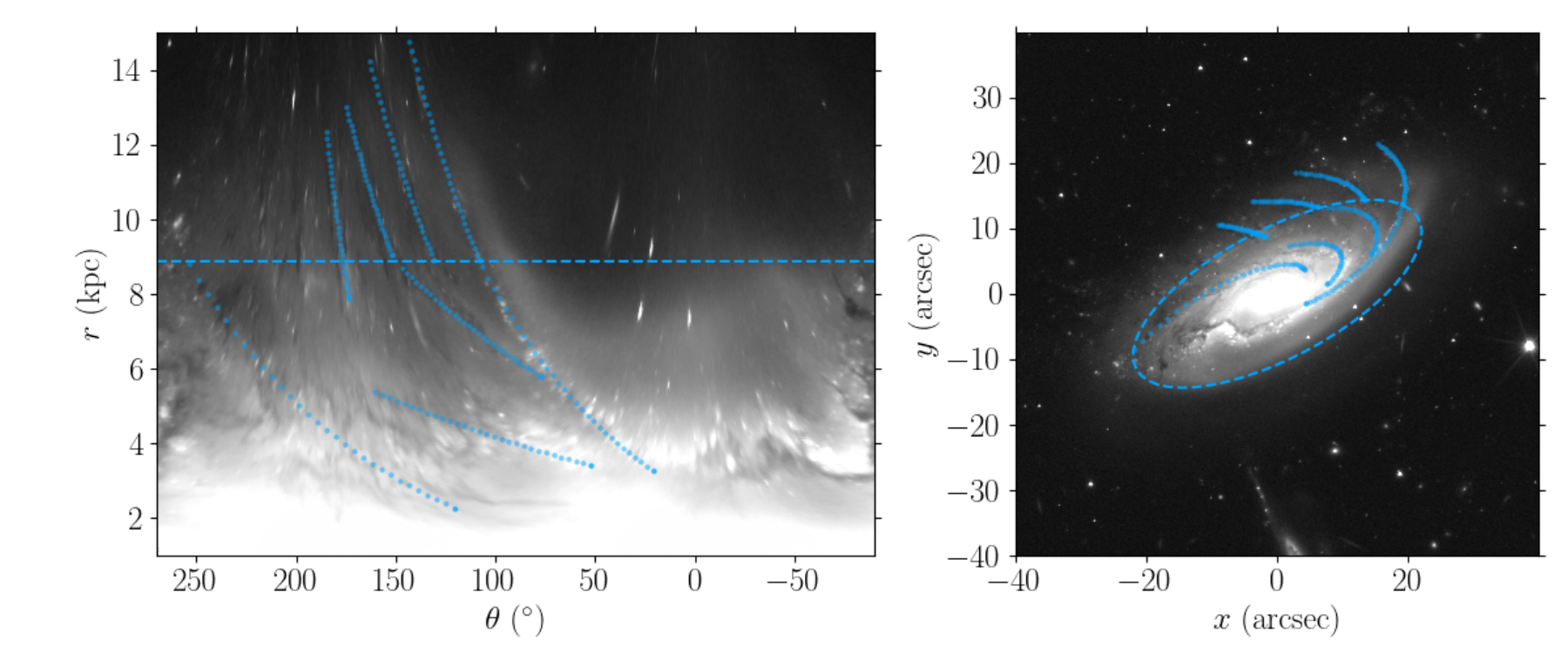}
\caption{The left panel shows the \IE image of UGC 2665 galaxy “unwrapped” in polar coordinates in terms of radial distance
from the centre of the galaxy ($r$) and azimuthal angle around the disc ($\theta$). The right panel shows the original image in greyscale. Logarithmic spiral arms have been drawn on the left-hand panel on the prominent dust lanes in cyan colour crosses. Spiral arms are shown projected back onto the galaxy disc on the right panel.} \label{figure:fig5}
\end{figure*}

\begin{figure*}[htbp!]
\centering
\includegraphics[width=1\textwidth]{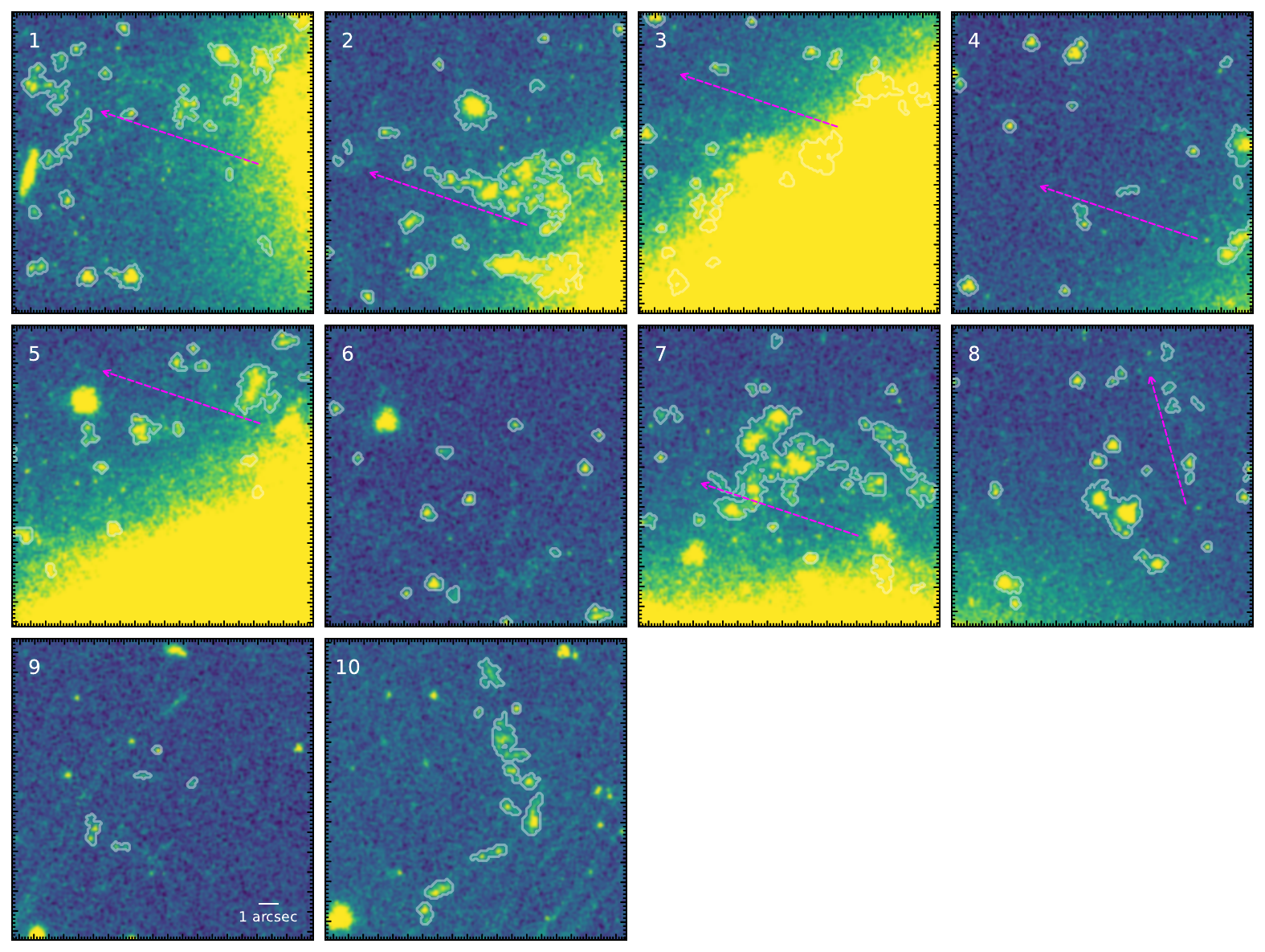}
\caption{Zoom in on the \IE imaging of stripped features for UGC 2665 galaxy. These correspond to the boxes marked in Fig. \ref{figure:fig3}. Each box has a size of $5.1\,{\rm kpc}$ $\times$ $4.4\,{\rm kpc}$ with the arcsec bar shown corresponding to $338\,{\rm pc}$ at cluster frame. The details of markers are as in Fig. \ref{figure:fig3}. Contours corresponding to the boundaries of the segments detected from \IE image in section 3.3 are overlaid.} \label{figure:fig6}
\end{figure*}

\begin{figure*}[htbp!]
\centering
\includegraphics[width=1\textwidth]{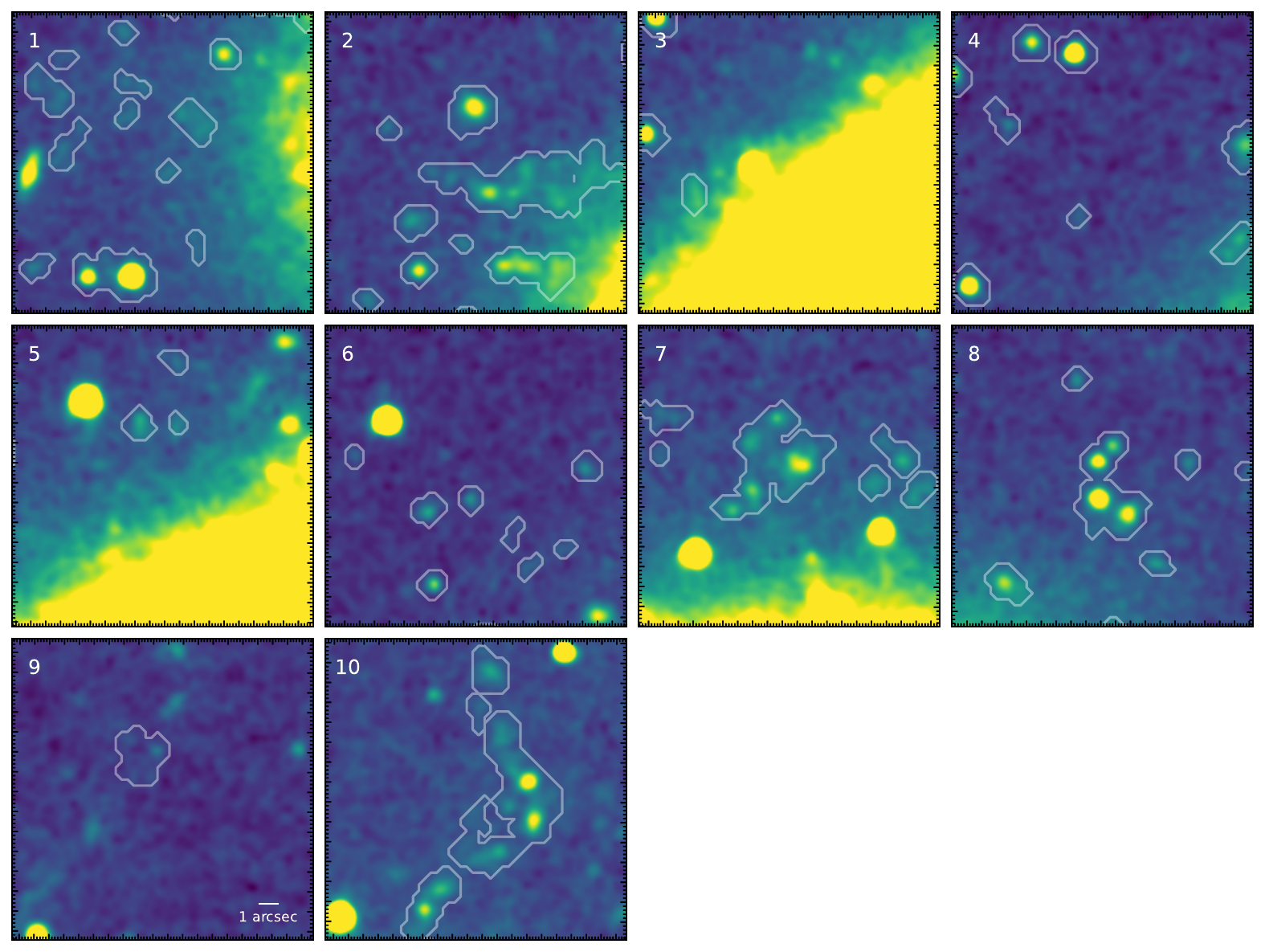}
\caption{Zoom in on the \YE band imaging of stripped features for UGC 2665 galaxy. These correspond to the boxes marked in Fig. \ref{figure:fig3}. Each box has a size of $5.1\,{\rm kpc}$ $\times$ $4.4\,{\rm kpc}$ with the arcsec bar shown corresponding to $338\,{\rm pc}$ at cluster frame. The details of markers are the same as in Fig. \ref{figure:fig3}. Contours corresponding to the boundaries of the segments detected from \YE image in section 3.3 are overlaid.} \label{figure:fig7}
\end{figure*}

\begin{figure*}[htbp!]
\centering
\includegraphics[width=1\textwidth]{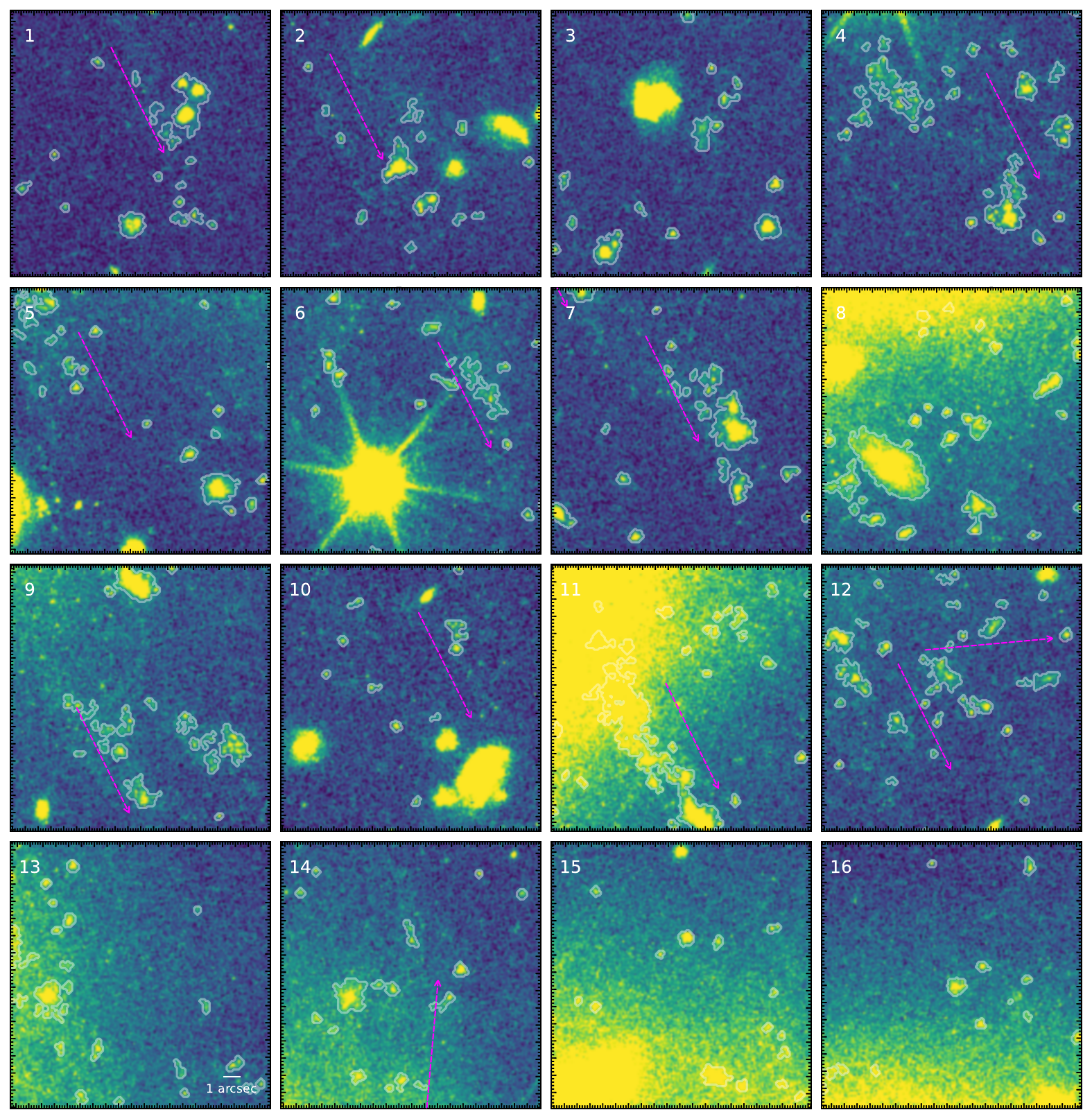}
\caption{Zoom in on the \IE imaging of stripped features for MCG +07-07-070. These correspond to the boxes marked in Fig. \ref{figure:fig4}. Each box is having a size of $5.1\,{\rm kpc}$ $\times$ $4.4\,{\rm kpc}$ with the arcsec bar shown corresponding to $338\,{\rm pc}$ at cluster frame. The details of markers are the same as in Fig. \ref{figure:fig4}. Contours corresponding to the boundaries of the segments detected from \IE image in section 3.3 are overlaid.} \label{figure:fig8}
\end{figure*}

\begin{figure*}[htbp!]
\centering
\includegraphics[width=1\textwidth]{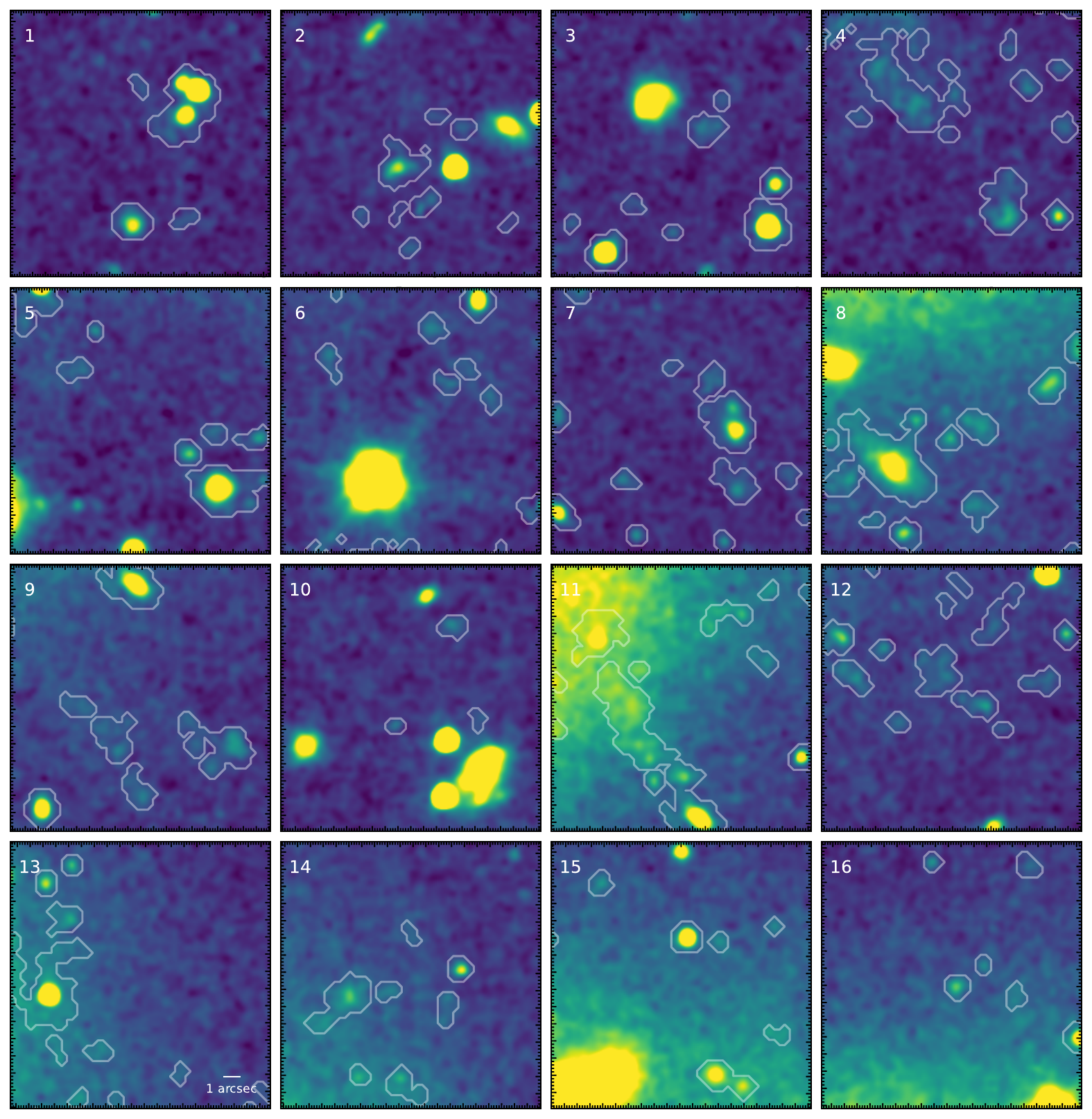}
\caption{Zoom in on the \YE band imaging of stripped features for MCG +07-07-070. These correspond to the boxes marked in Fig. \ref{figure:fig4}. Each box has a size of $5.1\,{\rm kpc}$ $\times$ $4.4\,{\rm kpc}$ with the arcsec bar shown corresponding to $338\,{\rm pc}$ at cluster frame. The details of markers are the same as in Fig. \ref{figure:fig4}. Contours corresponding to the boundaries of the segments detected from \YE image in section 3.3 are overlaid.} \label{figure:fig9}
\end{figure*}

\begin{table*}[htbp!]
\caption{Details on the two galaxies undergoing RPS in the Perseus cluster. Spectroscopic redshift of the  galaxies, stellar mass ($M_\star$) and half-light radius ($R_{\rm e}$) from \citet{Cuillandre_2024a}. Projected distance from the cluster centre ($D_{\rm centre}$) and velocity offset of galaxies from \citet{Roberts_2022}.}
\begin{center}
\smallskip
\label{table:T1}
\smallskip
\begin{tabular}{lcccccccc} %23
\hline
 Galaxy           & RA(J2000)  & Dec(J2000)  & Redshift & Stellar mass &  $R_{\rm e}$  & $D_{\rm centre}$ & $D_{\rm centre}$ & Velocity\\
 name             & h:m:s  & \degree:\arcmin:\arcsec & $z$ & $M_\odot$ &  kpc & kpc & $r_{200}$ & km s$^{-1}$\\
\hline
UGC 2665 & 03:19:27.37 &	$+$41:38:07.10  & 0.0258 & 1.78 $\times$ 10$^{10}$ &   3.94  & 183 & 0.10 & 2441 \\
MCG +07-07-070 & 03:20:22.02 & $+$41:38:26.60 & 0.0130 & 2.14 $\times$ 10$^{9}$  & 4.80  & 220 & 0.12 & 1618\\
\hline
\end{tabular}
\end{center}
\end{table*}

\section{Results}

\subsection{Identifying the dominant perturbing mechanism: Gravitational or ram-pressure stripping}

Galaxies infalling into rich clusters can undergo gravitational interactions with the cluster potential as well as with other galaxies in the immediate vicinity. The gravitational interactions can create tidal features that form tails, shells, and plumes around the galaxies (see galaxy images in \citealt{Bilek_2020,Bilek_2022}). These features, however, are normally very diffuse, with low surface brightness, formed from the stars that are tidally pulled out of the galaxies over a large region. On the other hand, RPS is a purely hydrodynamic process in which the gas is first stripped, sometimes appearing as filamentary cometary structures with or without clumpy and compact regions, where new episodes of star formation can occur (see \citealt{Poggianti_2016,Poggianti_2019,Durret_2021}). We examine the primary mechanism responsible for the features observed in the case of two galaxies, using imaging observations from \textit{Euclid} and the associated data products generated as part of ERO observations of the Perseus cluster.

\subsubsection{Gravitational interaction with the whole cluster}

The galaxy cluster can exert gravitational perturbations on infalling galaxies. The constituent matter of a radially infalling galaxy is subjected to the internal acceleration of the galaxy itself ($a_{\rm gal}$), and the two components of the acceleration from galaxy cluster potential: the radial acceleration ($a_{\rm rad\,cluster}$), the gradient of which tends to enhance the elongation of a galaxy along the galaxy-galaxy cluster direction, and the generally much weaker transverse acceleration ($a_{\rm trans\,cluster}$), the gradient of which tends to contract a galaxy  in the perpendicular directions \citep{Henriksen_1996}. When the cluster $a_{\rm rad\,cluster}$ overcomes the $a_{\rm gal}$, the perturbation can remove matter from the galactic disc. We compute the radial acceleration $a_{\rm rad\,cluster}$ exerted by the cluster
 on the galaxies using the following equation \citep{Henriksen_1996}.

\be
a_{\rm rad\,cluster}=M(r)\;G\;[1/r^{2} - 1/(R+r)^{2}]\;
\ee

The mass of the cluster within the radius $r$, $M(r)$, where $r$ is the distance of the galaxy from the cluster centre, is computed using a concentration parameter $c$ = 6 and a Navarro–Frenk–White radial density profile \citep{Navarro_1997}, as described in Eq. 9 of \citet{Boselli_2022}. $G$ is the gravitational constant. We adopt the position of NGC 1275, the central type-D giant elliptical galaxy, as the centre of the Perseus cluster \citep{Roberts_2022}. UGC 2665 is observed to be located at a projected distance of $183\,{\rm kpc}$ from the centre of the cluster (0.10\,$r_{200}$) with a velocity offset $\sim$ 2441 km s$^{-1}$ from the Perseus cluster redshift. MCG +07-07-070 is also found very close to the cluster centre, with a projected distance of $\sim$ $220\,{\rm kpc}$ (0.12\,$r_{200}$) and a velocity offset with respect to the cluster $\sim$ 1618 km s$^{-1}$ \citep{Roberts_2022}. The proximity of both galaxies to the cluster centre, along with their high-velocity offsets relative to the cluster, suggests that they were recently accreted and are likely experiencing their first infall. Table\,\ref{table:T1} gives details on the two galaxies in the Perseus cluster taken from \citet{Cuillandre_2024a}. We used the distance from the cluster centre ($r$), the effective radius ($R_{\rm e}$) given in Table\,\ref{table:T1} and the dynamical mass of the Perseus cluster at location of galaxy $M(r)$ to compute the $a_{\rm rad\,cluster}$ using Eq. (1). \\

We compute $a_{\rm gal}$ of the galaxy using the following equation.

\be
a_{\rm gal}=G\;m_{\rm dyn}/R^{2}\;
\ee

The dynamical mass of the galaxy ($m_{\rm dyn}$) is estimated using the $M_\star$ values given in Table\,\ref{table:T1}. We used the method  described in \citet{Behroozi_2013} where the halo mass versus stellar mass/halo mass relation can be used to compute the median stellar mass for a given halo mass. We note that the inverse of which does not give the average halo mass for a given stellar mass because of the scatter in the relation. Our goal is to get a rough estimate of the $m_{\rm dyn}$ of the galaxy from the $M_\star$. We therefore computed the $m_{\rm dyn}$ by considering the 0.01 dex scatter on the $M_{\rm \star}$/$m_{\rm dyn}$ and put a lower and upper limit for the $m_{\rm dyn}$. Within these limits we discuss whether the gravitational effects dominate over hydrodynamic effects on these galaxies. The UGC 2665 has $m_{\rm dyn}$ $\sim$ 40\,$M_\star$ (lower limit $\sim$ 29\,$M_\star$,upper limit $\sim$ 67 \,$M_\star$), and the MCG +07-07-070 has $\sim$ 12.5\,$M_\star$ (lower limit $\sim$ 11\,$M_\star$,upper limit $\sim$ 14 \,$M_\star$).

When $a_{\rm rad\,cluster}$/$a_{\rm gal}$ $>$ 1, the cluster potential can create gravitational perturbations able to remove matter from the infalling galaxy. The computed $a_{\rm rad\,cluster}$/$a_{\rm gal}$ values for both galaxies are given in Table\,\ref{table:T2}. We should consider these values as upper limits, as we used projected distances from the centre of the cluster to compute $a_{\rm rad\,cluster}$. Both galaxies are found to have $a_{\rm gal}$ greater than $a_{\rm rad\,cluster}$. The cluster potential may have a greater effect on MCG +07-07-070, which has a slightly lower mass than UGC 2665. The truncation radius ($R_{\rm trunc}$), beyond which matter can be removed by the cluster potential, is computed for both galaxies  as given in \citet{Binney_2008}. 

\be
R_{\rm trunc}=r\;[m_{\rm dyn}/M(r)]^{1/3}\;
\ee
where $r$, $M(r)$ from Eq. (1) and $m_{\rm dyn}$ from Eq. (2). The truncation radius is $\sim$ $37\,{\rm kpc}$ for UGC 2665, while it is $\sim$ $14\,{\rm kpc}$ for MCG +07-07-070. The truncation radius might range in between $33\,{\rm kpc}$ < $R_{\rm trunc}$ < $44\,{\rm kpc}$ for UGC 2665 and $13\,{\rm kpc}$ < $R_{\rm trunc}$ < $14\,{\rm kpc}$ for MCG +07-07-070 considering the dispersion on the $m_{\rm dyn}$. The isophotal radius measured at $26\,{\rm mag}\,{\rm arcsec}^{-2}$ from the \IE image is $13.64\pm0.01\,{\rm kpc}$ for UGC 2665 and $13.12\pm0.01\,{\rm kpc}$ for MCG+07-07-070 \citep{Cuillandre_2024a}. The truncation radius values are larger than the isophotal radius of the galaxies (comparable to MCG+07-07-070 lower limit), implying that it is not possible for the cluster potential to pull out material from the galaxy.

\subsubsection{Gravitational interaction with nearby companions}

We now search in NASA/IPAC Extragalactic Database (NED)\footnote{https://ned.ipac.caltech.edu/} for any nearby galaxy that could have gravitationally interacted with these two objects. We search within an area covered by a circle with a 20 arcminute ($\sim$ $405\,{\rm kpc}$) radius centered on the galaxy and a velocity separation of 2000 km s$^{-1}$ ($\sim$ 2 $\times$ velocity dispersion of Perseus). The galaxies are having spectroscopic redshift information from the catalog of \citet{Kang_2024} down to $r$-band apparent  magnitude ($r_{Petro,0}$) $\sim$ 20.5. Stellar masses of the galaxies are taken from \citet{Cuillandre_2024a}. In the case of galaxy UGC 2665 ($M_\star$= 1.78 $\times$ 10$^{10}$ $M_\odot$), 67 galaxies meet these criteria, with the nearest two galaxies being WISEA J031917.76+413839.6 ($M_\star$ = 3.64 $M_\odot$ $\times$ 10$^{10}$ $M_\odot$) located at a projected distance of $\sim$ $40\,{\rm kpc}$ to the west with a velocity difference of $\delta v$ = 1577\,km s$^{-1}$ and WISEA J031937.46+413758.3 ($M_\star$ = 1.48 $M_\odot$ $\times$ 10$^{10}$ $M_\odot$), located at a projected distance of $\sim$ $40\,{\rm kpc}$ to the east with a velocity difference of $\delta v$ = 831\,km s$^{-1}$. The two closest galaxies in velocity space are WISEA J031848.10+412622.8 ($\delta v$ = 3\,km s$^{-1}$) located at a projected distance of $\sim$ $281\,{\rm kpc}$ and PUDG R24 ($M_\star$ = 3.91 $\times$ 10$^{8}$ $M_\odot$), $\delta v$ = 30\,km s$^{-1}$), an ultra diffuse galaxy at a projected distance $\sim$ $288\,{\rm kpc}$ \citep{Gannon_2022}. The radial acceleration exerted by the nearest object on the perturbed galaxy can be calculated using the same formalism discussed earlier in this section, as given in Eq. (1), but with the cluster mass replaced by the neighbouring galaxy mass and using the distance to the neighbour ($r$) and the effective radius of the galaxy ($R_{\rm e}$). We checked the $a_{\rm rad\,neighbour}$/$a_{\rm gal}$ for all the neighbouring galaxies within the search radius of 405 kpc centered on UGC 2665 and found that the highest value is 0.004 which is negligible.

There are 131 galaxies that meet this criteria for MCG +07-07-070 ($M_\star$= 2.14 $\times$ 10$^{9}$ $M_\odot$) with the nearest galaxy NGC 1281 ($M_\star$ = 5 $\times$ 10$^{10}$ $M_\odot$) located at a projected distance of $\sim$ $61\,{\rm kpc}$ with a velocity difference $\delta v$ = 315.0\,km s$^{-1}$. The closest galaxy in velocity space is WISEA J031943.81+412725.1 ($M_\star$ = 1.26 $\times$ 10$^{10}$ $M_\odot$, $\delta v$ = 2\,km s$^{-1}$), located at a projected distance of $\sim$ $276\,{\rm kpc}$. We checked the $a_{\rm rad\,neighbour}$/$a_{\rm gal}$ for all the neighbouring galaxies within the search radius of 405 kpc centered on MCG +07-07-070 and found that the highest value is 0.09. The values are much smaller when considering the nearest galaxy in velocity space. Additionally, we note that a high-velocity encounter with another nearby galaxy can only remove mass from a galaxy when the distance between them is shorter than the typical length of the tidal tail and the galaxy's mass is $\geq$ 1.33 $\times$ dynamical mass of the main galaxy \citep{Boselli_2022}. We could not detect any massive galaxies at the distances of the tail length to both galaxies. Given the high-velocity dispersion (1040\,km s$^{-1}$) of the cluster and the relatively high velocity of the galaxies, it is unlikely that the galaxies are undergoing gravitational interactions with other galaxies at the cluster centre that could lead to the observed features. Another possibility is that interactions with a group of galaxies can collectively perturb the galaxies. In velocity space, we searched for galaxies within a 20 arcmin radius centered on both galaxies, but we did not find any grouped systems that could have interacted with the galaxy. However, as discussed in \citet{Boselli_2023}, gravitational interactions between the galaxies could have occurred in the past, when they were falling into the cluster outskirts, potentially perturbing the stars and the ISM. Galaxies could have undergone preprocessing in groups and got dispersed while falling into the cluster \citep{Dressler_2004}. This can then help strip the gas efficiently as the galaxy moves to the denser regions of the ICM. We recall here that the ram-pressure scales with the square of the velocity of the galaxy times the density of the ICM. We cannot entirely dismiss the possibility of gravitational interactions from any of these neighbouring galaxies, and galaxy groups in the past, as well as from the faint nearby galaxies with no confirmed spectroscopic redshifts. The gravitational perturbations caused by other companions can combine with those due to the cluster potential well (galaxy harassment), making the perturbing process much more efficient \citep{Moore_1996}. There can be galaxies in the Perseus cluster that may be detected in \textit{Euclid} imaging but are not listed in NED due to being beyond the magnitude limits of available spectroscopic surveys. Such galaxies can also exert gravitational influence. In the case of UGC 2665, this is relevant, since the galaxy that appears to be interacting from the southern direction does not have confirmed redshift information.

We now analyze the \IE imaging data to identify signatures of gravitational interactions, such as tidal tails, shells, and plumes around the galaxies, as well as asymmetries on the galaxy disc. The colour composite images of the galaxies shown in Figs. \ref{figure:fig1} and \ref{figure:fig2} show several interesting details. The southern edge of the disc of UGC 2665 is consistent with a smooth stellar halo. In the case of MCG +07-07-070, there is diffuse emission extended along the end of the spiral arms. Isophotal analysis was performed on the \IE images of the galaxies down to surface brightness levels of $30\,{\rm mag}\,{\rm arcsec}^{-2}$. The isophotes generated for these levels are overlaid on the galaxy images in Figs. \ref{figure:fig3} and \ref{figure:fig4}. The galaxy images are generally symmetric, suggesting that the underlying population that contributes to the bulk of the galaxy's stellar mass is not perturbed. The contamination from neighbouring stars and galaxies creates artifacts that appear as connections in the surface brightness levels. In the case of UGC 2665, there is a connection with the galaxy in the south as shown in Fig. \ref{figure:fig3}. There are features emanating from the disc of the galaxies likely due to RPS but there are no faint diffuse extended features expected from tidal interactions. Deep optical imaging is expected to reveal features from any gravitational interactions in the form of isophotal asymmetries for the galaxy disc 
 \citep{Duc_2015,Boselli_2023}. We note these features can be present below the surface brightness limits, as simulations suggest the presence of such features at very low surface brightness of $33\,{\rm mag}\,{\rm arcsec}^{-2}$ \citep{Mancillas_2019}. However, observations of a large sample of galaxies gathered with MegaCam at the CFHT demonstrate that tidal features are rare at a surface brightness limit below $27.5\,{\rm mag}\,{\rm arcsec}^{-2}$ \citep{Sola_2022}. The asymmetric structures seen in the images are mainly filamentary and clumpy with a cometary shape.
 As shown in Figs. \ref{figure:fig3} and \ref{figure:fig4}, these features have gradual surface brightness variation and appear to be escaping from the stellar discs detected with a limiting surface brightness level of $30\,{\rm mag}\,{\rm arcsec}^{-2}$ in the \IE images.

\subsubsection{Ram pressure stripping}

RPS occurs when the hydrodynamic pressure exerted on the cold ISM of an infalling galaxy moving at a velocity ($V$) into a galaxy cluster with an ICM density ($\rho_{\rm ICM}$) surpasses the internal gravitational force holding the cold ISM to the galaxy disc. This can be expressed as;

\be
P_{\rm ram} = \rho_{\rm ICM}\;V{^2} \geq 2 \; \pi\; G \;\Sigma_{\rm star}\;\Sigma_{\rm gas} \; 
\ee
where $\Sigma_{\rm star}$ and $\Sigma_{\rm gas}$ represent the stellar and gas mass surface density. Following the mathematical formalism given in \citet{Boselli_2022} we compute $\rho_{\rm ICM}$ for the region around the two galaxies in the Perseus cluster. We used the radial variation of electron density ($n_{\rm e}$) in the Perseus cluster from \citet{Churazov_2003} to estimate the (projected) value at the galaxy location ($n_{\rm e} =$ 10$^{-2}$ cm$^{-3}$) within the cluster. $\Sigma_{\rm star}$ and $\Sigma_{\rm gas}$ are computed for both galaxies within their effective radii using Eq. (20) of \citet{Boselli_2022}.

Since we do not have an estimate of the measured gas mass for the two galaxies, we used Eq. 25 from \citet{Boselli_2022} to compute the total gas mass for each galaxy from the given stellar mass, using the luminosity-dependent $X_{\rm CO}$ conversion factor provided in \citet{Boselli_2022}. The computed values of the ram pressure and the gravitational anchoring forces for two galaxies are given in Table\,\ref{table:T2}. It is clear from Table\,\ref{table:T2} that ram pressure exceeds the gravitational anchoring of gas for both galaxies (UGC 2665 by a factor $\sim$ 61 and for MCG +07-07-070  by a factor $\sim$ 430). We caution that the location of the galaxies within the cluster (and hence the ICM density) is estimated based on the projected distance from the cluster centre. The ram-pressure computation provides an upper limit, which can decrease significantly depending on the true distance from the centre, where the ICM density is correspondingly lower.

\begin{table*}[htbp!]
\caption{Details on the ratio of radial acceleration due to cluster potential to the centripetal acceleration of galaxy ($a_{\rm rad\,clust}$/$a_{\rm gal}$), the ratio of radial acceleration due to neighbouring galaxy to the centripetal acceleration of galaxy ($a_{\rm rad\,neighbour}$/$a_{\rm gal}$) (with lower and upper limits given in brackets) and ram-pressure ($\rho_{\rm ICM} V{^2}$) and gravitation anchoring forces (2 $\pi$ $G$ $\Sigma_{\rm star} \Sigma_{\rm gas}$) for the two galaxies.}
\begin{center}
\smallskip
\label{table:T2}
\smallskip
\begin{tabular}{lccccc} %23
\hline
 Galaxy   & $a_{\rm rad\,cluster}$/$a_{\rm gal}$ & $a_{\rm rad\,neighbour}$/$a_{\rm gal}$         & $\rho_{\rm ICM} V{^2}$  & 2 $\pi$ $G$ $\Sigma_{\rm star} \Sigma_{\rm gas}$  \\
name     &  &        & dyn cm$^{-2}$  & dyn cm$^{-2}$ \\
\hline
UGC 2665 & 0.011 (0.007,0.015)  &    0.00008 (0.00005,0.00012)     &1.15 $\times$ 10$^{-9}$  &  1.90 $\times$ 10$^{-11}$ \\
MCG +07-07-070 & 0.40 (0.35,0.45) &   0.0014 (0.0012,0.0016)   &5.03 $\times$ 10$^{-10}$  &  1.17 $\times$ 10$^{-12}$\\
\hline
\end{tabular}
\end{center}
\end{table*}

\subsubsection{Multifrequency analysis}

The $2\,\sigma$, $4\,\sigma$ and $6\,\sigma$ contours \footnote{$\sigma$ is the Sky RMS} of NUV (blue) and H$\alpha$ (red) are overlaid on both galaxies in Figs. \ref{figure:fig3} and \ref{figure:fig4}. Both the NUV and H$\alpha$ emission are due to the presence of recent star formation and follow the filamentary features seen in \textit{Euclid} imaging. \citet{Roberts_2022} presents the LOFAR $144\,{\rm MHz}$ radio continuum imaging of the two galaxies, which reveals the presence of extended, cometary tails extending from the galaxy disc. This low-frequency, non-thermal radio continuum is from synchrotron emission given off by cosmic ray electrons that were accelerated by supernova within the galaxy disc. We overlay the $3\,\sigma$, $6\,\sigma$ and $12\,\sigma$ $144\,{\rm MHz}$ radio continuum contours (black) over the \IE imaging of the two galaxies in Figs. \ref{figure:fig3} and \ref{figure:fig4}. We note that the LOFAR imaging is at a lower spatial resolution of $\ang{;;6}$ compared to the near-infrared, optical, and UV imaging presented here. The features detected from \IE imaging of the two galaxies are almost co-aligned with the radio-continuum contours, with a slight displacement in the case of MCG +07-07-070. 
 
Contributions to radio continuum emission can come from physical mechanisms other than star formation \citep{vanWeeren_2019,Hardcastle_2020}. These include past and recent activity from an active galactic nucleus (AGN) at the galaxy centre, as well as radio relics from recent mergers found within the galaxy cluster. The presence of an AGN can produce jets and create lobe-like features detectable in LOFAR $144\,{\rm MHz}$ radio continuum imaging. According to the WHAN diagram \citep{CidFernandes_2011}, which combines the equivalent width (EW) of the H$\alpha$ emission line with the emission line ratio EW(\ion{N}{ii})/EW(H$\alpha$), UGC2665 is classified as having a weak AGN, with an energy budget insufficient to produce jet/lobe-like features \citep{Meusinger_2020}. Another possibility is radio relics from cluster mergers that can be detected through radio continuum imaging. The orientation and appearance of the radio continuum contours are more likely to be associated with galaxies than with the cluster. The radio continuum emission therefore is associated with the stripped tails of the galaxy and traces cosmic ray electrons accelerated by supernovae. Cosmic ray electrons in the stripped tails can undergo synchrotron aging, characterized by a steep spectral index that can be flattened by the presence of star formation \citep{Roberts_2022}. The displacement of radio continuum contours from the features seen in \IE imaging for MCG +07-07-070 can be explained by the detection of regions with steep spectral index in 144 MHz radio continuum imaging. The radio mini-halo detected around NGC 1275, the central cluster galaxy, at 144 MHz overlaps with UGC 2665, which can also contribute to radio continuum emission detected from the RPS tail of the galaxy \citep{Roberts_2022}.

The correspondence between the broadband optical emission from these features and the presence of a radio continuum tail strongly supports the RPS origin. First, the gas is stripped from the galaxy, forming extended tails where star formation occurs \textit{in situ}, visible as optical emission. The direction of the tails can indicate a ram-pressure origin, since galaxies falling into a cluster for the first time on radial orbits tend to have tails oriented opposite to their velocity vector. The orientation of the stripped tails can help in understanding the orbital dynamics of the galaxy within the cluster \citep{Smith_2022, Salinas_2024,George_2024}. However, this is complicated by the effect of galaxy rotation when moving at high speeds within the galaxy cluster environment.

\subsubsection{Ram-pressure stripping as the dominant perturbing mechanism}

In summary, we emphasize that the morphology of the features observed in \textit{Euclid} imaging of both galaxies does not display any diffuse tidal tail or shells, which would be expected in recent gravitational perturbations, down to the limiting surface brightness of $30.1\,{\rm mag}\,{\rm arcsec}^{-2}$ of \textit{Euclid} \IE (and similarly in the slightly shallower \YE,\JE,\HE imaging). The features have a filamentary and clumpy structure, with a cometary shape, which is expected in star formation occurring in the tails of galaxies undergoing RPS. These features are extended and have a low surface brightness connection to the galaxy’s disc. In the case of MCG +07-07-070, the one-sided features from the galaxy's disc resemble the fireballs detected in the stripped tails of galaxies in nearby clusters \citep{Yoshida_2008,Hester_2010,Yoshida_2012,Smith_2010,Kenney_2014,Jachym_2014,Jachym_2019,Giunchi_2023b}. The two galaxies have also extended radio continuum tails of cometary shape that are co-aligned with the features detected in \IE imaging. The radio continuum tails have the expected morphology of ram pressure stripping, where the stripped gas is displaced in the ICM, while the \textit{in situ} formed stars are not affected. This is particularly clear in the case of MCG +07-07-070, where the radio continuum detected tail is displaced from the features detected through \IE imaging. Galaxies experiencing their first infall should have stripped tails pointing away from the cluster centre, which can then be redirected towards the centre after a pericentre passage. We observe that the tail of UGC 2665 is oriented away from the cluster centre, whereas the tail of MCG +07-07-070 is oriented toward the cluster centre, as shown in Figs. \ref{figure:fig1} and \ref{figure:fig2}. As demonstrated in Table\,\ref{table:T2}, the location of both galaxies in the Perseus cluster is far enough from the centre to avoid gravitational perturbations from the cluster potential, and ram pressure dominates over the gravitational anchoring forces of the cold gas, which can potentially strip the gas. Theoretical arguments, supported by observational evidence from \textit{Euclid} and multi-wavelength data from UVIT, CFHT H$\alpha$, and LOFAR, suggest that the two galaxies are currently experiencing a RPS event. Still, the effect of gravitational interactions in the past with other galaxies cannot be fully ruled out. Such an interaction can perturb the gas content of the galaxy, loosening the tightly bound disc gas, thereby making the RPS process more efficient \citep{Cortese_2021}.

\begin{figure*}[h]
\centering
   \subfloat[]{\includegraphics[width=0.4\textwidth]{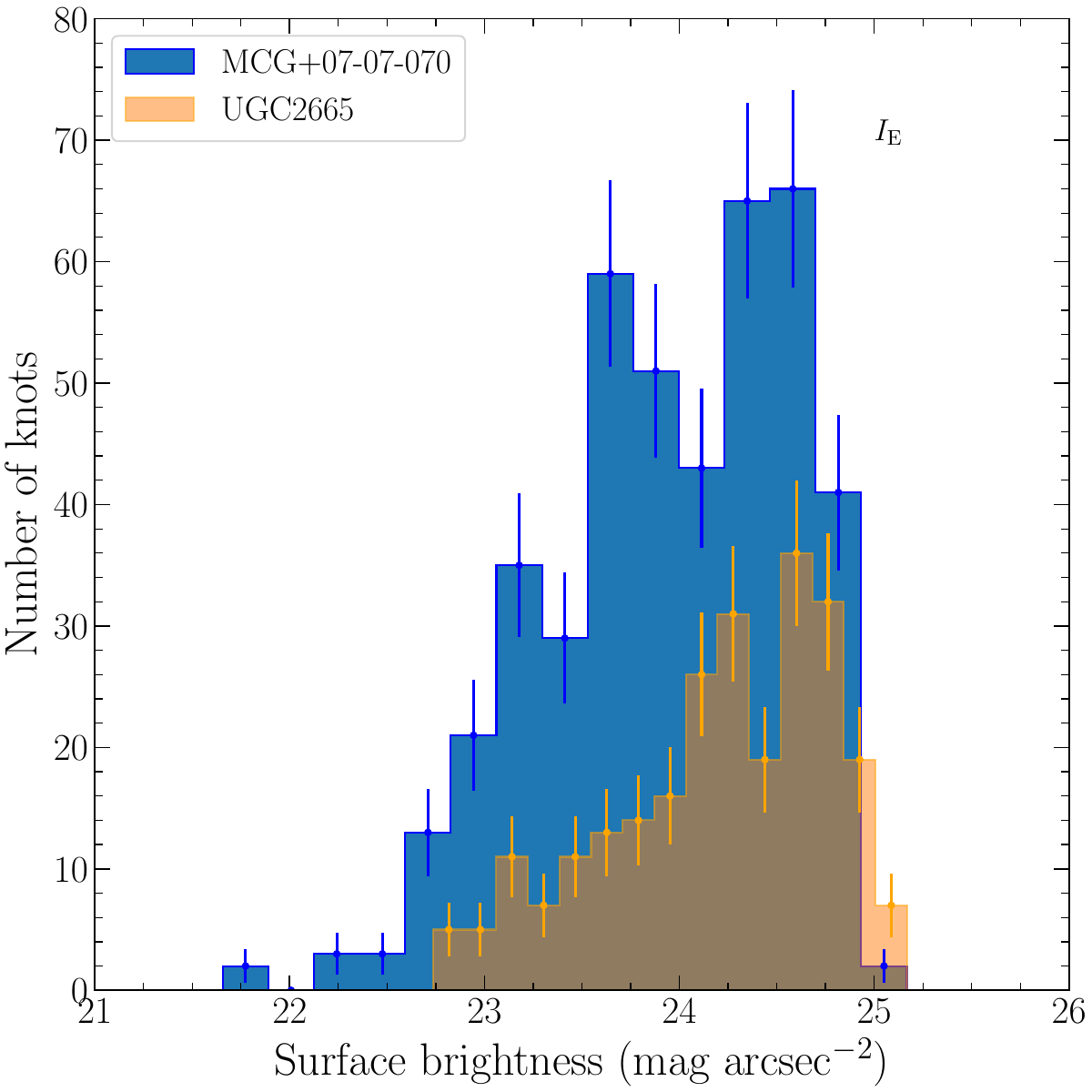}}
  \hspace{0.1cm}
  \subfloat[]{\includegraphics[width=0.4\textwidth]{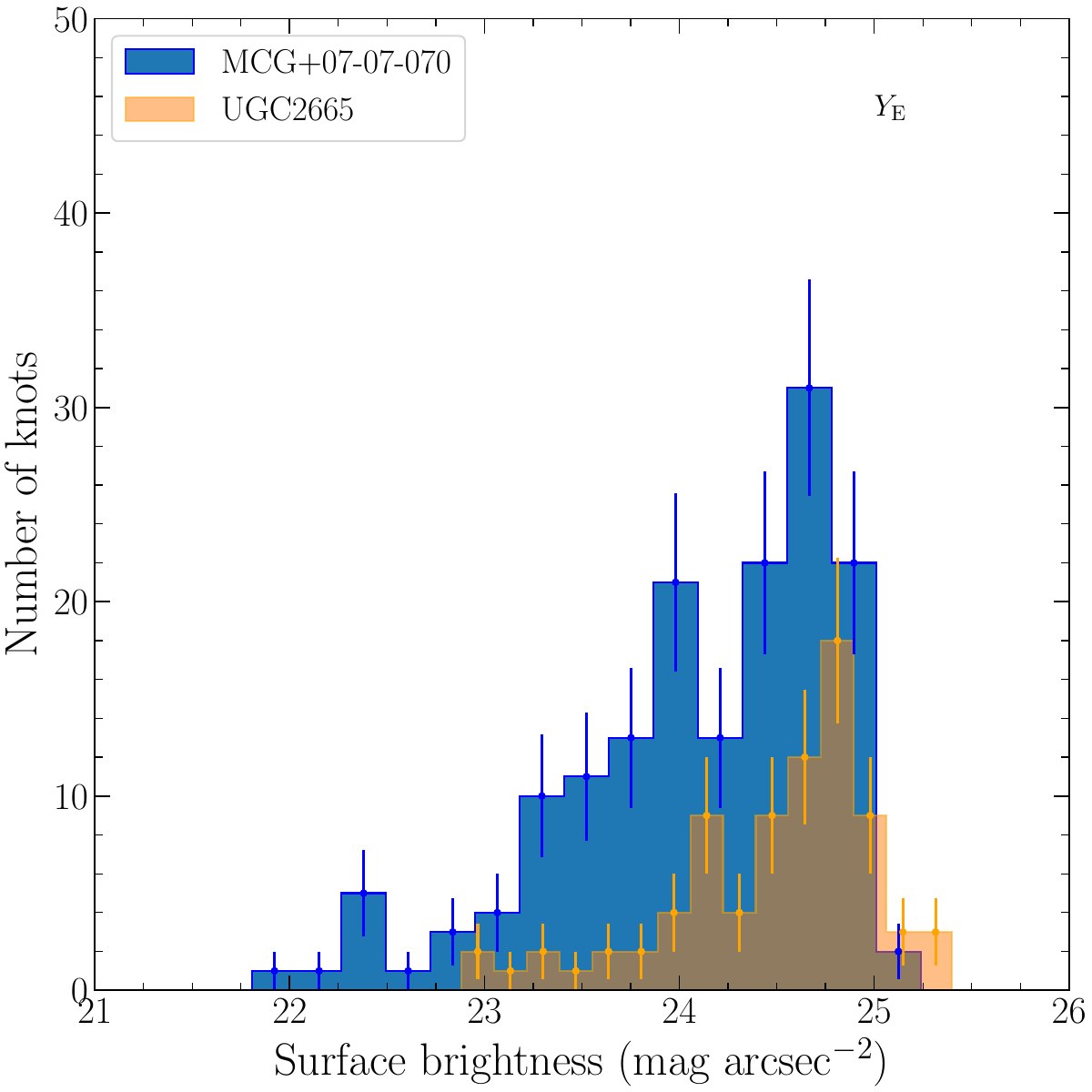}}
  \hspace{0.1cm}
    \subfloat[]{\includegraphics[width=0.4\textwidth]{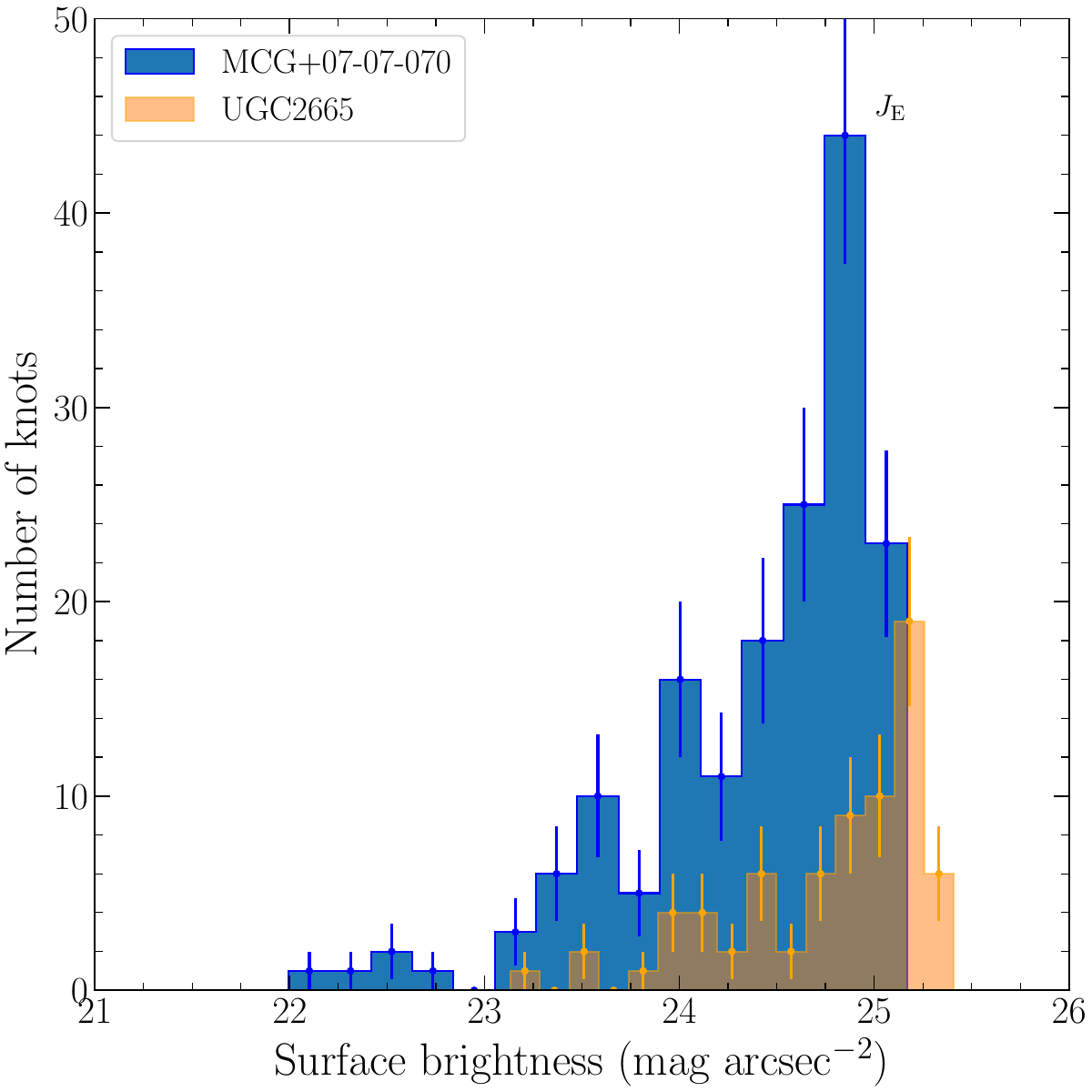}}
  \hspace{0.1cm}
    \subfloat[]{\includegraphics[width=0.4\textwidth]{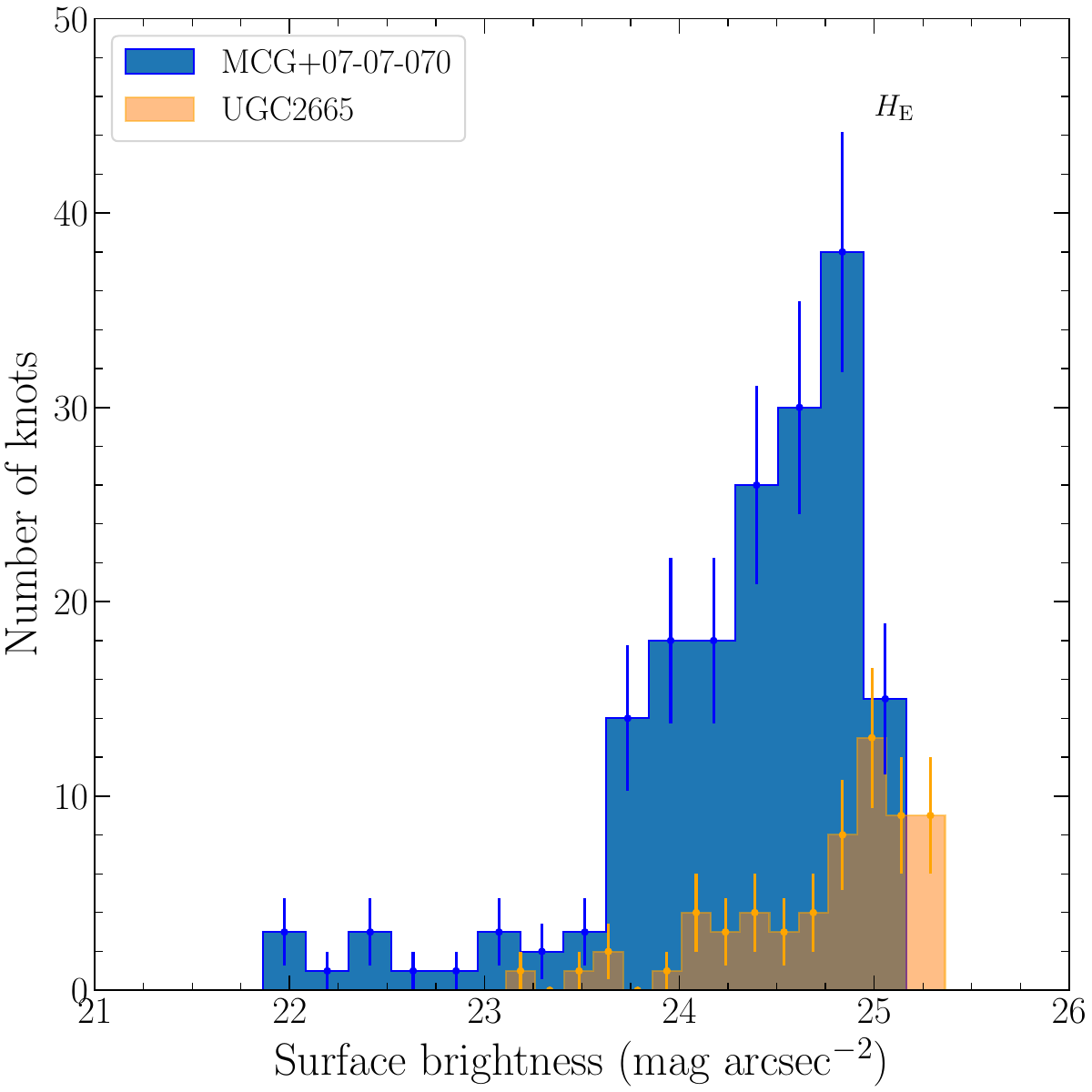}}
  \hspace{0.1cm}
   \caption{Distribution of surface brightness of segments from the stripped tails detected from \IE,\YE,\JE,\HE images of UGC 2665 and MCG$+$07$-$07$-$070. Poisson errors are shown with an error bar.
   } \label{figure:fig10}
\end{figure*}

\subsection{Ram-pressure stripping in optical and near-infrared imaging}

Having established the ram-pressure origin of the features seen in the \textit{Euclid} imaging data of the two galaxies, we now conduct a combined analysis of the stripped tails of the galaxies using both optical (\IE) and near-infrared (\YE) imaging data. The colour composite images of the galaxies, created from optical and infrared imaging shown in Figs.~\ref{figure:fig1} and ~\ref{figure:fig2}, display filamentary dusty and stellar structures escaping from their stellar discs, similar to those observed in a few cluster objects with available HST data \citep{Kenney_2015,Abramson_2016,Cramer_2019}. We will now discuss the morphological features detected in the \textit{Euclid} imaging for both galaxies.

UGC 2665 is a spiral galaxy, morphologically classified in NED as Scd, which lacks a bulge. The galaxy is observed almost edge-on, with a measured axis ratio of 0.444 from \IE imaging, corresponding to an inclination of $\sim$ 64\degree \citep{Cuillandre_2024a}. The galaxy displays features suggesting star formation is occurring in the stripped tail, visible in an edge-on direction. There are features in the galaxy that are due to dust lanes. The dust lanes on the galaxy disc are visible in \IE imaging, but not in \YE and \HE imaging, as shown in the colour composite image in Fig. \ref{figure:fig1}. The dust present in the galaxy disc can also be stripped along with the gas. There are knots that appear to emanate from the spiral arms but are superimposed on the galaxy's disc due to projection. The features outside the galaxy close to the disc are clumpy, elongated, and follow the direction of stripping.

The galaxy displays a distinct ‘‘unwinding’’ effect in the outer spiral arms of the disc, as shown in Fig. \ref{figure:fig5}. This is an effect observed in several RPS galaxies in the literature, both in observations \citep{Bellhouse_2021,Vulcani_2022} and simulations \citep{Schulz_2001,Roediger_2014,Steinhauser_2016}. To better visualize the effect of ram-pressure on the spiral arms of the galaxy, the \IE image of the galaxy is reprojected into polar coordinates, with radial distance
from the galaxys' centre ($r$) and azimuthal angle around the
disc ($\theta$) in the plane of the galaxy. The galaxy is corrected for inclination using the axial ratio from \IE imaging, by scaling the distances along the dimension of the minor axis. Each pixel is reprojected into polar coordinates according to its radial distance and azimuthal position. To preserve the area of each pixel in polar space, the reprojection is carried out for each corner of the pixel individually, mapping the shape of each pixel from a square in the cartesian image plane to a polygon in polar coordinates. This is particularly important in regions close to the galactic centre, where some pixels span a large range of azimuthal angles. The left panel of the figure shows the galaxy “unwrapped” in polar coordinates, whilst the right panel shows the original image. On the left-hand panel, logarithmic spiral arms have been drawn to highlight the prominent dust lanes. These spiral arms are shown reprojected back onto the galaxy disc on the right panel. The pitch angle (the angle between a spiral arm and the tangent to the circle on the plane of the disc) can be used as a measure of the tightness of the spiral arm. The pitch angle, $\sim$ 0, lies along the tangent, while {90\degree} lies along the normal; tightly wound spiral arms have lower pitch angles, and steeper, more loosely wound spiral arms have higher pitch angles. The median pitch angle of the most prominent spiral arms in the galaxy is $\sim$ {\ang{22} within 2 $\times$ the effective radius of the disc (marked by the dashed line/ellipse in Fig. \ref{figure:fig5}), and $\sim$ {\ang{39} outside this radius.
 Higher pitch angles in the outer regions of the disc compared with the inner disc suggest that the spiral arms are effectively being
unwound. In the trailing region of the galaxy's disc, the spiral arms are extremely extended, and they become steeper towards the outskirts of the disc and tails. As demonstrated by simulations in \citet{Bellhouse_2021}, this is consistent with ram-pressure “unwinding” the gas component of the spiral arms. The stars formed \textit{in situ} outside the galaxy's disc exhibit a steeper pitch angle than those formed within the disc, yet still maintain the general shape of the spiral arms. The presence of this pattern further confirms the RPS operating in this galaxy. The pattern and amount of unwinding in this galaxy suggest that it is consistent with ram-pressure stripping, as seen in galaxies confirmed by integral-field spectroscopy, and that gravitational processes are not necessary to explain the curved tails. Furthermore, the presence of unwinding in this manner tells us that the galaxy is likely to be in an early stage of infall. According to \citet{Bellhouse_2021}, the visual unwinding stage is relatively short-lived, lasting up to 0.5 Gyr before the pattern is washed out by the ICM wind into the tail of the galaxy.

The unique, low surface brightness optimised imaging data from \textit{Euclid} allows us to explore stripped tails at faint levels. We now perform a morphological analysis of the low surface brightness features observed along the stripped tails using optical, near-infrared, and other wavelength imaging data. \textit{Euclid} \IE imaging of the galaxy is used to construct a surface brightness map, with features associated with stripping marked in cyan boxes, as shown in Figs.~ \ref{figure:fig3} and ~\ref{figure:fig4}. The stripped features in boxes 1 to 10, marked on the \IE and \YE images, are shown in detail in Figs.~\ref{figure:fig6} and ~\ref{figure:fig7}. The RPS features seen in \IE imaging are seen in \YE imaging as well, though at a slightly lower spatial resolution. Fig. \ref{figure:fig3} shows the overlaid contours of ultraviolet, H$\alpha$ and $144\,{\rm MHz}$ radio continuum. The H$\alpha$ emission clearly displays a stripping pattern, with the emission from the bright features outside the galaxy detected in \IE imaging. H$\alpha$ emission in star-forming regions is due to the recombination of hydrogen that is ionized by O and early-B stars with ages $\leq$ 10 Myr \citep{Kennicutt_1998,Kennicutt_2012}. The NUV N245M contours appear to cover the detected features, suggesting the presence of very recent star formation (with ages $\leq$ 200--300 Myr) in the stripped tails. The NUV flux is coming from the photospheres of A--F spectral type stars with age $\sim$ 300 Myr \citep{Boselli_2009}. The H$\alpha$ and NUV N245M regions exhibit a spatial correlation, indicating very recent star formation (less than 10 Myr) in these areas. We note that the H$\alpha$ and NUV imaging is at a different depth compared to \textit{Euclid}, and therefore, we do not make any claims about the features detected along the low surface brightness tails. Radio continuum contours clearly reveal the morphology of the gas in the galaxy's disc and the intracluster medium, with prominent cometary tail features indicative of RPS. The feature highlighted in box 10 is the farthest from the disc and is likely associated with the RPS of the galaxy. Note that the bright streaks in box 10 are due to a foreground nearby bright star in the field. Diffuse emission and clumpy features are observed in the direction of stripping, prominently seen in boxes 1 to 10.\\

MCG +07-07-070 is a peculiar galaxy classified in NED as SBR(pec). The galaxy is seen face-on with a measured axis ratio of 0.721 from \IE imaging that corresponds to an inclination of $\sim$ 43\degree. The stripped material in MCG +07-07-070 is oriented towards the centre of the cluster, as seen in projection, which suggests it may be moving away from the cluster centre after a recent pericentric passage. The features are marked in cyan boxes numbered 1 to 16 in the \IE image of the galaxy, as shown in Fig. \ref{figure:fig4}. The galaxy \IE image is overlaid with H$\alpha$ and $144\,{\rm MHz}$ radio continuum contours. The NUV N245M contours are confined to the galaxy's disc in projection, as is the H$\alpha$, except for a few regions on the stripped tail ends. The faint features outside the disc of the galaxy do not show NUV N245M and H$\alpha$ emission. The H$\alpha$ morphology in the disc appears truncated with respect to that of the underlying stellar disc. Star-formation truncation is an aftereffect of RPS, where the outer disc gas is stripped, leaving the gas confined to the central regions, hosting star formation.  \citep{Koopmann_2004a,Koopmann_2004b,Koopmann_2006,Boselli_2006,Cortese_2012,Fossati_2013,Fritz_2017,Vulcani_2020}. The RPS scenario is expected to cause a stronger truncation of H$\alpha$ than NUV because the stripping occurs from the outside in. If there is an age effect, between 10 and 100 Myr after the start of the stripping, the H$\alpha$ disc will be completely truncated, while the NUV disc will only be marginally affected \citep{Boselli_2006}.
Mondelin et al. (submitted) provide more details on the truncation occurring in these galaxies based on the ERO of the Perseus field.
We note that MCG +07-07-070 H$\alpha$ observations can be affected by the galaxy’s peculiar velocity, such that the H$\alpha$ filter does not completely cover the emission line. Radio continuum contours show the cometary tail features expected from RPS and are slightly displaced from the features seen in \IE imaging. The stripped features from the \IE images are shown in detail in Fig. \ref{figure:fig8} and shown for the \YE image in Fig. \ref{figure:fig9}. The features exhibit a “fireball” structure (see Appendix for a description of this scenario), characterized by head clumps at the leading edge and diffuse emission at the trailing edge. The orientation of the stripping with respect to the line of sight likely explains this pattern, with the diffuse emission revealing the extent of star formation within the tails. The features marked with arrows in boxes 1, 2, 4, 5, 6, 7, 10, and 12 are not all parallel, as expected when the galaxy rotates while moving within the cluster. The stripped tail direction represents the composite vector of the stripping direction and the galaxy's rotation.

In general, we detect the features seen in the \IE imaging, albeit at a slightly lower resolution, in the \YE imaging of the two galaxies.
 The near-infrared emissions from the galaxies are dominated by an old-evolved population of stars. However, this is not the case for stripped tails, which can include contributions from the stellar continuum of cool main-sequence stars and red supergiants that have recently left the massive end of the main sequence. We demonstrate this in the next section based on colour information derived from the \textit{Euclid} optical and near-infrared imaging data. We note that the morphology of the optical and near-infrared detected features is very similar, with the peak emission from the knots coinciding.

\subsection{Surface brightness of stripped features: optical, infrared  analysis}

\textit{Euclid} imaging is optimized for high angular resolution and low surface brightness, enabling the detection of a larger extent of features created by RPS. We detected these features with \texttt{SourceXtractor++} \citep{Bertin_2020,Kummel_2022}. This new and extended implementation of a source detection algorithm has the advantage of performing photometric measurements in multiple bands based solely on the World Coordinate System, unlike \texttt{SExtractor2} \citep{Bertin_1996}, which requires alignment of all measurement images. 

\begin{table*}[htbp!]
    \caption{Parameters used for \texttt{SourceXtractor++} run of $u$,\IE,\YE,\JE,\HE images of the two galaxies.}
\smallskip
\label{table:T3}
\begin{center}
\smallskip
\begin{tabular}{lccccc} %23
\hline
Parameter & $u$ & $\IE$ & $\YE$ & $\JE$ & $\HE$ \\
\hline
       segmentation-filter &$2.0$ & $2.0$ & $2.0$& $2.0$& $2.0$ \\
       detection-threshold UGC 2665/MCG +07-07-070 & 1.5/1.5 &  1.5/1.5 & 1.0/1.3 & 1.0/1.3 & 1.0/1.3 \\
       detection-minimum-area  & 15& 10 & 5 & 5 & 5 \\
\hline
\end{tabular}
\end{center}
\end{table*}

To compute the surface brightness of the stripped features, we made independent \texttt{SourceXtractor++} runs on the \IE, \YE, \JE, and \HE images. Table\,\ref{table:T3} gives details on the parameters used for the run. As discussed in \citet{Cuillandre_2024a}, these ERO images have already had their backgrounds removed.
\begin{itemize}
\item we verified that the images used in this analysis have a reasonable background subtraction in our regions of interest and decided not to perform additional background determination;
\item we used the associated weight image for detection and photometry in each band;
\item we optimized the detection result with the built-in detection filter [2.0\,pixel  full width half maximum (FWHM)]; 
\item we selected the detection threshold and minimum area in each band independently to keep the false positive detection rate low ($<1.0$\%). We note that this is different for the two galaxies in NISP bands.
\end{itemize}
The average surface brightness distribution of the stripped features %shown in Fig. \ref{figure:fig10} 
were then computed using the {\texttt auto\_mag} brightness, which is measured in an elliptical aperture derived from the object's surface brightness distribution, and the segmentation area, which is the area above the detection threshold.
 
We utilized the segmentation map from the \IE image, which is deeper than near-infrared images, to isolate the features associated with stripping. The features associated with stripping are identified by visually inspecting the segmentation map overlaid on the \IE images of the two galaxies. These are the features that appear in the marked cyan boxes shown in Figs.~\ref{figure:fig3} and ~\ref{figure:fig4}.
We note that this process is done visually and may leave behind faint, irregular, and distant galaxies, contaminating the detection of genuine features. We checked the possibility of contamination by computing the number density of objects within the limiting magnitude of the detected knots in \IE, \YE, \JE, and \HE band images of the galaxies. We calculated the number density, measured in objects/arcmin${^2}$, within the boxes marked along the tails of the galaxies in Figs.~\ref{figure:fig3} and ~\ref{figure:fig4}, as well as in a few regions away from the galaxy. The number density is found to be 3.6/2.2/2.2/1.9 times higher along the stripped tails of UGC 2665 and 4/3.16/2/2.1 times higher along the stripped tails of MCG +07-07-070 in \IE/\YE/\JE/\HE bands. This suggests that objects cluster along features in the boxes, which are very likely part of the stripped tails. However, a robust confirmation requires spectroscopic redshift information. We selected those features that show diffuse emission associated with stripping. We excluded stars and features that have the morphology of a likely background galaxy from the selection. After measuring the surface brightness of stripped features from \IE images of the galaxies, we created segmentation maps for stripped tails independently using \YE, \JE, and \HE imaging data. The distribution of surface brightness of the features is shown in Fig. \ref{figure:fig10}. The limiting surface brightness of these features changes with the imaging band which in turn is dependent on the sensitivity of the imaging data. Table\,\ref{table:T4} gives the number of detected segments associated with the stripped features of the two galaxies. Table\,\ref{table:T5} lists the surface brightness and corresponding area of the faintest feature detected for two galaxies. We note that the faintest features have a surface brightness of $25.17\,{\rm mag}\,{\rm arcsec}^{-2}$ for MCG +07-07-070 and UGC 2665, as seen in \IE imaging. The surface brightness changes to $25.17\,{\rm mag}\,{\rm arcsec}^{-2}$ for MCG +07-07-070 and $25.36\,{\rm mag}\,{\rm arcsec}^{-2}$ for UGC 2665 based on \HE imaging data. To better understand the nature of the selected features, we provide the segmentation map of these features, detected independently from the \IE imaging in Figs.~\ref{figure:fig15} and \ref{figure:fig16} and at a NISP resolution in the \YE imaging of the two galaxies in Figs.~\ref{figure:fig17} and \ref{figure:fig18}. We also overlay the contours corresponding to the boundaries of the segments in Figs.~\ref{figure:fig6}, \ref{figure:fig7}, \ref{figure:fig8} and \ref{figure:fig9}. Our goal here is to determine the faintest features that can be resolved within the stripped tails with \textit{Euclid} imaging. This information can be used to plan for the study of stripped features from \textit{Euclid} DR1 imaging data.

\begin{table}[htbp!]
\caption{Number of the features detected independently from \IE,\YE,\JE,\HE images of the two galaxies.}
\smallskip
\label{table:T4}
\begin{center}
\smallskip
\begin{tabular}{lcccc} %23
\hline
Galaxy           & $N_{\IE}$ & $N_{\YE}$  & $N_{\JE}$ & $N_{\HE}$  \\
\hline
UGC 2665          &  252   &  81    &  72   &     62  \\
MCG +07-07-070    &  433   &  160   &  166  &     176  \\
\hline
\end{tabular}
\end{center}
\end{table}

\begin{table*}[htbp!]
\caption{Surface brightness and the corresponding area of the faintest features detected from \IE,\YE,\JE,\HE images of the two galaxies.}
\begin{center}
\smallskip
\label{table:T5}
\smallskip
\begin{tabular}{lcccccccc} %23
\hline
Galaxy           & $\mu_{\IE}$    & Area & $\mu_{\YE}$    & Area  & $\mu_{\JE}$    &  Area & $\mu_{\HE}$   & Area  \\
                 &${\rm mag}\,{\rm arcsec}^{-2}$ & arcsec$^{2}$ &${\rm mag}\,{\rm arcsec}^{-2}$ & arcsec$^{2}$  &${\rm mag}\,{\rm arcsec}^{-2}$ &  arcsec$^{2}$ &${\rm mag}\,{\rm arcsec}^{-2}$ & arcsec$^{2}$ \\
\hline
UGC 2665          &  25.17 $\pm$ 0.12  & 0.20   & 25.40   $\pm$ 0.38   &  0.54 & 25.41 $\pm$ 0.31  &  0.63  & 25.36  $\pm$ 0.20  & 1.53 \\
MCG +07-07-070    &  25.17 $\pm$ 0.10  & 0.48  &  25.24 $\pm$  0.35  & 0.54    & 25.17  $\pm$ 0.15 & 1.98   &  25.17  $\pm$ 0.14 & 1.98 \\
\hline
\end{tabular}
\end{center}
\end{table*}

\subsection{Optical- near-infrared colour of the stripped features}

We measured the \textit{Euclid} colour of the low surface brightness clumpy features detected for the two galaxies. Measuring object colours in imaging data with large differences in resolution and pixel scales requires special attention. \texttt{SourceXtractor++} projects the elliptical apertures determined on the detection image onto the other measurement images, measuring the brightness of the objects within. The image with the lowest resolution is used as the detection image, which allows the flux of each object to be measured in the same sky area for the better-resolved bands, thereby largely eliminating the impact of resolution on photometry. We used \HE as the detection band to measure the colours of the stripped features in the \textit{Euclid} bands.

We removed contamination from foreground stars after matching the position of knots with the \textit{Gaia} DR3 catalog of stars in the field. We note that the limiting magnitude for \textit{Gaia} $G$ is $\sim$ 21 in the regions of the galaxies, where the Perseus ERO has a point source depth with a 5$\sigma$ PSF magnitude of 28.0, 25.2, 25.4, and 25.3 in \IE, \YE, \JE, and \HE, respectively and with the exceptional angular resolution in \IE star complexes down to the scales of $50\,{\rm pc}$ can be resolved. This suggests that fainter stars may be present in these fields, detectable through \textit{Euclid} imaging, which could potentially contaminate the detected features. The extinction due to MilkyWay in the direction of the galaxies is computed for each band using the procedure described in \citet{Cuillandre_2024a}. The colours computed for these features are then used to constrain the age of the underlying stellar population in the stripped tails. We quantify the resolved knots found in the low surface brightness features identified from the segmentation map. We consider these knots, which lack redshift information, to be \textit{in situ} star-forming regions, as they fall on low surface brightness stripped features. Additionally, the regions are also covered by the emission detected through UVIT NUV narrow band imaging, as shown in Figs. \ref{figure:fig3} and \ref{figure:fig4}.

The zoom-in \IE images show that the identified knots have diffuse emission associated with them in the direction of stripping. These knots are selected from the regions outside the galaxy in the direction of stripping. Knots that appear to follow the stripping pattern are visible in projection, overlapping with the galaxy's disc. This is particularly true of UGC 2665, and we exclude these knots from consideration due to their susceptibility to contamination from the galaxy's disc flux. The selected knots on the stripped tails of the galaxies are used to construct a \textit{Euclid} colour-colour diagram. Figs. \ref{figure:fig11} and \ref{figure:fig12} show the colour-colour plots of the detected knots in UGC 2665 (green) and MCG +07-07-070 (grey), created using $\IE-\YE$, $\YE-\HE$, and $\YE-\JE$ colour combinations. We only show the knots that exceed the resolution limit ($\ang{;;0.50}$) of the \HE imaging data, with a $3\,\sigma$  detection. There are 37 knots detected for UGC 2665 and 138 knots detected for MCG +07- 07-070. The colour values for single stellar populations of ages ranging from 0 to 300 Myr were generated using \citet{BC_2003}  stellar population models (BC03), Padova 94 isochrones, and a Kroupa initial mass function (IMF, \citet{Kroupa_2001}), for solar metallicity, corresponding to a redshift of 0.01, are overlaid. The corresponding age range (<$300\,{\rm Myr}$) is shown in the colour bar scale and the colour distribution of the knots is shown in the side panels. The measured flux from these knots can be affected by extinction due to dust within the system and in the foreground from the Galaxy. We performed MilkyWay extinction correction for the magnitude values, but the intrinsic extinction can affect these colour values plotted here. The modeled colour values for different stellar population ages exhibit non-monotonous behavior, mainly because integrated stellar population colours shift to blue before the first massive stars become red supergiants (which takes about $10\,{\rm Myr}$), then to the very red colours of red supergiant dominated populations, and later to slightly less red colours when these supergiants become less luminous. The rotation of massive stars also affects this, as it determines the coolest temperature a star of a given mass can reach, and additionally, the spectra of massive stars are uncertain. This makes it difficult to estimate the ages of the underlying stellar population from \textit{Euclid} colour-colour plots alone. However, as we demonstrate here, we can determine whether the colours are compatible with a young age across an age/metallicity range. The \IE bandpass covers a broad wavelength range, which is sensitive to spectral energy variations for different stellar population ages. We could effectively discriminate the contribution of different age stellar populations to the spectral energy distribution by using $u, g, i, z$ imaging data. The resolution of the CFHT data is lower than that of the \textit{Euclid} imaging data. We used the $u$ band image with the lowest resolution of $\approx\,\ang{;;1.46}$ for detection and identification of the knots. We applied the segmentation map to the $g, r, i,$ and $z$ imaging data of both galaxies after running \texttt{SE++} on the $u$ band imaging data using the parameters given in Table\,\ref{table:T3}. There are 11 knots detected for UGC 2665 and 39 knots detected for MCG +07-07-070 that coincide with the regions seen in the \textit{Euclid} imaging. The magnitudes are measured and corrected for MilkyWay extinction. The $u - r$, $g - i$ plot of the detected knots for UGC 2665 (in green) and MCG +07-07-070 (in grey) are shown in Fig. \ref{figure:fig13}.  The colour values for single stellar population ages for $u - r$, $g - i$ colours over the range 0--$1000\,{\rm Myr}$ are overlaid. Most knots have colours consistent with hosting recent star formation, but a few have red colours. There can be several reasons for this behavior of the detected knots like source contamination and flux attenuation due to dust. CFHT ground-based imaging of these knots can be contaminated by background or foreground sources.Dust at the stripped tail sites, where these knots are located, can weaken blue band ($u$ and $g$) flux, leading to a strong degeneracy when interpreting the ages of the underlying stellar population. We examined how colours change due to dust extinction in the models we used in Figs. \ref{figure:fig11}, \ref{figure:fig12}, and \ref{figure:fig13} assuming $A_v=1$ mag (median value of $A_v=0.5$ mag along the stripped tails of GASP galaxies \citep{Poggianti_2019}), and the extinction law of \citet{Cardelli_1989} with $R_v=3.1$ for full range of solar metallicity model colours. The differential extinction values for different colours are sufficiently close to be represented by a single vector in the figures.
 $E(\IE-\YE)$ = 0.42 for the bluest model and  0.35 for the reddest model, $E(\YE-\HE)$ = 0.20 over the full range of model colours, $E(\YE-\JE)$ = 0.115  over the full range of model colours, $E(u-r)$ has a non-monotonic behaviour as a function of $(u-r)$ but stays between 0.685 and 0.705. $E(g-i)$ = 0.56 for the bluest model and 0.50 for the reddest model. We note that the adopted extinction does not explain the dispersion in the \textit{Euclid} colour plots. We note that we haven't included the nebular emission in our models, which can explain quite a part of the dispersion seen in \textit{Euclid} colour plots. Detailed modeling of the dust content and nebular emission is required to understand the true nature of the underlying stellar population in these knots. 

\subsection{The smallest knots detected in the stripped tails}

We demonstrate \textit{Euclid}’s capability to detect small-scale knots in the tails of RPS galaxies using higher resolution optical \IE imaging. The high spatial resolution imaging enables small-size knots to be detected within the stripped tails, which can be used to understand the smallest scales at which gas collapses in the stripped tails. Young, hot stars in star-forming complexes ionise dense molecular clouds, creating H\textsc{ii} regions with typical sizes of $\sim$ 0.01--$1000\,{\rm pc}$, found in both Galactic and extragalactic environments \citep{Hunt_2009}. 
We perform source detection using \texttt{SExtractor2} which is applied on a ring-filtered \IE frame (with inner and outer radii of 4 and 8 pixels) \citep{Bertin_1996}.
We used the following parameter values, DETECT\_MINAREA=3, DETECT\_THRESH=1.5, 
 ANALYSIS\_THRESH=1.5, DEBLEND\_NTHRESH=32, and DEBLEND\_MINCONT=0.0005 for the \texttt{SExtractor2} run on the \IE images of the galaxies. We performed aperture photometry for the detected sources using \texttt{photutils}, with an aperture diameter of four times the FWHM of the PSF, which was about 7 pixels in all the filters. The aperture magnitudes were corrected for the fraction of the light beyond the aperture. Additionally, the background was estimated for an annulus around each object with an inner diameter of 10 times the FWHM of PSF and a thickness of 20 pixels.
We selected the knots at the resolution limit ($\ang{;;0.16}$) that fall on the low surface brightness features within the boxes overlaid in Figs. \ref{figure:fig3} and \ref{figure:fig4}. We show only the detected knots at the resolution limit of the \IE imaging data. These knots are unresolved and have an upper size limit of 2$\times \ang{;;0.16}$ $\sim$ $108\,{\rm pc}$. The \IE magnitudes detected from the tails of UGC 2665 (in orange) and MCG +07-07-070 (in blue) are shown in Fig. \ref{figure:fig14}. There are 42 knots for UGC 2665 and 46 knots for MCG +07-07-070. The detected knots are of size $\leq$ $108\,{\rm pc}$ with a limiting magnitude of 27.8 mag for UGC 2665 and 30.2 mag for MCG +07-07-070. It is worth noting that, at these faint magnitudes, contamination from foreground stars is possible, and the true nature of these knots associated with the stripped tails remains to be fully established. Eight knots in UGC 2665 and five in MCG +07-07-070 have H$\alpha$ emission associated with them, but none show any NUV emission. The reason we fail to detect all these knots in NUV and H$\alpha$ may be that \textit{Euclid} is more sensitive to faint features than UVIT/CFHT and has a higher angular resolution, which is crucial for avoiding flux dilution of point sources like those analyzed in this work.  
Similar-sized star-forming knots (50--$100\,{\rm pc}$) have been reported in the tails of RPS galaxies in the Coma and Virgo clusters \citep{Cramer_2019,Boselli_2021}.

\textit{Euclid}'s exceptional angular resolution enables identification of star-forming complexes as small as $108\,{\rm pc}$ at the distance of the Perseus cluster, providing further insight into the impact of perturbation on the stripped material in two galaxies. Figs. \ref{figure:fig11} reveals a significant limitation in studying the physical properties of these features at high spatial resolution and low surface brightness levels: the broadband filters of \textit{Euclid} are insufficient to identify the dominant stellar population and determine the age.

\begin{figure}
\centering
\includegraphics[width=0.45\textwidth]{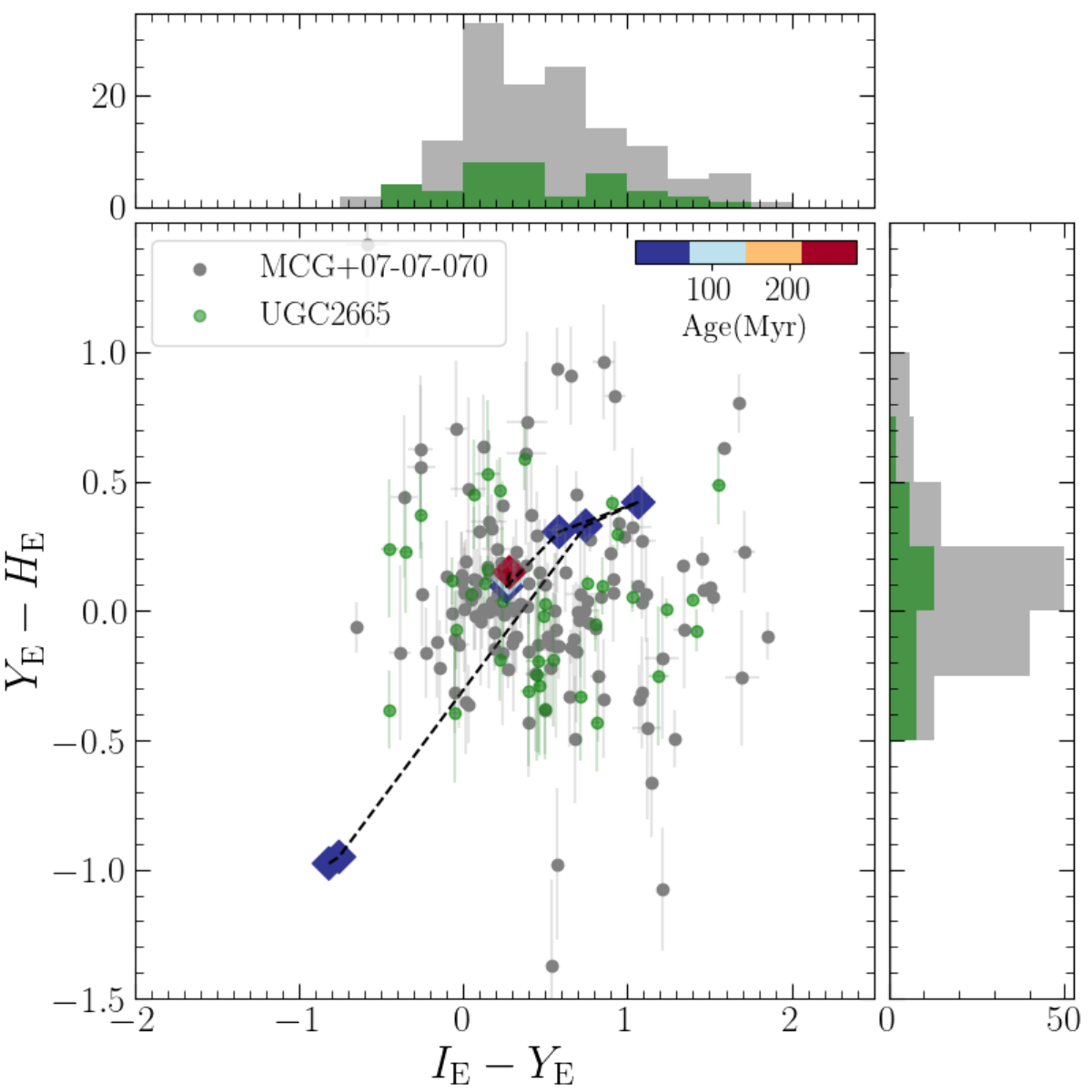}
\caption{Colour-colour plot of the detected knots created using \IE-\YE,\YE-\HE colour combination for UGC 2665 and MCG +07-07-070. We show the colour of the low surface brightness features detected from the segmentation map over the stripped tails of the galaxies. The colour values for different single stellar population ages generated using BC03 stellar population models, Padova 94 isochrones, and Kroupa IMF for solar metallicity are shown in filled circles connected with dotted line. The corresponding age is shown in the colour-bar scale. The \IE-\YE and \YE-\HE colour distribution of the knots are shown in side panels.}\label{figure:fig11}
\end{figure}

\begin{figure}
\centering
\includegraphics[width=0.45\textwidth]{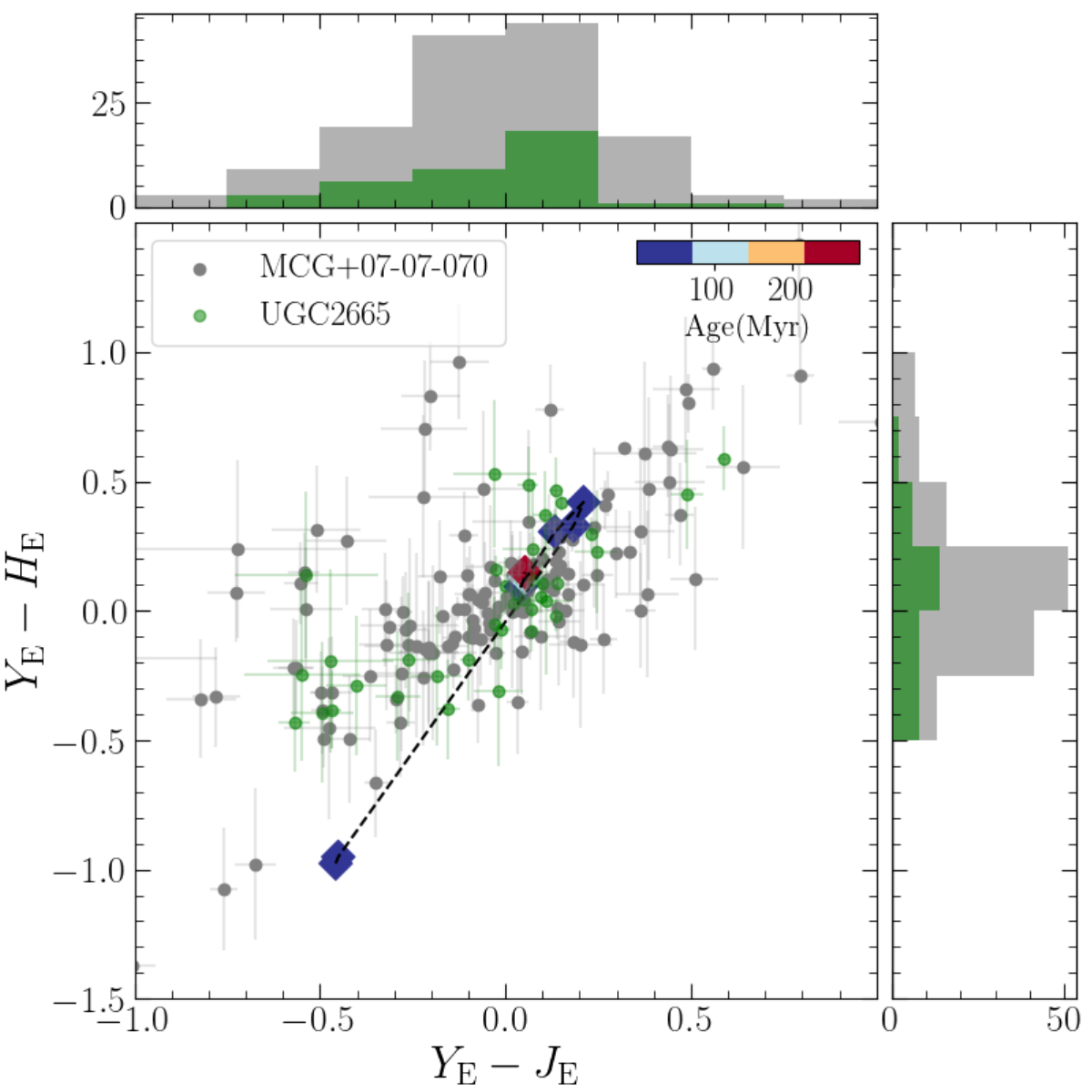}
\caption{Colour-colour plot of the detected knots created using \YE-\JE,\YE-\HE colour combination for UGC 2665 and MCG +07-07-070. Details are same as in Fig.~\ref{figure:fig11}.}\label{figure:fig12}
\end{figure}

\begin{figure}
\centering
\includegraphics[width=0.45\textwidth]{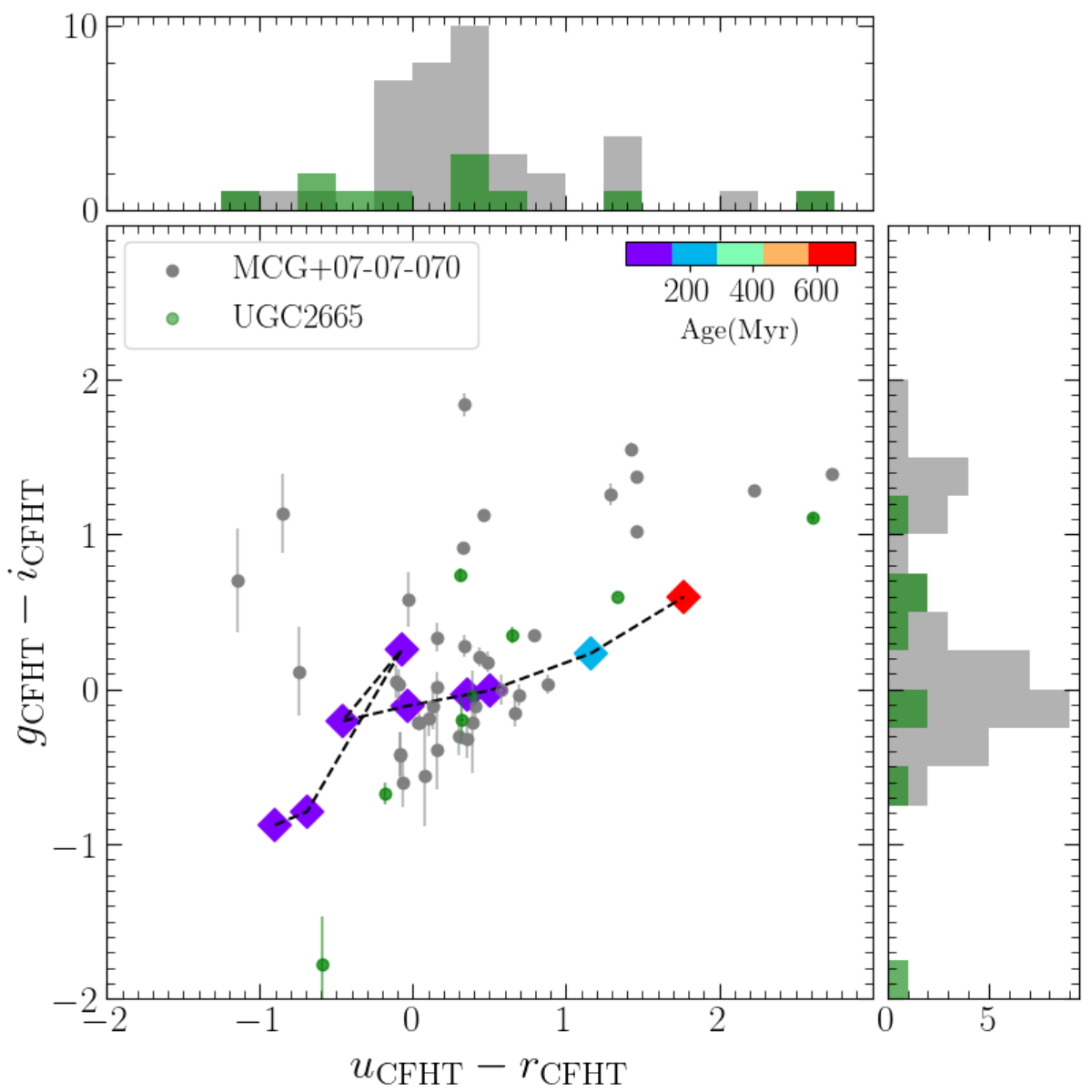}
\caption{Colour-colour plot of the detected knots created using $u-r$, $g-i$ colour combination for UGC 2665 and MCG +07-07-070. Details are same as in Fig.~\ref{figure:fig11}.}\label{figure:fig13}
\end{figure}

\begin{figure}
\centering
\includegraphics[width=0.4\textwidth]{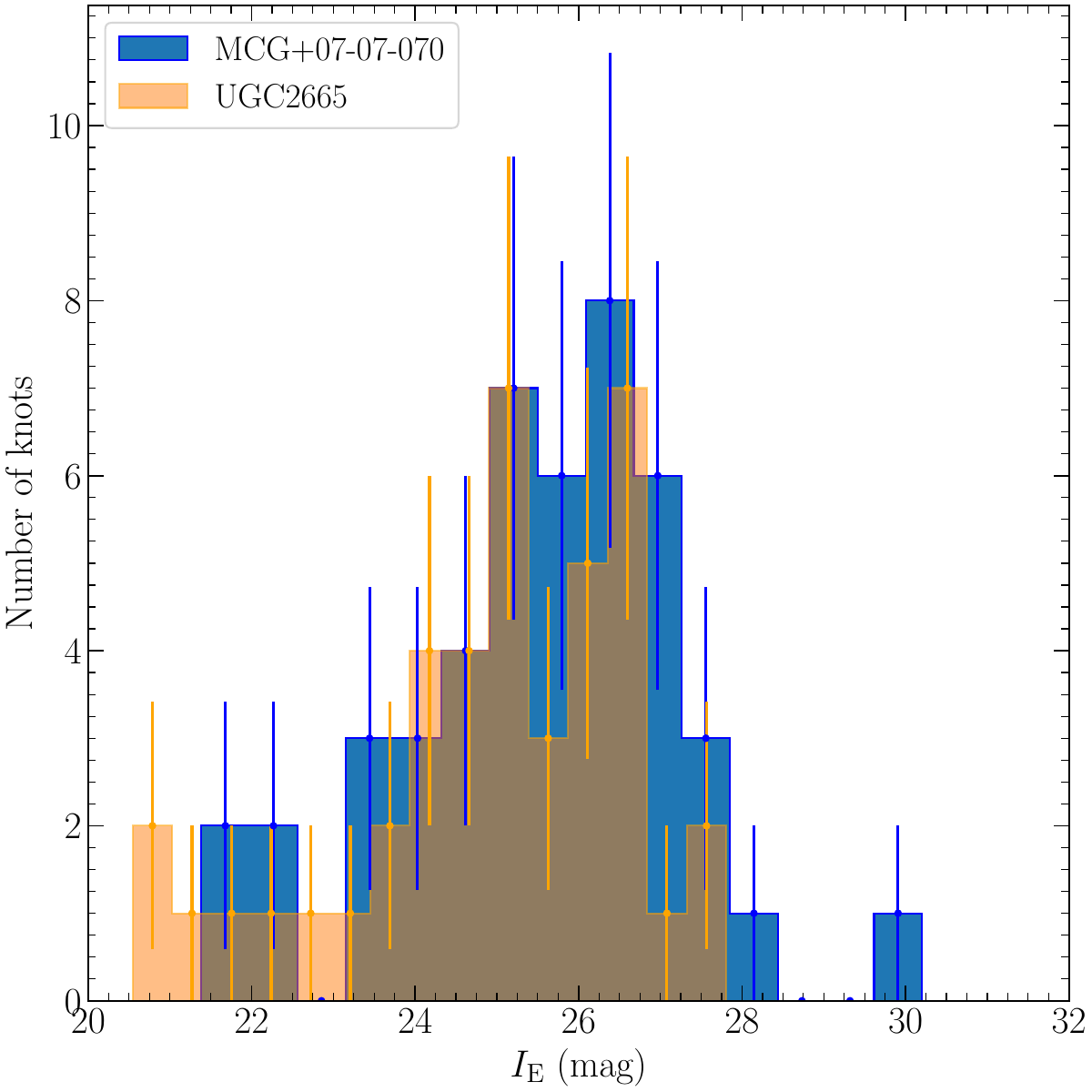}
\caption{Distribution of the magnitudes of smallest knots detected from \IE image of UGC 2665 and MCG$+$07$-$07$-$070. Poisson errors are shown with an error bar.} \label{figure:fig14}
\end{figure}

\section{Discussion}

The \textit{Euclid} optical and near-infrared imaging of the Perseus cluster field reveals that RPS is the dominant perturbing mechanism for galaxies UGC 2665 and MCG +07-07-070. The cometary-shaped features with clumps likely host star-forming knots in the stripped tails outside the galaxies, as observed in the low surface brightness optimized \IE, \YE, \JE, \HE imaging data. Both galaxies show cometary tails in low-frequency radio continuum imaging observations. These tails have a morphology expected from RPS. In the case of UGC 2665, dust lanes are visible on the galaxy's disc in \IE imaging, but they almost disappear in the \YE imaging data. This galaxy's appearance is very similar to that of NGC 4522, another object in the Virgo cluster undergoing RPS and hosting dust lanes \citep{Kenney_1999}. The dust lane features resemble those observed in the HST imaging of spiral galaxy NGC 4921, which is undergoing RPS in the Coma cluster, as seen at high angular resolution \citep{Kenney_2015}. The dust lanes appear to follow the stripping pattern, suggesting that dust has been stripped away with the gas, as observed in other galaxies in Virgo and Coma galaxy clusters \citep{Abramson_2016,Longobardi_2020}. As dust and gas are well mixed within the ISM, they can be removed together during the hydrodynamic interactions experienced due to RPS. The dust lanes in the disc of the galaxy UGC 2665 follow the spiral arms, which appear to have an unwinding pattern in the direction of stripping. The orientations of dust lanes along with the presence of cometary tails in radio continuum observations are strong evidence supporting RPS in this galaxy. We have identified the candidate star-forming knots that appear to be connected to the main body of the galaxy by low surface brightness features, with the direction of the features indicating that these are likely due to ongoing star formation in the ram-pressure stripped tails of the galaxy. This is further corroborated using the NUV N245M imaging data of the galaxies. We identified those knots that fall within the NUV N245M contours as star-forming knots in the stripped tails of the galaxies. We caution here that there can be background or foreground objects contaminating the knots, and in the absence of a redshift estimation, we cannot quantify them as real knots associated with the RPS. The presence of associated diffuse emission, oriented towards the knots as shown in Figs.~\ref{figure:fig6} and \ref{figure:fig8}, suggests that these are formed \textit{in situ} in the stripped tails of the galaxies. The positions of detected knots from the stripped tails of UGC 2665 and MCG +07-07-070 in the $u - r$, $g - i$ colour-colour plane are consistent with hosting recent star formation. \\

The current study showcases \textit{Euclid}'s ability to resolve faint star-forming knots in optical and near-infrared at the smallest possible scales across a wide region around galaxies undergoing RPS in the Perseus cluster. The imaging data, with its excellent image quality (FWHM $\approx\,\ang{;;0.16}$) and high sensitivity to low surface brightness features ($\mu$ $\sim$ $30.1\,{\rm mag}\,{\rm arcsec}^{-2}$), enables us to detect the furthest extent of star-forming regions associated with stripping at different distances from the galaxy's disc at the highest spatial resolution. The extent of star formation in the stripped tails can be identified by measuring the length of the low surface brightness diffuse features. The longer the length, the more extended the regions of ongoing star formation, and hence the larger the strength of RPS. The detected diffuse features are expected to be longer for galaxies in clusters that experience higher ram pressure effects \citep{Kenney_2004}.\\

The \textit{Euclid} ERO of the Perseus cluster field is taken with 4 ROS, whereas the normal EWS will have only a single ROS. The ERO data are therefore slightly deeper than the normal EWS, for which the expected limiting surface brightness for diffuse emission is $29.8\,{\rm mag}\,{\rm arcsec}^{-2}$ in \IE \citep{Scaramella_2022}. The EWS will provide uniform optical and near-infrared imaging data for a 14\,000\,deg$^2$ area of the sky, offering a spatial resolution and low surface brightness that enables the detection of low surface brightness features associated with perturbed galaxies within different regions of galaxy clusters, galaxy groups, and even in the field environment. This will enable studies of star formation quenching mechanisms at different redshifts and will enable us to understand the dominant process when the Universe was much younger and denser. The dominant perturbation mechanism leading to star formation quenching can be identified using low surface brightness imaging data over a wide field of view around the galaxies. This information enables us to quantify the fraction of RPS tails in galaxies, where star formation occurs, and allows for comparisons with other multi-wavelength data, tuned models, and hydrodynamic simulations. The occurrence of star formation in the stripped tails is intriguing, particularly since it happens in the very hostile environment of hot, dense ICM. There are observations of galaxies with stripped tails of cold, ionised, or hot gas, but with little or no associated star formation (eg: \citet{Vollmer_2012}, also see Table 2 of \citet{Boselli_2022} for a list of RPS galaxies in local Universe cluster). The reason for this low star formation efficiency along the stripped tails of some galaxies is not understood. Studies of the stripped tails, both with and without star formation, can provide important clues about the required conditions for the progression of star formation in these environments. It is essential to detect the extent and understand the nature of star formation in the stripped tails of these galaxies. Understanding \textit{in situ} star formation at pc scales in the stripped tails of galaxies is crucial to determine the scales at which gas collapses in the hostile environment of hot and dense ICM. By combining this with cold gas (H\textsc{i} \& H$_{2}$) measurements, we can gain insight into the efficiency of star formation in these environments. \textit{Euclid}’s low surface brightness optimized imaging can identify galaxies with faint signatures of RPS near the disc, characterized by little or no star formation, as well as those with significant star formation along the tails. Using sensitive radio telescopes like MeerKAT to obtain H\textsc{i} data, we can compare the H\textsc{i} tail with the \textit{Euclid} imaging data to measure the extent of stellar emission along the stripped tails of these galaxies. Analyzing a large statistical sample of RPS galaxies across various environments and redshifts allows us to gain insights into the evolution of star formation and general trends in star formation efficiency within stripped tails, which can be compared to models \citep{Taylor_2005,Burkhart_2016} and simulations \citep{Kronberger_2008,Tonnesen_2012,Steyrleithner_2020,Boselli_2021}.

\section{Summary}

Using high-resolution optical and near-infrared imaging observations optimized for low surface brightness, we demonstrate how \textit{Euclid} can uniquely study galaxy evolution in dense environments, as exemplified by two galaxies in the Perseus cluster. These galaxies display filamentary structures suggesting an external perturbation. \textit{Euclid} ERO of the Perseus cluster, combined with multi-wavelength data including the NUV, H$\alpha$, and radio continuum imaging, we demonstrate that the dominant perturbing mechanism for these galaxies is RPS. \textit{Euclid}'s low surface brightness optimised imaging capability has made it possible to detect the diffuse features visible in optical and near-infrared along the stripped tails of these galaxies. We have detected features associated with the stripped tails of the galaxies, with 252 detected for UGC 2665 and 433 detected for MCG +07-07-070 from \IE imaging, which are associated with the diffuse emission. The detected features have a limiting surface brightness of $25.17\,{\rm mag}\,{\rm arcsec}^{-2}$, covering an area of 0.20 arcsec$^2$ for UGC 2665, and $25.17\,{\rm mag}\,{\rm arcsec}^{-2}$, covering an area of 0.48 arcsec$^2$ for MCG +07-07-070. The detected features show good correspondence in morphology between optical and infrared (\YE,\JE,\HE) at the smallest spatial scales possible with \textit{Euclid} imaging. \textit{Euclid} colours alone are insufficient for studying stellar population ages in unresolved star-forming regions. We constructed the \textit{Euclid} $\IE-\YE$, $\YE-\HE$, and CFHT $u-r$, $g-i$ colour-colour plane and used single stellar population models to demonstrate that the position of these detected features can be explained by recent star formation. Some features on the higher age model grid may be due to the presence of dust stripped along with gas in these regions. We detect 42 knots for UGC 2665 and 46 knots for MCG +07-07-070 at the resolution limit of the \IE imaging data. These unresolved knots have a size of $\leq$ $108\,{\rm pc}$ with limiting magnitudes of 27.8 mag for UGC 2665 and 30.2 mag for MCG +07-07-070.

The study of two galaxies undergoing RPS, with $\sim$ $30.1\,{\rm mag}\,{\rm arcsec}^{-2}$ surface brightness features at scales of $\sim$ $108\,{\rm pc}$, showcases \textit{Euclid}'s potential to detect and characterize galaxies undergoing morphological transformation in dense environments. The wide and deep survey by \textit{Euclid} will provide a unique dataset of optical and infrared imaging, covering a large region and enabling the detection of features around a statistically significant number of galaxies of varying masses at different environments, a feat previously impossible.

\begin{acknowledgements}
\AckERO  
\AckEC
E.S is grateful to the Leverhulme Trust for funding under the grant number RPG-2021-205. Based on observations obtained with MegaPrime/MegaCam, a joint project of CFHT and CEA/DAPNIA, at the Canada-France-Hawaii Telescope (CFHT) which is operated by the National Research Council (NRC) of Canada, the Institut National des Science de l'Univers of the Centre National de la Recherche Scientifique (CNRS) of France, and the University of Hawaii. The observations at the Canada-France-Hawaii Telescope were performed with care and respect from the summit of Maunakea which is a significant cultural and historic site.
This work presents results from the European Space Agency (ESA) space mission Gaia. Gaia data are being processed by the Gaia Data Processing and Analysis Consortium (DPAC). Funding for the DPAC is provided by national institutions, in particular the institutions participating in the Gaia MultiLateral Agreement (MLA).

\end{acknowledgements}

\bibliography{paper}}

\begin{appendix}
\onecolumn 

\section{\label{appendix:fireball}Fireball features for galaxy MCG +07-07-070}

\textit{Euclid}'s low surface brightness optimised imaging capability has made it possible to detect the diffuse features visible in optical and near-infrared along the stripped tails of MCG +07-07-070 galaxy. The bright knots and the diffuse feature seen in Fig. \ref{figure:fig8} for galaxy MCG +07-07-070 are thought to have formed due to the “fireball” scenario described by \citet{Kenney_2014} and summarised as follows: A gas-rich spiral galaxy falls for the first time at high velocity into a galaxy cluster and can undergo strong RPS. The stripped gas clouds can be moving in the opposite direction of the motion of the galaxy. As the gas cloud accelerates and moves with the galaxy, new stars are forming within it, while slightly older stars become decoupled. This is possible since stars are not affected by RPS and new stars, once formed, are bound only to the gravitational potential of the galaxy. This scenario results in a knot-like region associated with the gas cloud, where the most recent star formation is taking place. A trail follows, where a slightly older population, which formed within the gas cloud, is present. The diffuse trail region should therefore contain a slightly more evolved population of stars compared to the head region (see \citealt{Kenney_2014} for a cartoon of the fireball model, Fig. 16.)

\section{\label{appendix:Segmap}Segmentation map of detected features}

\begin{figure*}[htbp!]
\centering
\includegraphics[width=1\textwidth]{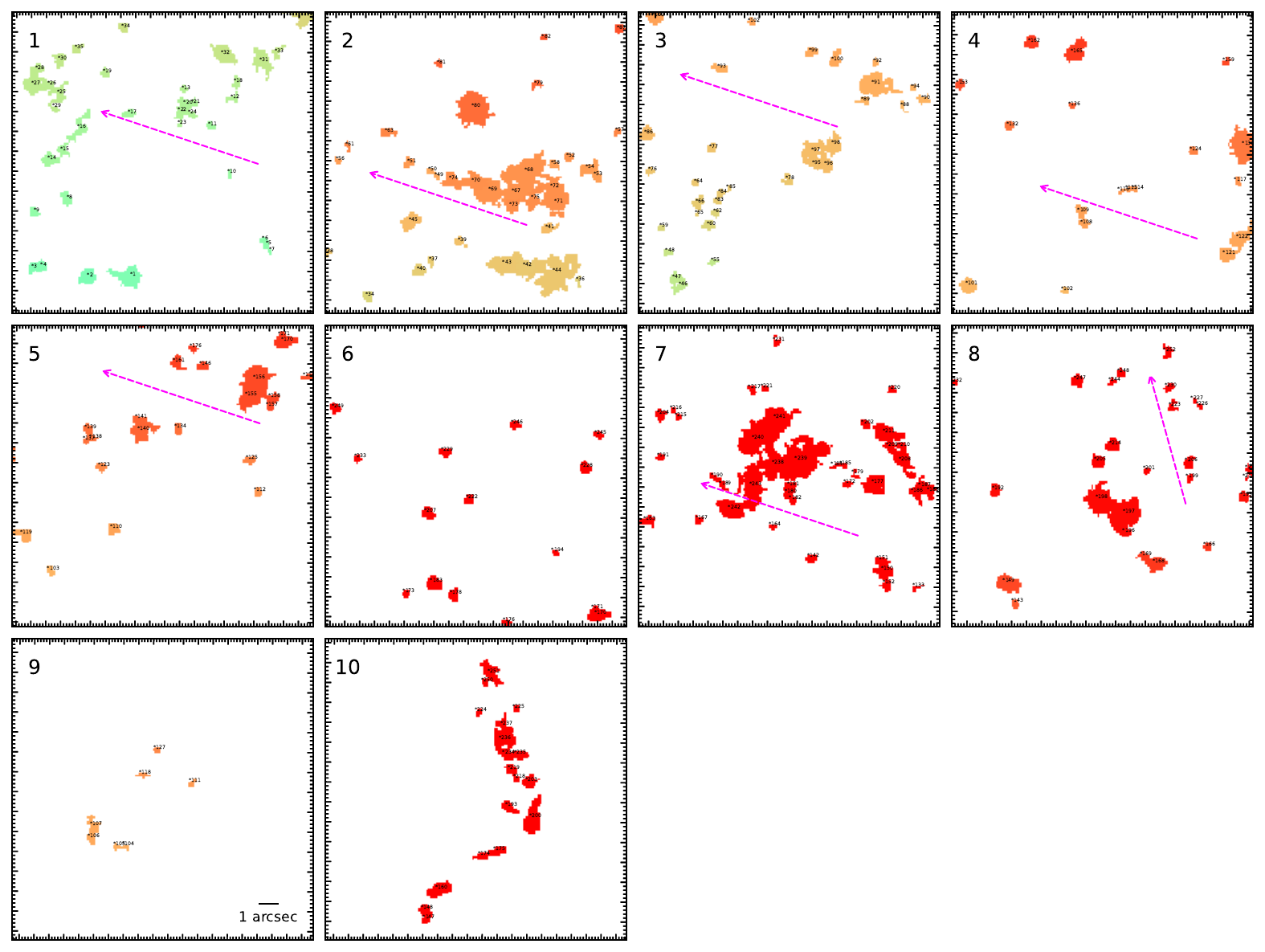}
\caption{Zoom in on the segmentation map of region selected from \IE imaging of stripped features for UGC 2665. Each box has a size of $5.1\,{\rm kpc}$ $\times$ $4.4\,{\rm kpc}$ with the arcsec bar shown corresponding to $338\,{\rm pc}$ at cluster frame. The details of markers are the same as in Fig. \ref{figure:fig3}. Segments are marked with individual identification number with colour scale to red shows the progressing  identification number.}
\label{figure:fig15}
\end{figure*}

\begin{figure*}[htbp!]
\centering
\includegraphics[width=1\textwidth]{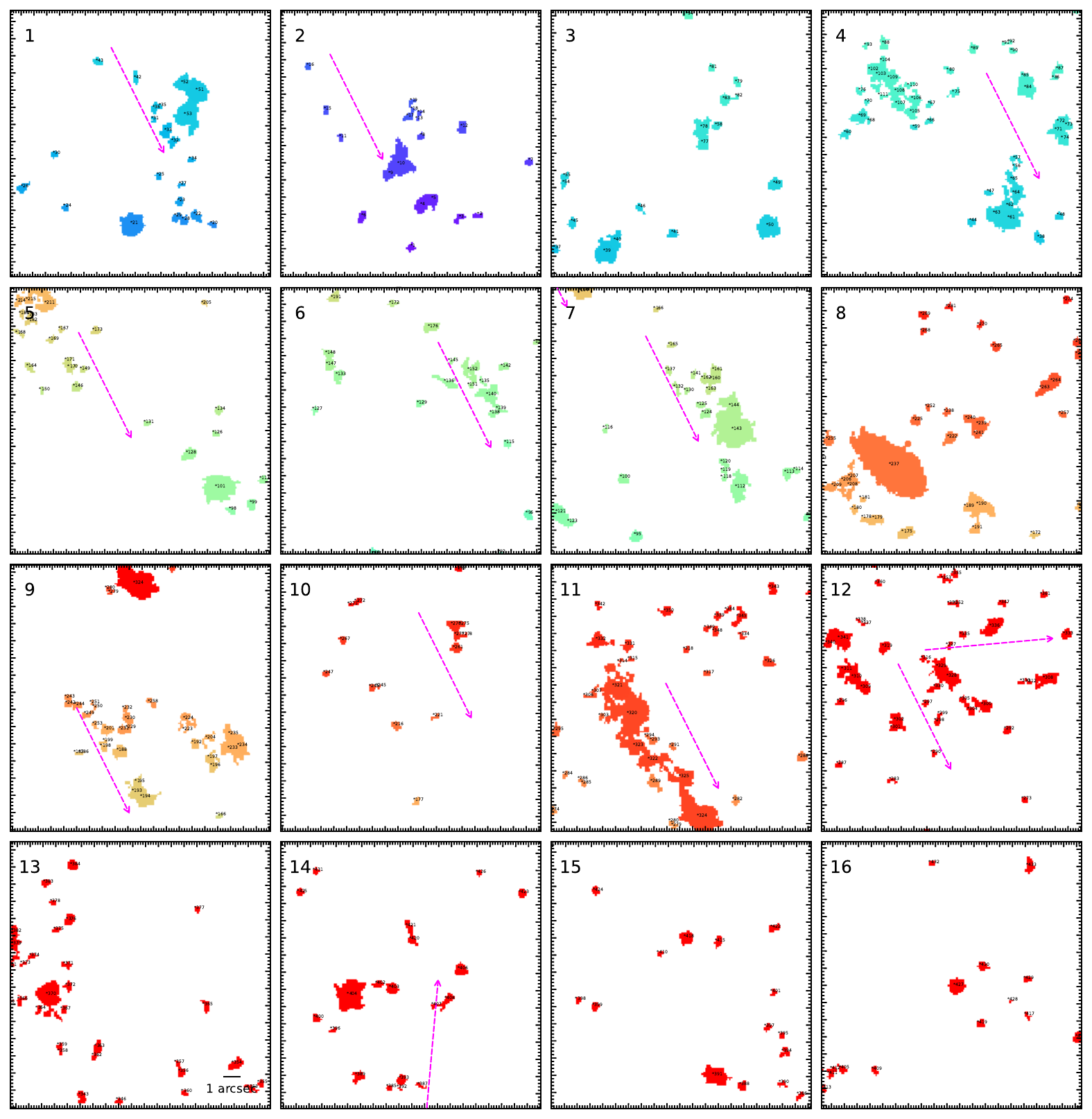}
\caption{Zoom in on the segmentation map of region selected from \IE imaging of stripped features for MCG +07-07-070. The details of markers are the same as in Fig. \ref{figure:fig4} and other details are same as in Fig.~\ref{figure:fig15}.}
\label{figure:fig16}
\end{figure*}

\begin{figure*}[htbp!]
\centering
\includegraphics[width=1\textwidth]{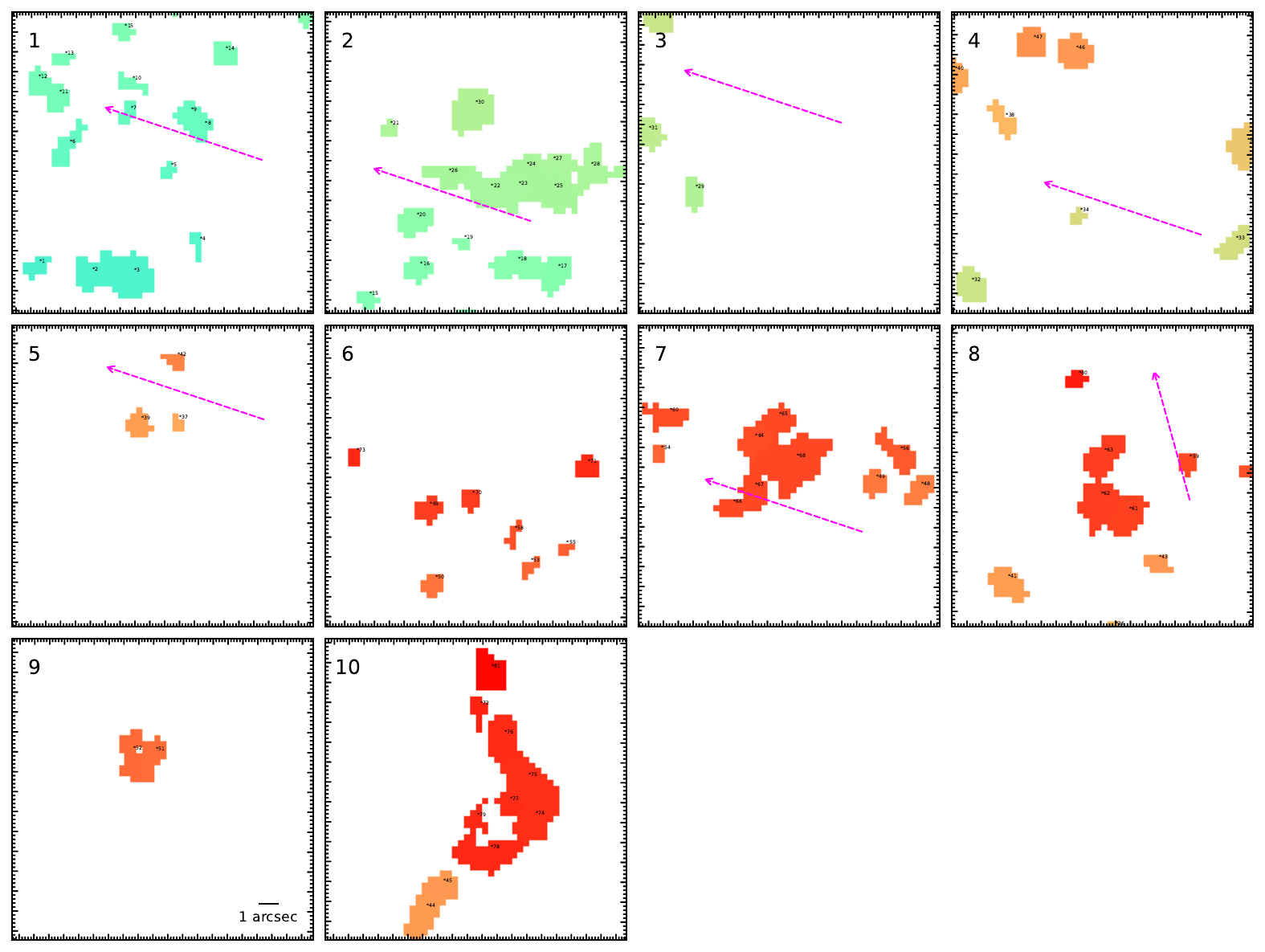}
\caption{Zoom in on the segmentation map of region selected from \YE imaging of stripped features for UGC 2665. The details of markers are the same as in Fig. \ref{figure:fig3} and other details are same as in Fig.~\ref{figure:fig15}.}
\label{figure:fig17}
\end{figure*}

\begin{figure*}[htbp!]
\centering
\includegraphics[width=1\textwidth]{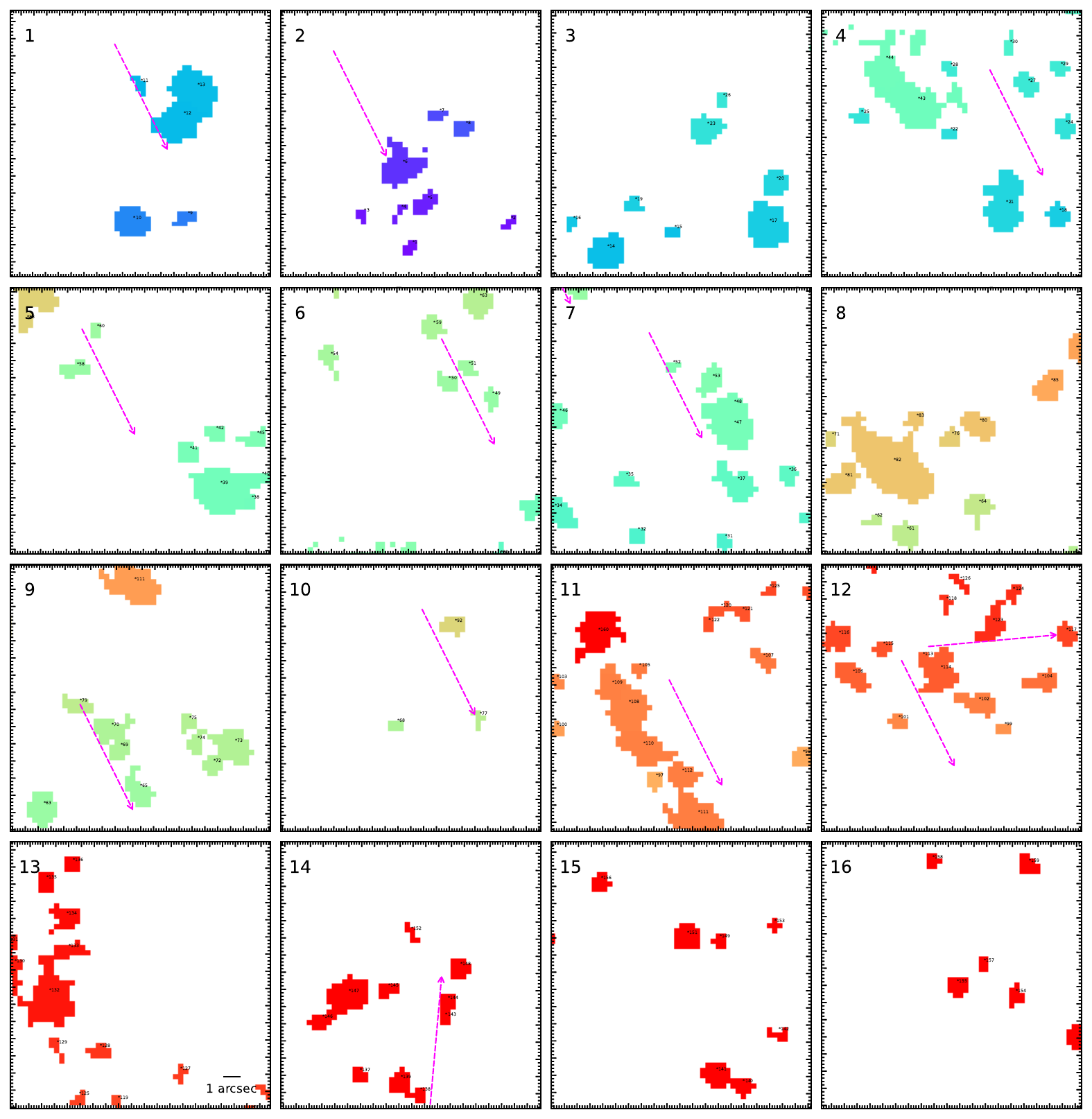}
\caption{Zoom in on the segmentation map of region selected from \YE imaging of stripped features for MCG +07-07-070. The details of markers are the same as in Fig. \ref{figure:fig4} and other details are same as in Fig.~\ref{figure:fig15}.}
\label{figure:fig18}
\end{figure*}

\end{appendix}

\end{document}

%% file: acronym.tex
\acrodef{AGN}{active galactic nucleus}
\acrodef{ASIC}{application specific integrated circuit}
\acrodef{BFE}{brighter-fatter effect}
\acrodef{BCDs}{blue compact dwarfs}
\acrodef{BCGs}{blue compact galaxies}
\acrodef{BCG}{brightest cluster galaxy}
\acrodef{CaLA}{camera-lens assembly}
\acrodef{CCD}{charge-coupled device}
\acrodef{CEA}{Comité Energie Atomique}
\acrodef{CoLA}{corrector-lens assembly}
\acrodef{CDS}{Correlated Double Sampling}
\acrodef{CFC}{cryo-flex cable}
\acrodef{CFHT}{Canada-France-Hawaii Telescope}
\acrodef{CGH}{computer-generated hologram}
\acrodef{CNES}{Centre National d'Etude Spacial}
\acrodef{CPPM}{Centre de Physique des Particules de Marseille}
\acrodef{CPU}{central processing unit}
\acrodef{CTE}{coefficient of thermal expansion}
\acrodef{DCU}{Detector Control Unit}
\acrodef{DES}{Dark Energy Survey}
\acrodef{DGL}{diffuse galactic light}
\acrodef{DPU}{Data Processing Unit}
\acrodef{DS}{Detector System}
\acrodef{EDS}{Euclid Deep Survey}
\acrodef{EE}{encircled energy}
\acrodef{ERO}{Early Release Observations}
\acrodef{ESA}{European Space Agency}
\acrodef{EWS}{Euclid Wide Survey}
\acrodef{FDIR}{Fault Detection, Isolation and Recovery}
\acrodef{FGS}{fine guidance sensor}
\acrodef{FOM}[FoM]{figure of merit}
\acrodef{FOV}[FoV]{field of view}
\acrodef{FPA}{focal plane array}
\acrodef{FWA}{filter-wheel assembly}
\acrodef{FWC}{full-well capacity}
\acrodef{FWHM}{full width at half maximum}
\acrodef{GC}{globular cluster}
\acrodef{GWA}{grism-wheel assembly}
\acrodef{H2RG}{HAWAII-2RG}
\acrodef{HST}{{\it Hubble} Space Telescope}
\acrodef{HSC}{Subaru-Hyper Suprime-Cam}
\acrodef{ISM}{interstellar medium}
\acrodef{IP2I}{Institut de Physique des 2 Infinis de Lyon}
\acrodef{JWST}{{\em James Webb} Space Telescope}
\acrodef{IAD}{ion-assisted deposition}
\acrodef{ICGC}{intracluster globular cluster}
\acrodef{ICU}{instrument control unit}
\acrodef{ICL}{Intra-cluster light}
\acrodef{ICM}{intra-cluster medium}
\acrodef{IGM}{intragalactic medium}
\acrodef{IMF}{initial mass function}
\acrodef{IPC}{inter-pixel capacitance}
\acrodef{LAM}{Laboratoire d'Astrophysique de Marseille}
\acrodef{LED}{light-emitting diode}
\acrodef{LF}{Luminosity function}
\acrodef{LSB}{low surface brightness}
\acrodef{LSST}{Legacy Survey of Space and Time}
\acrodef{MACC}{Multiple Accumulated}
\acrodef{MLI}{multi-layer insulation}
\acrodef{MMU}{Mass Memory Unit}
\acrodef{MPE}{Max-Planck-Institut für extraterrestrische Physik}
\acrodef{MPIA}{Max-Planck-Institut für Astronomie}
\acrodef{MW}{Milky Way}
\acrodef{NA}{numerical aperture}
\acrodef{NASA}{National Aeronautic and Space Administration}
\acrodef{JPL}{NASA Jet Propulsion Laboratory}
\acrodef{MZ-CGH}{multi-zonal computer-generated hologram}
\acrodef{NGVS}{Next Generation Virgo Survey}
\acrodef{NI-CU}{NISP calibration unit}
\acrodef{NI-OA}{near-infrared optical assembly}
\acrodef{NI-GWA}{NISP Grism Wheel Assembly}
\acrodef{NIR}{near-infrared}
\acrodef{NISP}{Near-Infrared Spectrometer and Photometer}
\acrodef{NSCs}{Nuclear star clusters}
\acrodef{PARMS}{plasma-assisted reactive magnetron sputtering}
\acrodef{PLM}{payload module}
\acrodef{PTFE}{polytetrafluoroethylene}
\acrodef{PV}{performance verification}
\acrodef{PWM}{pulse-width modulation}
\acrodef{PSF}{point spread function}
\acrodef{QE}{quantum efficiency}
\acrodef{RGB}{red-green-blue}
\acrodef{RMS}{root mean square}
\acrodef{ROI}[RoI]{region of interest}
\acrodef{ROIC}{readout-integrated chip}
\acrodef{ROS}{reference observing sequence}
\acrodef{SBF}{surface brightness fluctuation}
\acrodef{SCA}{sensor chip array}
\acrodef{SCE}{sensor chip electronic}
\acrodef{SCS}{sensor chip system}
\acrodef{SED}{spectral energy distribution}
\acrodef{SDSS}{Sloan Digital Sky Survey}
\acrodef{SGS}{science ground segment}
\acrodef{SHS}{Shack-Hartmann sensor}
\acrodef{SMF}{stellar mass function}
\acrodef{SNR}[SNR]{signal-to-noise ratio}
\acrodef{SED}{spectral energy distribution}
\acrodef{SiC}{silicon carbide}
\acrodef{SVM}{service module}
\acrodef{UCDs}{ultra compact dwarfs}
\acrodef{UDGs}{ultra diffuse galaxies}
\acrodef{UNIONS}{Ultraviolet Near Infrared Optical Northern Survey}
\acrodef{VGC}{Virgo cluster catalogue}
\acrodef{VIS}{visible imager}
\acrodef{WD}{white dwarf}
\acrodef{WCS}{world coordinate system}
\acrodef{WFE}{wavefront error}
\acrodef{ZP}{zero point}

%% file: authorsandaffiliations.tex
% Injected 20 May 2024 17:23
%%%% please do not edit the author list -- contact ECEB Bureau for changes
% \newcommand{\orcid}[1]{} %% if already defined in aa.cls: comment, or use renewcommand	
%%%% Version Monday 24th of March 2025 02:26:35 PM UT												
%%%% Please do not edit the author list -- contact ECEB Bureau for changes
%\newcommand{\orcid}[1]{} %% if already defined in aa.cls: comment, or use renewcommand			   
\author{Koshy~George\orcid{0000-0002-1734-8455}\thanks{\email{Koshy.George@physik.lmu.de}}\inst{\ref{aff1}}
\and A.~Boselli\inst{\ref{aff2}}
\and J.-C.~Cuillandre\orcid{0000-0002-3263-8645}\inst{\ref{aff3}}
\and M.~K\"ummel\orcid{0000-0003-2791-2117}\inst{\ref{aff1}}
\and A.~Lan\c{c}on\orcid{0000-0002-7214-8296}\inst{\ref{aff4}}
\and C.~Bellhouse\orcid{0000-0002-6179-8007}\inst{\ref{aff5}}
\and T.~Saifollahi\orcid{0000-0002-9554-7660}\inst{\ref{aff4}}
\and M.~Mondelin\orcid{0009-0004-5954-0930}\inst{\ref{aff3}}
\and M.~Bolzonella\orcid{0000-0003-3278-4607}\inst{\ref{aff6}}
\and P.~Joseph\orcid{0000-0003-1409-1903}\inst{\ref{aff7},\ref{aff8}}
\and I.~D.~Roberts\orcid{0000-0002-0692-0911}\inst{\ref{aff9}}
\and R.~J.~van~Weeren\orcid{0000-0002-0587-1660}\inst{\ref{aff10}}
\and Q.~Liu\orcid{0000-0002-7490-5991}\inst{\ref{aff10}}
\and E.~Sola\orcid{0000-0002-2814-3578}\inst{\ref{aff11}}
\and M.~Urbano\orcid{0000-0001-5640-0650}\inst{\ref{aff4}}
\and M.~Baes\orcid{0000-0002-3930-2757}\inst{\ref{aff12}}
\and R.~F.~Peletier\orcid{0000-0001-7621-947X}\inst{\ref{aff13}}
\and M.~Klein\inst{\ref{aff14}}
\and C.~T.~Davies\orcid{0000-0003-0015-263X}\inst{\ref{aff14}}
\and I.~A.~Zinchenko\orcid{0000-0002-2944-2449}\inst{\ref{aff1}}
\and J.~G.~Sorce\orcid{0000-0002-2307-2432}\inst{\ref{aff15},\ref{aff16}}
\and M.~Poulain\orcid{0000-0002-7664-4510}\inst{\ref{aff17}}
\and N.~Aghanim\orcid{0000-0002-6688-8992}\inst{\ref{aff16}}
\and B.~Altieri\orcid{0000-0003-3936-0284}\inst{\ref{aff18}}
\and A.~Amara\inst{\ref{aff19}}
\and S.~Andreon\orcid{0000-0002-2041-8784}\inst{\ref{aff20}}
\and N.~Auricchio\orcid{0000-0003-4444-8651}\inst{\ref{aff6}}
\and C.~Baccigalupi\orcid{0000-0002-8211-1630}\inst{\ref{aff21},\ref{aff22},\ref{aff23},\ref{aff24}}
\and M.~Baldi\orcid{0000-0003-4145-1943}\inst{\ref{aff25},\ref{aff6},\ref{aff26}}
\and A.~Balestra\orcid{0000-0002-6967-261X}\inst{\ref{aff27}}
\and S.~Bardelli\orcid{0000-0002-8900-0298}\inst{\ref{aff6}}
\and P.~Battaglia\orcid{0000-0002-7337-5909}\inst{\ref{aff6}}
\and A.~Biviano\orcid{0000-0002-0857-0732}\inst{\ref{aff22},\ref{aff21}}
\and D.~Bonino\orcid{0000-0002-3336-9977}\inst{\ref{aff28}}
\and E.~Branchini\orcid{0000-0002-0808-6908}\inst{\ref{aff29},\ref{aff30},\ref{aff20}}
\and M.~Brescia\orcid{0000-0001-9506-5680}\inst{\ref{aff31},\ref{aff32}}
\and J.~Brinchmann\orcid{0000-0003-4359-8797}\inst{\ref{aff33},\ref{aff34}}
\and S.~Camera\orcid{0000-0003-3399-3574}\inst{\ref{aff35},\ref{aff36},\ref{aff28}}
\and G.~Ca\~nas-Herrera\orcid{0000-0003-2796-2149}\inst{\ref{aff37},\ref{aff38},\ref{aff10}}
\and V.~Capobianco\orcid{0000-0002-3309-7692}\inst{\ref{aff28}}
\and C.~Carbone\orcid{0000-0003-0125-3563}\inst{\ref{aff39}}
\and J.~Carretero\orcid{0000-0002-3130-0204}\inst{\ref{aff40},\ref{aff41}}
\and S.~Casas\orcid{0000-0002-4751-5138}\inst{\ref{aff42}}
\and M.~Castellano\orcid{0000-0001-9875-8263}\inst{\ref{aff43}}
\and G.~Castignani\orcid{0000-0001-6831-0687}\inst{\ref{aff6}}
\and S.~Cavuoti\orcid{0000-0002-3787-4196}\inst{\ref{aff32},\ref{aff44}}
\and K.~C.~Chambers\orcid{0000-0001-6965-7789}\inst{\ref{aff45}}
\and A.~Cimatti\inst{\ref{aff46}}
\and C.~Colodro-Conde\inst{\ref{aff47}}
\and G.~Congedo\orcid{0000-0003-2508-0046}\inst{\ref{aff48}}
\and C.~J.~Conselice\orcid{0000-0003-1949-7638}\inst{\ref{aff49}}
\and L.~Conversi\orcid{0000-0002-6710-8476}\inst{\ref{aff50},\ref{aff18}}
\and Y.~Copin\orcid{0000-0002-5317-7518}\inst{\ref{aff51}}
\and F.~Courbin\orcid{0000-0003-0758-6510}\inst{\ref{aff52},\ref{aff53}}
\and H.~M.~Courtois\orcid{0000-0003-0509-1776}\inst{\ref{aff54}}
\and M.~Cropper\orcid{0000-0003-4571-9468}\inst{\ref{aff55}}
\and A.~Da~Silva\orcid{0000-0002-6385-1609}\inst{\ref{aff56},\ref{aff57}}
\and H.~Degaudenzi\orcid{0000-0002-5887-6799}\inst{\ref{aff58}}
\and G.~De~Lucia\orcid{0000-0002-6220-9104}\inst{\ref{aff22}}
\and A.~M.~Di~Giorgio\orcid{0000-0002-4767-2360}\inst{\ref{aff59}}
\and H.~Dole\orcid{0000-0002-9767-3839}\inst{\ref{aff16}}
\and M.~Douspis\orcid{0000-0003-4203-3954}\inst{\ref{aff16}}
\and F.~Dubath\orcid{0000-0002-6533-2810}\inst{\ref{aff58}}
\and X.~Dupac\inst{\ref{aff18}}
\and S.~Dusini\orcid{0000-0002-1128-0664}\inst{\ref{aff60}}
\and S.~Escoffier\orcid{0000-0002-2847-7498}\inst{\ref{aff61}}
\and M.~Farina\orcid{0000-0002-3089-7846}\inst{\ref{aff59}}
\and F.~Faustini\orcid{0000-0001-6274-5145}\inst{\ref{aff43},\ref{aff62}}
\and S.~Ferriol\inst{\ref{aff51}}
\and S.~Fotopoulou\orcid{0000-0002-9686-254X}\inst{\ref{aff63}}
\and M.~Frailis\orcid{0000-0002-7400-2135}\inst{\ref{aff22}}
\and E.~Franceschi\orcid{0000-0002-0585-6591}\inst{\ref{aff6}}
\and S.~Galeotta\orcid{0000-0002-3748-5115}\inst{\ref{aff22}}
\and B.~Gillis\orcid{0000-0002-4478-1270}\inst{\ref{aff48}}
\and C.~Giocoli\orcid{0000-0002-9590-7961}\inst{\ref{aff6},\ref{aff26}}
\and J.~Gracia-Carpio\inst{\ref{aff64}}
\and A.~Grazian\orcid{0000-0002-5688-0663}\inst{\ref{aff27}}
\and F.~Grupp\inst{\ref{aff64},\ref{aff1}}
\and S.~V.~H.~Haugan\orcid{0000-0001-9648-7260}\inst{\ref{aff65}}
\and W.~Holmes\inst{\ref{aff66}}
\and I.~M.~Hook\orcid{0000-0002-2960-978X}\inst{\ref{aff67}}
\and F.~Hormuth\inst{\ref{aff68}}
\and A.~Hornstrup\orcid{0000-0002-3363-0936}\inst{\ref{aff69},\ref{aff70}}
\and P.~Hudelot\inst{\ref{aff71}}
\and K.~Jahnke\orcid{0000-0003-3804-2137}\inst{\ref{aff72}}
\and M.~Jhabvala\inst{\ref{aff73}}
\and E.~Keih\"anen\orcid{0000-0003-1804-7715}\inst{\ref{aff74}}
\and S.~Kermiche\orcid{0000-0002-0302-5735}\inst{\ref{aff61}}
\and A.~Kiessling\orcid{0000-0002-2590-1273}\inst{\ref{aff66}}
\and B.~Kubik\orcid{0009-0006-5823-4880}\inst{\ref{aff51}}
\and M.~Kunz\orcid{0000-0002-3052-7394}\inst{\ref{aff75}}
\and H.~Kurki-Suonio\orcid{0000-0002-4618-3063}\inst{\ref{aff76},\ref{aff77}}
\and A.~M.~C.~Le~Brun\orcid{0000-0002-0936-4594}\inst{\ref{aff78}}
\and D.~Le~Mignant\orcid{0000-0002-5339-5515}\inst{\ref{aff2}}
\and S.~Ligori\orcid{0000-0003-4172-4606}\inst{\ref{aff28}}
\and P.~B.~Lilje\orcid{0000-0003-4324-7794}\inst{\ref{aff65}}
\and V.~Lindholm\orcid{0000-0003-2317-5471}\inst{\ref{aff76},\ref{aff77}}
\and I.~Lloro\orcid{0000-0001-5966-1434}\inst{\ref{aff79}}
\and G.~Mainetti\orcid{0000-0003-2384-2377}\inst{\ref{aff80}}
\and D.~Maino\inst{\ref{aff81},\ref{aff39},\ref{aff82}}
\and E.~Maiorano\orcid{0000-0003-2593-4355}\inst{\ref{aff6}}
\and O.~Mansutti\orcid{0000-0001-5758-4658}\inst{\ref{aff22}}
\and O.~Marggraf\orcid{0000-0001-7242-3852}\inst{\ref{aff83}}
\and K.~Markovic\orcid{0000-0001-6764-073X}\inst{\ref{aff66}}
\and M.~Martinelli\orcid{0000-0002-6943-7732}\inst{\ref{aff43},\ref{aff84}}
\and N.~Martinet\orcid{0000-0003-2786-7790}\inst{\ref{aff2}}
\and F.~Marulli\orcid{0000-0002-8850-0303}\inst{\ref{aff85},\ref{aff6},\ref{aff26}}
\and R.~Massey\orcid{0000-0002-6085-3780}\inst{\ref{aff86}}
\and S.~Maurogordato\inst{\ref{aff87}}
\and E.~Medinaceli\orcid{0000-0002-4040-7783}\inst{\ref{aff6}}
\and S.~Mei\orcid{0000-0002-2849-559X}\inst{\ref{aff88},\ref{aff89}}
\and Y.~Mellier\inst{\ref{aff90},\ref{aff71}}
\and M.~Meneghetti\orcid{0000-0003-1225-7084}\inst{\ref{aff6},\ref{aff26}}
\and E.~Merlin\orcid{0000-0001-6870-8900}\inst{\ref{aff43}}
\and G.~Meylan\inst{\ref{aff91}}
\and J.~J.~Mohr\orcid{0000-0002-6875-2087}\inst{\ref{aff92}}
\and A.~Mora\orcid{0000-0002-1922-8529}\inst{\ref{aff93}}
\and M.~Moresco\orcid{0000-0002-7616-7136}\inst{\ref{aff85},\ref{aff6}}
\and L.~Moscardini\orcid{0000-0002-3473-6716}\inst{\ref{aff85},\ref{aff6},\ref{aff26}}
\and R.~Nakajima\orcid{0009-0009-1213-7040}\inst{\ref{aff83}}
\and C.~Neissner\orcid{0000-0001-8524-4968}\inst{\ref{aff94},\ref{aff41}}
\and R.~C.~Nichol\orcid{0000-0003-0939-6518}\inst{\ref{aff19}}
\and S.-M.~Niemi\orcid{0009-0005-0247-0086}\inst{\ref{aff37}}
\and J.~W.~Nightingale\orcid{0000-0002-8987-7401}\inst{\ref{aff95}}
\and C.~Padilla\orcid{0000-0001-7951-0166}\inst{\ref{aff94}}
\and S.~Paltani\orcid{0000-0002-8108-9179}\inst{\ref{aff58}}
\and F.~Pasian\orcid{0000-0002-4869-3227}\inst{\ref{aff22}}
\and K.~Pedersen\inst{\ref{aff96}}
\and W.~J.~Percival\orcid{0000-0002-0644-5727}\inst{\ref{aff9},\ref{aff97},\ref{aff98}}
\and V.~Pettorino\inst{\ref{aff37}}
\and S.~Pires\orcid{0000-0002-0249-2104}\inst{\ref{aff3}}
\and G.~Polenta\orcid{0000-0003-4067-9196}\inst{\ref{aff62}}
\and M.~Poncet\inst{\ref{aff99}}
\and L.~A.~Popa\inst{\ref{aff100}}
\and L.~Pozzetti\orcid{0000-0001-7085-0412}\inst{\ref{aff6}}
\and F.~Raison\orcid{0000-0002-7819-6918}\inst{\ref{aff64}}
\and R.~Rebolo\orcid{0000-0003-3767-7085}\inst{\ref{aff47},\ref{aff101},\ref{aff102}}
\and A.~Renzi\orcid{0000-0001-9856-1970}\inst{\ref{aff103},\ref{aff60}}
\and J.~Rhodes\orcid{0000-0002-4485-8549}\inst{\ref{aff66}}
\and G.~Riccio\inst{\ref{aff32}}
\and E.~Romelli\orcid{0000-0003-3069-9222}\inst{\ref{aff22}}
\and M.~Roncarelli\orcid{0000-0001-9587-7822}\inst{\ref{aff6}}
\and E.~Rossetti\orcid{0000-0003-0238-4047}\inst{\ref{aff25}}
\and R.~Saglia\orcid{0000-0003-0378-7032}\inst{\ref{aff1},\ref{aff64}}
\and Z.~Sakr\orcid{0000-0002-4823-3757}\inst{\ref{aff104},\ref{aff105},\ref{aff106}}
\and D.~Sapone\orcid{0000-0001-7089-4503}\inst{\ref{aff107}}
\and B.~Sartoris\orcid{0000-0003-1337-5269}\inst{\ref{aff1},\ref{aff22}}
\and J.~A.~Schewtschenko\orcid{0000-0002-4913-6393}\inst{\ref{aff48}}
\and M.~Schirmer\orcid{0000-0003-2568-9994}\inst{\ref{aff72}}
\and P.~Schneider\orcid{0000-0001-8561-2679}\inst{\ref{aff83}}
\and A.~Secroun\orcid{0000-0003-0505-3710}\inst{\ref{aff61}}
\and G.~Seidel\orcid{0000-0003-2907-353X}\inst{\ref{aff72}}
\and M.~Seiffert\orcid{0000-0002-7536-9393}\inst{\ref{aff66}}
\and S.~Serrano\orcid{0000-0002-0211-2861}\inst{\ref{aff108},\ref{aff109},\ref{aff110}}
\and C.~Sirignano\orcid{0000-0002-0995-7146}\inst{\ref{aff103},\ref{aff60}}
\and G.~Sirri\orcid{0000-0003-2626-2853}\inst{\ref{aff26}}
\and L.~Stanco\orcid{0000-0002-9706-5104}\inst{\ref{aff60}}
\and J.~Steinwagner\orcid{0000-0001-7443-1047}\inst{\ref{aff64}}
\and P.~Tallada-Cresp\'{i}\orcid{0000-0002-1336-8328}\inst{\ref{aff40},\ref{aff41}}
\and A.~N.~Taylor\inst{\ref{aff48}}
\and I.~Tereno\orcid{0000-0002-4537-6218}\inst{\ref{aff56},\ref{aff111}}
\and S.~Toft\orcid{0000-0003-3631-7176}\inst{\ref{aff112},\ref{aff113}}
\and R.~Toledo-Moreo\orcid{0000-0002-2997-4859}\inst{\ref{aff114}}
\and F.~Torradeflot\orcid{0000-0003-1160-1517}\inst{\ref{aff41},\ref{aff40}}
\and I.~Tutusaus\orcid{0000-0002-3199-0399}\inst{\ref{aff105}}
\and E.~A.~Valentijn\inst{\ref{aff13}}
\and L.~Valenziano\orcid{0000-0002-1170-0104}\inst{\ref{aff6},\ref{aff115}}
\and J.~Valiviita\orcid{0000-0001-6225-3693}\inst{\ref{aff76},\ref{aff77}}
\and T.~Vassallo\orcid{0000-0001-6512-6358}\inst{\ref{aff1},\ref{aff22}}
\and G.~Verdoes~Kleijn\orcid{0000-0001-5803-2580}\inst{\ref{aff13}}
\and A.~Veropalumbo\orcid{0000-0003-2387-1194}\inst{\ref{aff20},\ref{aff30},\ref{aff29}}
\and Y.~Wang\orcid{0000-0002-4749-2984}\inst{\ref{aff116}}
\and J.~Weller\orcid{0000-0002-8282-2010}\inst{\ref{aff1},\ref{aff64}}
\and G.~Zamorani\orcid{0000-0002-2318-301X}\inst{\ref{aff6}}
\and F.~M.~Zerbi\inst{\ref{aff20}}
\and E.~Zucca\orcid{0000-0002-5845-8132}\inst{\ref{aff6}}
\and C.~Burigana\orcid{0000-0002-3005-5796}\inst{\ref{aff117},\ref{aff115}}
\and L.~Gabarra\orcid{0000-0002-8486-8856}\inst{\ref{aff118}}
\and J.~Mart\'{i}n-Fleitas\orcid{0000-0002-8594-569X}\inst{\ref{aff93}}
\and V.~Scottez\inst{\ref{aff90},\ref{aff119}}}
										   
%%%% please do not edit the affiliation list -- contact ECEB Bureau for changes
\institute{Universit\"ats-Sternwarte M\"unchen, Fakult\"at f\"ur Physik, Ludwig-Maximilians-Universit\"at M\"unchen, Scheinerstrasse 1, 81679 M\"unchen, Germany\label{aff1}
\and
Aix-Marseille Universit\'e, CNRS, CNES, LAM, Marseille, France\label{aff2}
\and
Universit\'e Paris-Saclay, Universit\'e Paris Cit\'e, CEA, CNRS, AIM, 91191, Gif-sur-Yvette, France\label{aff3}
\and
Universit\'e de Strasbourg, CNRS, Observatoire astronomique de Strasbourg, UMR 7550, 67000 Strasbourg, France\label{aff4}
\and
School of Physics and Astronomy, University of Nottingham, University Park, Nottingham NG7 2RD, UK\label{aff5}
\and
INAF-Osservatorio di Astrofisica e Scienza dello Spazio di Bologna, Via Piero Gobetti 93/3, 40129 Bologna, Italy\label{aff6}
\and
Indian Institute of Astrophysics, Koramangala II Block, Bangalore 560034, India\label{aff7}
\and
Department of Physics and Electronics, CHRIST (Deemed to be University), Bangalore 560029, India\label{aff8}
\and
Waterloo Centre for Astrophysics, University of Waterloo, Waterloo, Ontario N2L 3G1, Canada\label{aff9}
\and
Leiden Observatory, Leiden University, Einsteinweg 55, 2333 CC Leiden, The Netherlands\label{aff10}
\and
Institute of Astronomy, University of Cambridge, Madingley Road, Cambridge CB3 0HA, UK\label{aff11}
\and
Sterrenkundig Observatorium, Universiteit Gent, Krijgslaan 281 S9, 9000 Gent, Belgium\label{aff12}
\and
Kapteyn Astronomical Institute, University of Groningen, PO Box 800, 9700 AV Groningen, The Netherlands\label{aff13}
\and
Ludwig-Maximilians-University, Schellingstrasse 4, 80799 Munich, Germany\label{aff14}
\and
Univ. Lille, CNRS, Centrale Lille, UMR 9189 CRIStAL, 59000 Lille, France\label{aff15}
\and
Universit\'e Paris-Saclay, CNRS, Institut d'astrophysique spatiale, 91405, Orsay, France\label{aff16}
\and
Space physics and astronomy research unit, University of Oulu, Pentti Kaiteran katu 1, FI-90014 Oulu, Finland\label{aff17}
\and
ESAC/ESA, Camino Bajo del Castillo, s/n., Urb. Villafranca del Castillo, 28692 Villanueva de la Ca\~nada, Madrid, Spain\label{aff18}
\and
School of Mathematics and Physics, University of Surrey, Guildford, Surrey, GU2 7XH, UK\label{aff19}
\and
INAF-Osservatorio Astronomico di Brera, Via Brera 28, 20122 Milano, Italy\label{aff20}
\and
IFPU, Institute for Fundamental Physics of the Universe, via Beirut 2, 34151 Trieste, Italy\label{aff21}
\and
INAF-Osservatorio Astronomico di Trieste, Via G. B. Tiepolo 11, 34143 Trieste, Italy\label{aff22}
\and
INFN, Sezione di Trieste, Via Valerio 2, 34127 Trieste TS, Italy\label{aff23}
\and
SISSA, International School for Advanced Studies, Via Bonomea 265, 34136 Trieste TS, Italy\label{aff24}
\and
Dipartimento di Fisica e Astronomia, Universit\`a di Bologna, Via Gobetti 93/2, 40129 Bologna, Italy\label{aff25}
\and
INFN-Sezione di Bologna, Viale Berti Pichat 6/2, 40127 Bologna, Italy\label{aff26}
\and
INAF-Osservatorio Astronomico di Padova, Via dell'Osservatorio 5, 35122 Padova, Italy\label{aff27}
\and
INAF-Osservatorio Astrofisico di Torino, Via Osservatorio 20, 10025 Pino Torinese (TO), Italy\label{aff28}
\and
Dipartimento di Fisica, Universit\`a di Genova, Via Dodecaneso 33, 16146, Genova, Italy\label{aff29}
\and
INFN-Sezione di Genova, Via Dodecaneso 33, 16146, Genova, Italy\label{aff30}
\and
Department of Physics "E. Pancini", University Federico II, Via Cinthia 6, 80126, Napoli, Italy\label{aff31}
\and
INAF-Osservatorio Astronomico di Capodimonte, Via Moiariello 16, 80131 Napoli, Italy\label{aff32}
\and
Instituto de Astrof\'isica e Ci\^encias do Espa\c{c}o, Universidade do Porto, CAUP, Rua das Estrelas, PT4150-762 Porto, Portugal\label{aff33}
\and
Faculdade de Ci\^encias da Universidade do Porto, Rua do Campo de Alegre, 4150-007 Porto, Portugal\label{aff34}
\and
Dipartimento di Fisica, Universit\`a degli Studi di Torino, Via P. Giuria 1, 10125 Torino, Italy\label{aff35}
\and
INFN-Sezione di Torino, Via P. Giuria 1, 10125 Torino, Italy\label{aff36}
\and
European Space Agency/ESTEC, Keplerlaan 1, 2201 AZ Noordwijk, The Netherlands\label{aff37}
\and
Institute Lorentz, Leiden University, Niels Bohrweg 2, 2333 CA Leiden, The Netherlands\label{aff38}
\and
INAF-IASF Milano, Via Alfonso Corti 12, 20133 Milano, Italy\label{aff39}
\and
Centro de Investigaciones Energ\'eticas, Medioambientales y Tecnol\'ogicas (CIEMAT), Avenida Complutense 40, 28040 Madrid, Spain\label{aff40}
\and
Port d'Informaci\'{o} Cient\'{i}fica, Campus UAB, C. Albareda s/n, 08193 Bellaterra (Barcelona), Spain\label{aff41}
\and
Institute for Theoretical Particle Physics and Cosmology (TTK), RWTH Aachen University, 52056 Aachen, Germany\label{aff42}
\and
INAF-Osservatorio Astronomico di Roma, Via Frascati 33, 00078 Monteporzio Catone, Italy\label{aff43}
\and
INFN section of Naples, Via Cinthia 6, 80126, Napoli, Italy\label{aff44}
\and
Institute for Astronomy, University of Hawaii, 2680 Woodlawn Drive, Honolulu, HI 96822, USA\label{aff45}
\and
Dipartimento di Fisica e Astronomia "Augusto Righi" - Alma Mater Studiorum Universit\`a di Bologna, Viale Berti Pichat 6/2, 40127 Bologna, Italy\label{aff46}
\and
Instituto de Astrof\'{\i}sica de Canarias, V\'{\i}a L\'actea, 38205 La Laguna, Tenerife, Spain\label{aff47}
\and
Institute for Astronomy, University of Edinburgh, Royal Observatory, Blackford Hill, Edinburgh EH9 3HJ, UK\label{aff48}
\and
Jodrell Bank Centre for Astrophysics, Department of Physics and Astronomy, University of Manchester, Oxford Road, Manchester M13 9PL, UK\label{aff49}
\and
European Space Agency/ESRIN, Largo Galileo Galilei 1, 00044 Frascati, Roma, Italy\label{aff50}
\and
Universit\'e Claude Bernard Lyon 1, CNRS/IN2P3, IP2I Lyon, UMR 5822, Villeurbanne, F-69100, France\label{aff51}
\and
Institut de Ci\`{e}ncies del Cosmos (ICCUB), Universitat de Barcelona (IEEC-UB), Mart\'{i} i Franqu\`{e}s 1, 08028 Barcelona, Spain\label{aff52}
\and
Instituci\'o Catalana de Recerca i Estudis Avan\c{c}ats (ICREA), Passeig de Llu\'{\i}s Companys 23, 08010 Barcelona, Spain\label{aff53}
\and
UCB Lyon 1, CNRS/IN2P3, IUF, IP2I Lyon, 4 rue Enrico Fermi, 69622 Villeurbanne, France\label{aff54}
\and
Mullard Space Science Laboratory, University College London, Holmbury St Mary, Dorking, Surrey RH5 6NT, UK\label{aff55}
\and
Departamento de F\'isica, Faculdade de Ci\^encias, Universidade de Lisboa, Edif\'icio C8, Campo Grande, PT1749-016 Lisboa, Portugal\label{aff56}
\and
Instituto de Astrof\'isica e Ci\^encias do Espa\c{c}o, Faculdade de Ci\^encias, Universidade de Lisboa, Campo Grande, 1749-016 Lisboa, Portugal\label{aff57}
\and
Department of Astronomy, University of Geneva, ch. d'Ecogia 16, 1290 Versoix, Switzerland\label{aff58}
\and
INAF-Istituto di Astrofisica e Planetologia Spaziali, via del Fosso del Cavaliere, 100, 00100 Roma, Italy\label{aff59}
\and
INFN-Padova, Via Marzolo 8, 35131 Padova, Italy\label{aff60}
\and
Aix-Marseille Universit\'e, CNRS/IN2P3, CPPM, Marseille, France\label{aff61}
\and
Space Science Data Center, Italian Space Agency, via del Politecnico snc, 00133 Roma, Italy\label{aff62}
\and
School of Physics, HH Wills Physics Laboratory, University of Bristol, Tyndall Avenue, Bristol, BS8 1TL, UK\label{aff63}
\and
Max Planck Institute for Extraterrestrial Physics, Giessenbachstr. 1, 85748 Garching, Germany\label{aff64}
\and
Institute of Theoretical Astrophysics, University of Oslo, P.O. Box 1029 Blindern, 0315 Oslo, Norway\label{aff65}
\and
Jet Propulsion Laboratory, California Institute of Technology, 4800 Oak Grove Drive, Pasadena, CA, 91109, USA\label{aff66}
\and
Department of Physics, Lancaster University, Lancaster, LA1 4YB, UK\label{aff67}
\and
Felix Hormuth Engineering, Goethestr. 17, 69181 Leimen, Germany\label{aff68}
\and
Technical University of Denmark, Elektrovej 327, 2800 Kgs. Lyngby, Denmark\label{aff69}
\and
Cosmic Dawn Center (DAWN), Denmark\label{aff70}
\and
Institut d'Astrophysique de Paris, UMR 7095, CNRS, and Sorbonne Universit\'e, 98 bis boulevard Arago, 75014 Paris, France\label{aff71}
\and
Max-Planck-Institut f\"ur Astronomie, K\"onigstuhl 17, 69117 Heidelberg, Germany\label{aff72}
\and
NASA Goddard Space Flight Center, Greenbelt, MD 20771, USA\label{aff73}
\and
Department of Physics and Helsinki Institute of Physics, Gustaf H\"allstr\"omin katu 2, 00014 University of Helsinki, Finland\label{aff74}
\and
Universit\'e de Gen\`eve, D\'epartement de Physique Th\'eorique and Centre for Astroparticle Physics, 24 quai Ernest-Ansermet, CH-1211 Gen\`eve 4, Switzerland\label{aff75}
\and
Department of Physics, P.O. Box 64, 00014 University of Helsinki, Finland\label{aff76}
\and
Helsinki Institute of Physics, Gustaf H{\"a}llstr{\"o}min katu 2, University of Helsinki, Helsinki, Finland\label{aff77}
\and
Laboratoire d'etude de l'Univers et des phenomenes eXtremes, Observatoire de Paris, Universit\'e PSL, Sorbonne Universit\'e, CNRS, 92190 Meudon, France\label{aff78}
\and
SKA Observatory, Jodrell Bank, Lower Withington, Macclesfield, Cheshire SK11 9FT, UK\label{aff79}
\and
Centre de Calcul de l'IN2P3/CNRS, 21 avenue Pierre de Coubertin 69627 Villeurbanne Cedex, France\label{aff80}
\and
Dipartimento di Fisica "Aldo Pontremoli", Universit\`a degli Studi di Milano, Via Celoria 16, 20133 Milano, Italy\label{aff81}
\and
INFN-Sezione di Milano, Via Celoria 16, 20133 Milano, Italy\label{aff82}
\and
Universit\"at Bonn, Argelander-Institut f\"ur Astronomie, Auf dem H\"ugel 71, 53121 Bonn, Germany\label{aff83}
\and
INFN-Sezione di Roma, Piazzale Aldo Moro, 2 - c/o Dipartimento di Fisica, Edificio G. Marconi, 00185 Roma, Italy\label{aff84}
\and
Dipartimento di Fisica e Astronomia "Augusto Righi" - Alma Mater Studiorum Universit\`a di Bologna, via Piero Gobetti 93/2, 40129 Bologna, Italy\label{aff85}
\and
Department of Physics, Institute for Computational Cosmology, Durham University, South Road, Durham, DH1 3LE, UK\label{aff86}
\and
Universit\'e C\^{o}te d'Azur, Observatoire de la C\^{o}te d'Azur, CNRS, Laboratoire Lagrange, Bd de l'Observatoire, CS 34229, 06304 Nice cedex 4, France\label{aff87}
\and
Universit\'e Paris Cit\'e, CNRS, Astroparticule et Cosmologie, 75013 Paris, France\label{aff88}
\and
CNRS-UCB International Research Laboratory, Centre Pierre Binetruy, IRL2007, CPB-IN2P3, Berkeley, USA\label{aff89}
\and
Institut d'Astrophysique de Paris, 98bis Boulevard Arago, 75014, Paris, France\label{aff90}
\and
Institute of Physics, Laboratory of Astrophysics, Ecole Polytechnique F\'ed\'erale de Lausanne (EPFL), Observatoire de Sauverny, 1290 Versoix, Switzerland\label{aff91}
\and
University Observatory, LMU Faculty of Physics, Scheinerstrasse 1, 81679 Munich, Germany\label{aff92}
\and
Aurora Technology for European Space Agency (ESA), Camino bajo del Castillo, s/n, Urbanizacion Villafranca del Castillo, Villanueva de la Ca\~nada, 28692 Madrid, Spain\label{aff93}
\and
Institut de F\'{i}sica d'Altes Energies (IFAE), The Barcelona Institute of Science and Technology, Campus UAB, 08193 Bellaterra (Barcelona), Spain\label{aff94}
\and
School of Mathematics, Statistics and Physics, Newcastle University, Herschel Building, Newcastle-upon-Tyne, NE1 7RU, UK\label{aff95}
\and
DARK, Niels Bohr Institute, University of Copenhagen, Jagtvej 155, 2200 Copenhagen, Denmark\label{aff96}
\and
Department of Physics and Astronomy, University of Waterloo, Waterloo, Ontario N2L 3G1, Canada\label{aff97}
\and
Perimeter Institute for Theoretical Physics, Waterloo, Ontario N2L 2Y5, Canada\label{aff98}
\and
Centre National d'Etudes Spatiales -- Centre spatial de Toulouse, 18 avenue Edouard Belin, 31401 Toulouse Cedex 9, France\label{aff99}
\and
Institute of Space Science, Str. Atomistilor, nr. 409 M\u{a}gurele, Ilfov, 077125, Romania\label{aff100}
\and
Consejo Superior de Investigaciones Cientificas, Calle Serrano 117, 28006 Madrid, Spain\label{aff101}
\and
Universidad de La Laguna, Departamento de Astrof\'{\i}sica, 38206 La Laguna, Tenerife, Spain\label{aff102}
\and
Dipartimento di Fisica e Astronomia "G. Galilei", Universit\`a di Padova, Via Marzolo 8, 35131 Padova, Italy\label{aff103}
\and
Institut f\"ur Theoretische Physik, University of Heidelberg, Philosophenweg 16, 69120 Heidelberg, Germany\label{aff104}
\and
Institut de Recherche en Astrophysique et Plan\'etologie (IRAP), Universit\'e de Toulouse, CNRS, UPS, CNES, 14 Av. Edouard Belin, 31400 Toulouse, France\label{aff105}
\and
Universit\'e St Joseph; Faculty of Sciences, Beirut, Lebanon\label{aff106}
\and
Departamento de F\'isica, FCFM, Universidad de Chile, Blanco Encalada 2008, Santiago, Chile\label{aff107}
\and
Institut d'Estudis Espacials de Catalunya (IEEC),  Edifici RDIT, Campus UPC, 08860 Castelldefels, Barcelona, Spain\label{aff108}
\and
Satlantis, University Science Park, Sede Bld 48940, Leioa-Bilbao, Spain\label{aff109}
\and
Institute of Space Sciences (ICE, CSIC), Campus UAB, Carrer de Can Magrans, s/n, 08193 Barcelona, Spain\label{aff110}
\and
Instituto de Astrof\'isica e Ci\^encias do Espa\c{c}o, Faculdade de Ci\^encias, Universidade de Lisboa, Tapada da Ajuda, 1349-018 Lisboa, Portugal\label{aff111}
\and
Cosmic Dawn Center (DAWN)\label{aff112}
\and
Niels Bohr Institute, University of Copenhagen, Jagtvej 128, 2200 Copenhagen, Denmark\label{aff113}
\and
Universidad Polit\'ecnica de Cartagena, Departamento de Electr\'onica y Tecnolog\'ia de Computadoras,  Plaza del Hospital 1, 30202 Cartagena, Spain\label{aff114}
\and
INFN-Bologna, Via Irnerio 46, 40126 Bologna, Italy\label{aff115}
\and
Infrared Processing and Analysis Center, California Institute of Technology, Pasadena, CA 91125, USA\label{aff116}
\and
INAF, Istituto di Radioastronomia, Via Piero Gobetti 101, 40129 Bologna, Italy\label{aff117}
\and
Department of Physics, Oxford University, Keble Road, Oxford OX1 3RH, UK\label{aff118}
\and
ICL, Junia, Universit\'e Catholique de Lille, LITL, 59000 Lille, France\label{aff119}}

%% file: paper.bbl
\begin{thebibliography}{148}
\expandafter\ifx\csname natexlab\endcsname\relax\def\natexlab#1{#1}\fi

\bibitem[{{Abramson} {et~al.}(2016){Abramson}, {Kenney}, {Crowl}, \& {Tal}}]{Abramson_2016}
{Abramson}, A., {Kenney}, J., {Crowl}, H., \& {Tal}, T. 2016, \aj, 152, 32

\bibitem[{{Abramson} \& {Kenney}(2014)}]{Abramson_2014}
{Abramson}, A. \& {Kenney}, J. D.~P. 2014, \aj, 147, 63

\bibitem[{{Agrawal}(2006)}]{Agrawal_2006}
{Agrawal}, P.~C. 2006, Advances in Space Research, 38, 2989

\bibitem[{{Aguerri} {et~al.}(2020){Aguerri}, {Girardi}, {Agulli}, {Negri}, {Dalla Vecchia}, \& {Dom{\'\i}nguez Palmero}}]{Aguerri_2020}
{Aguerri}, J.~A.~L., {Girardi}, M., {Agulli}, I., {et~al.} 2020, \mnras, 494, 1681

\bibitem[{{Barnes}(2004)}]{Barnes_2004}
{Barnes}, J.~E. 2004, \mnras, 350, 798

\bibitem[{{Barnes} \& {Hernquist}(1992)}]{Barnes_1992}
{Barnes}, J.~E. \& {Hernquist}, L. 1992, \araa, 30, 705

\bibitem[{{Behroozi} {et~al.}(2013){Behroozi}, {Wechsler}, \& {Conroy}}]{Behroozi_2013}
{Behroozi}, P.~S., {Wechsler}, R.~H., \& {Conroy}, C. 2013, \apj, 770, 57

\bibitem[{{Bellhouse} {et~al.}(2017){Bellhouse}, {Jaff{\'e}}, {Hau}, {McGee}, {Poggianti}, {Moretti}, {Gullieuszik}, {Bettoni}, {Fasano}, {D'Onofrio}, {Fritz}, {Omizzolo}, {Sheen}, \& {Vulcani}}]{Bellhouse_2017}
{Bellhouse}, C., {Jaff{\'e}}, Y.~L., {Hau}, G.~K.~T., {et~al.} 2017, \apj, 844, 49

\bibitem[{{Bellhouse} {et~al.}(2021){Bellhouse}, {McGee}, {Smith}, {Poggianti}, {Jaff{\'e}}, {Kraljic}, {Franchetto}, {Fritz}, {Vulcani}, {Tonnesen}, {Roediger}, {Moretti}, {Gullieuszik}, \& {Shin}}]{Bellhouse_2021}
{Bellhouse}, C., {McGee}, S.~L., {Smith}, R., {et~al.} 2021, \mnras, 500, 1285

\bibitem[{{Bertin} \& {Arnouts}(1996)}]{Bertin_1996}
{Bertin}, E. \& {Arnouts}, S. 1996, \aaps, 117, 393

\bibitem[{{Bertin} {et~al.}(2020){Bertin}, {Schefer}, {Apostolakos}, {{\'A}lvarez-Ayll{\'o}n}, {Dubath}, \& {K{\"u}mmel}}]{Bertin_2020}
{Bertin}, E., {Schefer}, M., {Apostolakos}, N., {et~al.} 2020, in Astronomical Society of the Pacific Conference Series, Vol. 527, Astronomical Data Analysis Software and Systems XXIX, ed. R.~{Pizzo}, E.~R. {Deul}, J.~D. {Mol}, J.~{de Plaa}, \& H.~{Verkouter}, 461

\bibitem[{{B{\'\i}lek} {et~al.}(2020){B{\'\i}lek}, {Duc}, {Cuillandre}, {Gwyn}, {Cappellari}, {Bekaert}, {Bonfini}, {Bitsakis}, {Paudel}, {Krajnovi{\'c}}, {Durrell}, \& {Marleau}}]{Bilek_2020}
{B{\'\i}lek}, M., {Duc}, P.-A., {Cuillandre}, J.-C., {et~al.} 2020, \mnras, 498, 2138

\bibitem[{{B{\'\i}lek} {et~al.}(2022){B{\'\i}lek}, {Fensch}, {Ebrov{\'a}}, {Nagesh}, {Famaey}, {Duc}, \& {Kroupa}}]{Bilek_2022}
{B{\'\i}lek}, M., {Fensch}, J., {Ebrov{\'a}}, I., {et~al.} 2022, \aap, 660, A28

\bibitem[{Binney \& Tremaine(2008)}]{Binney_2008}
Binney, J. \& Tremaine, S. 2008, Galactic Dynamics: Second Edition, revised, 2 edn. (Princeton University Press)

\bibitem[{{Boissier} {et~al.}(2012){Boissier}, {Boselli}, {Duc}, {Cortese}, {van Driel}, {Heinis}, {Voyer}, {Cucciati}, {Ferrarese}, {C{\^o}t{\'e}}, {Cuillandre}, {Gwyn}, \& {Mei}}]{Boissier_2012}
{Boissier}, S., {Boselli}, A., {Duc}, P.~A., {et~al.} 2012, \aap, 545, A142

\bibitem[{{Bolzonella} {et~al.}(2010){Bolzonella}, {Kova{\v{c}}}, {Pozzetti}, {Zucca}, {Cucciati}, {Lilly}, {Peng}, {Iovino}, {Zamorani}, {Vergani}, {Tasca}, {Lamareille}, {Oesch}, {Caputi}, {Kampczyk}, {Bardelli}, {Maier}, {Abbas}, {Knobel}, {Scodeggio}, {Carollo}, {Contini}, {Kneib}, {Le F{\`e}vre}, {Mainieri}, {Renzini}, {Bongiorno}, {Coppa}, {de la Torre}, {de Ravel}, {Franzetti}, {Garilli}, {Le Borgne}, {Le Brun}, {Mignoli}, {Pell{\'o}}, {Perez-Montero}, {Ricciardelli}, {Silverman}, {Tanaka}, {Tresse}, {Bottini}, {Cappi}, {Cassata}, {Cimatti}, {Guzzo}, {Koekemoer}, {Leauthaud}, {Maccagni}, {Marinoni}, {McCracken}, {Memeo}, {Meneux}, {Porciani}, {Scaramella}, {Aussel}, {Capak}, {Halliday}, {Ilbert}, {Kartaltepe}, {Salvato}, {Sanders}, {Scarlata}, {Scoville}, {Taniguchi}, \& {Thompson}}]{Bolzonella_2010}
{Bolzonella}, M., {Kova{\v{c}}}, K., {Pozzetti}, L., {et~al.} 2010, \aap, 524, A76

\bibitem[{{Bolzonella} {et~al.}(2000){Bolzonella}, {Miralles}, \& {Pell{\'o}}}]{Bolzonella_2000}
{Bolzonella}, M., {Miralles}, J.~M., \& {Pell{\'o}}, R. 2000, \aap, 363, 476

\bibitem[{{Boselli} {et~al.}(2009){Boselli}, {Boissier}, {Cortese}, {Buat}, {Hughes}, \& {Gavazzi}}]{Boselli_2009}
{Boselli}, A., {Boissier}, S., {Cortese}, L., {et~al.} 2009, \apj, 706, 1527

\bibitem[{{Boselli} {et~al.}(2016){Boselli}, {Cuillandre}, {Fossati}, {Boissier}, {Bomans}, {Consolandi}, {Anselmi}, {Cortese}, {C{\^o}t{\'e}}, {Durrell}, {Ferrarese}, {Fumagalli}, {Gavazzi}, {Gwyn}, {Hensler}, {Sun}, \& {Toloba}}]{Boselli_2016}
{Boselli}, A., {Cuillandre}, J.~C., {Fossati}, M., {et~al.} 2016, \aap, 587, A68

\bibitem[{{Boselli} {et~al.}(2018){Boselli}, {Fossati}, {Cuillandre}, {Boissier}, {Boquien}, {Buat}, {Burgarella}, {Consolandi}, {Cortese}, {C{\^o}t{\'e}}, {C{\^o}t{\'e}}, {Durrell}, {Ferrarese}, {Fumagalli}, {Gavazzi}, {Gwyn}, {Hensler}, {Koribalski}, {Roediger}, {Roehlly}, {Russeil}, {Sun}, {Toloba}, {Vollmer}, \& {Zavagno}}]{Boselli_2018}
{Boselli}, A., {Fossati}, M., {Cuillandre}, J.~C., {et~al.} 2018, \aap, 615, A114

\bibitem[{{Boselli} {et~al.}(2022){Boselli}, {Fossati}, \& {Sun}}]{Boselli_2022}
{Boselli}, A., {Fossati}, M., \& {Sun}, M. 2022, \aapr, 30, 3

\bibitem[{{Boselli} \& {Gavazzi}(2006)}]{Boselli_2006}
{Boselli}, A. \& {Gavazzi}, G. 2006, \pasp, 118, 517

\bibitem[{{Boselli} \& {Gavazzi}(2014)}]{Boselli_2014}
{Boselli}, A. \& {Gavazzi}, G. 2014, \aapr, 22, 74

\bibitem[{{Boselli} {et~al.}(2002){Boselli}, {Lequeux}, \& {Gavazzi}}]{Boselli_2002}
{Boselli}, A., {Lequeux}, J., \& {Gavazzi}, G. 2002, \aap, 384, 33

\bibitem[{{Boselli} {et~al.}(2021){Boselli}, {Lupi}, {Epinat}, {Amram}, {Fossati}, {Anderson}, {Boissier}, {Boquien}, {Consolandi}, {C{\^o}t{\'e}}, {Cuillandre}, {Ferrarese}, {Galbany}, {Gavazzi}, {G{\'o}mez-L{\'o}pez}, {Gwyn}, {Hensler}, {Hutchings}, {Kuncarayakti}, {Longobardi}, {Peng}, {Plana}, {Postma}, {Roediger}, {Roehlly}, {Schimd}, {Trinchieri}, \& {Vollmer}}]{Boselli_2021}
{Boselli}, A., {Lupi}, A., {Epinat}, B., {et~al.} 2021, \aap, 646, A139

\bibitem[{{Boselli} {et~al.}(2023){Boselli}, {Serra}, {de Gasperin}, {Vollmer}, {Amram}, {Edler}, {Fossati}, {Consolandi}, {C{\^o}t{\'e}}, {Cuillandre}, {Ferrarese}, {Gwyn}, {Postma}, {Boquien}, {Braine}, {Combes}, {Gavazzi}, {Hensler}, {Miville-Deschenes}, {Murgia}, {Roediger}, {Roehlly}, {Smith}, {Zhang}, \& {Zabel}}]{Boselli_2023}
{Boselli}, A., {Serra}, P., {de Gasperin}, F., {et~al.} 2023, \aap, 676, A92

\bibitem[{{Brinchmann} {et~al.}(2004){Brinchmann}, {Charlot}, {White}, {Tremonti}, {Kauffmann}, {Heckman}, \& {Brinkmann}}]{Brinchmann_2004}
{Brinchmann}, J., {Charlot}, S., {White}, S.~D.~M., {et~al.} 2004, \mnras, 351, 1151

\bibitem[{{Bruzual} \& {Charlot}(2003)}]{BC_2003}
{Bruzual}, G. \& {Charlot}, S. 2003, \mnras, 344, 1000

\bibitem[{{Burkhart} \& {Loeb}(2016)}]{Burkhart_2016}
{Burkhart}, B. \& {Loeb}, A. 2016, \apjl, 824, L7

\bibitem[{{Cardelli} {et~al.}(1989){Cardelli}, {Clayton}, \& {Mathis}}]{Cardelli_1989}
{Cardelli}, J.~A., {Clayton}, G.~C., \& {Mathis}, J.~S. 1989, \apj, 345, 245

\bibitem[{{Chung} {et~al.}(2009){Chung}, {van Gorkom}, {Kenney}, {Crowl}, \& {Vollmer}}]{Chung_2009}
{Chung}, A., {van Gorkom}, J.~H., {Kenney}, J. D.~P., {Crowl}, H., \& {Vollmer}, B. 2009, \aj, 138, 1741

\bibitem[{{Churazov} {et~al.}(2003){Churazov}, {Forman}, {Jones}, \& {B{\"o}hringer}}]{Churazov_2003}
{Churazov}, E., {Forman}, W., {Jones}, C., \& {B{\"o}hringer}, H. 2003, \apj, 590, 225

\bibitem[{{Cid Fernandes} {et~al.}(2011){Cid Fernandes}, {Stasi{\'n}ska}, {Mateus}, \& {Vale Asari}}]{CidFernandes_2011}
{Cid Fernandes}, R., {Stasi{\'n}ska}, G., {Mateus}, A., \& {Vale Asari}, N. 2011, \mnras, 413, 1687

\bibitem[{{Cortese} {et~al.}(2012){Cortese}, {Boissier}, {Boselli}, {Bendo}, {Buat}, {Davies}, {Eales}, {Heinis}, {Isaak}, \& {Madden}}]{Cortese_2012}
{Cortese}, L., {Boissier}, S., {Boselli}, A., {et~al.} 2012, \aap, 544, A101

\bibitem[{{Cortese} {et~al.}(2021){Cortese}, {Catinella}, \& {Smith}}]{Cortese_2021}
{Cortese}, L., {Catinella}, B., \& {Smith}, R. 2021, \pasa, 38, e035

\bibitem[{{Cortese} {et~al.}(2007){Cortese}, {Marcillac}, {Richard}, {Bravo-Alfaro}, {Kneib}, {Rieke}, {Covone}, {Egami}, {Rigby}, {Czoske}, \& {Davies}}]{Cortese_2007}
{Cortese}, L., {Marcillac}, D., {Richard}, J., {et~al.} 2007, \mnras, 376, 157

\bibitem[{{Cowie} \& {Songaila}(1977)}]{Cowie_1977}
{Cowie}, L.~L. \& {Songaila}, A. 1977, \nat, 266, 501

\bibitem[{{Cramer} {et~al.}(2019){Cramer}, {Kenney}, {Sun}, {Crowl}, {Yagi}, {J{\'a}chym}, {Roediger}, \& {Waldron}}]{Cramer_2019}
{Cramer}, W.~J., {Kenney}, J.~D.~P., {Sun}, M., {et~al.} 2019, \apj, 870, 63

\bibitem[{{Cropper} {et~al.}(2016){Cropper}, {Pottinger}, {Niemi}, {Azzollini}, {Denniston}, {Szafraniec}, {Awan}, {Mellier}, {Berthe}, {Martignac}, {Cara}, {Di Giorgio}, {Sciortino}, {Bozzo}, {Genolet}, {Cole}, {Philippon}, {Hailey}, {Hunt}, {Swindells}, {Holland}, {Gow}, {Murray}, {Hall}, {Skottfelt}, {Amiaux}, {Laureijs}, {Racca}, {Salvignol}, {Short}, {Lorenzo Alvarez}, {Kitching}, {Hoekstra}, {Massey}, \& {Israel}}]{Cropper_2016}
{Cropper}, M., {Pottinger}, S., {Niemi}, S., {et~al.} 2016, in Society of Photo-Optical Instrumentation Engineers (SPIE) Conference Series, Vol. 9904, Space Telescopes and Instrumentation 2016: Optical, Infrared, and Millimeter Wave, ed. H.~A. {MacEwen}, G.~G. {Fazio}, M.~{Lystrup}, N.~{Batalha}, N.~{Siegler}, \& E.~C. {Tong}, 99040Q

\bibitem[{{Cropper} {et~al.}(2014){Cropper}, {Pottinger}, {Niemi}, {Denniston}, {Cole}, {Szafraniec}, {Mellier}, {Berth{\'e}}, {Martignac}, {Cara}, {di Giorgio}, {Sciortino}, {Paltani}, {Genolet}, {Fourmand}, {Charra}, {Guttridge}, {Winter}, {Endicott}, {Holland}, {Gow}, {Murray}, {Hall}, {Amiaux}, {Laureijs}, {Racca}, {Salvignol}, {Short}, {Lorenzo Alvarez}, {Kitching}, {Hoekstra}, \& {Massey}}]{Cropper_2014}
{Cropper}, M., {Pottinger}, S., {Niemi}, S.~M., {et~al.} 2014, in Society of Photo-Optical Instrumentation Engineers (SPIE) Conference Series, Vol. 9143, Space Telescopes and Instrumentation 2014: Optical, Infrared, and Millimeter Wave, ed. J.~M. {Oschmann}, Jr., M.~{Clampin}, G.~G. {Fazio}, \& H.~A. {MacEwen}, 91430J

\bibitem[{{Cuillandre} {et~al.}(2024{\natexlab{a}}){Cuillandre}, {Bertin}, {Bolzonella}, {et~al.}}]{Cuillandre_2024b}
{Cuillandre}, J.-C., {Bertin}, E., {Bolzonella}, M., {et~al.} 2024{\natexlab{a}}, \aap, accepted, arXiv:2405.13496

\bibitem[{{Cuillandre} {et~al.}(2024{\natexlab{b}}){Cuillandre}, {Bolzonella}, {Boselli}, {et~al.}}]{Cuillandre_2024a}
{Cuillandre}, J.-C., {Bolzonella}, M., {Boselli}, A., {et~al.} 2024{\natexlab{b}}, \aap, accepted, arXiv:2405.13501

\bibitem[{{Daddi} {et~al.}(2007){Daddi}, {Dickinson}, {Morrison}, {Chary}, {Cimatti}, {Elbaz}, {Frayer}, {Renzini}, {Pope}, {Alexander}, {Bauer}, {Giavalisco}, {Huynh}, {Kurk}, \& {Mignoli}}]{Daddi_2007}
{Daddi}, E., {Dickinson}, M., {Morrison}, G., {et~al.} 2007, \apj, 670, 156

\bibitem[{{Dressler}(2004)}]{Dressler_2004}
{Dressler}, A. 2004, in Clusters of Galaxies: Probes of Cosmological Structure and Galaxy Evolution, ed. J.~S. {Mulchaey}, A.~{Dressler}, \& A.~{Oemler}, 206

\bibitem[{{Duc} {et~al.}(2015){Duc}, {Cuillandre}, {Karabal}, {Cappellari}, {Alatalo}, {Blitz}, {Bournaud}, {Bureau}, {Crocker}, {Davies}, {Davis}, {de Zeeuw}, {Emsellem}, {Khochfar}, {Krajnovi{\'c}}, {Kuntschner}, {McDermid}, {Michel-Dansac}, {Morganti}, {Naab}, {Oosterloo}, {Paudel}, {Sarzi}, {Scott}, {Serra}, {Weijmans}, \& {Young}}]{Duc_2015}
{Duc}, P.-A., {Cuillandre}, J.-C., {Karabal}, E., {et~al.} 2015, \mnras, 446, 120

\bibitem[{{Durret} {et~al.}(2021){Durret}, {Chiche}, {Lobo}, \& {Jauzac}}]{Durret_2021}
{Durret}, F., {Chiche}, S., {Lobo}, C., \& {Jauzac}, M. 2021, \aap, 648, A63

\bibitem[{{Ebeling} {et~al.}(2014){Ebeling}, {Stephenson}, \& {Edge}}]{Ebeling_2014}
{Ebeling}, H., {Stephenson}, L.~N., \& {Edge}, A.~C. 2014, \apjl, 781, L40

\bibitem[{{Elbaz} {et~al.}(2007){Elbaz}, {Daddi}, {Le Borgne}, {Dickinson}, {Alexander}, {Chary}, {Starck}, {Brandt}, {Kitzbichler}, {MacDonald}, {Nonino}, {Popesso}, {Stern}, \& {Vanzella}}]{Elbaz_2007}
{Elbaz}, D., {Daddi}, E., {Le Borgne}, D., {et~al.} 2007, \aap, 468, 33

\bibitem[{{Elmegreen}(2010)}]{Elmegreen_2010}
{Elmegreen}, B.~G. 2010, in IAU Symposium, Vol. 266, Star Clusters: Basic Galactic Building Blocks Throughout Time and Space, ed. R.~{de Grijs} \& J.~R.~D. {L{\'e}pine}, 3--13

\bibitem[{{Elmegreen} {et~al.}(2014){Elmegreen}, {Elmegreen}, {Adamo}, {Aloisi}, {Andrews}, {Annibali}, {Bright}, {Calzetti}, {Cignoni}, {Evans}, {Gallagher}, {Gouliermis}, {Grebel}, {Hunter}, {Johnson}, {Kim}, {Lee}, {Sabbi}, {Smith}, {Thilker}, {Tosi}, \& {Ubeda}}]{Elmegreen_2014}
{Elmegreen}, D.~M., {Elmegreen}, B.~G., {Adamo}, A., {et~al.} 2014, \apjl, 787, L15

\bibitem[{{Euclid Collaboration: Cropper} {et~al.}(2024){Euclid Collaboration: Cropper}, {Al Bahlawan}, {Amiaux}, {et~al.}}]{Cropper_2024}
{Euclid Collaboration: Cropper}, M., {Al Bahlawan}, A., {Amiaux}, J., {et~al.} 2024, \aap, accepted, arXiv:2405.13492

\bibitem[{{Euclid Collaboration: Jahnke} {et~al.}(2024){Euclid Collaboration: Jahnke}, {Gillard}, {Schirmer}, {et~al.}}]{Jahnke_2024}
{Euclid Collaboration: Jahnke}, K., {Gillard}, W., {Schirmer}, M., {et~al.} 2024, \aap, accepted, arXiv:2405.13493

\bibitem[{{Euclid Collaboration: Mellier} {et~al.}(2024){Euclid Collaboration: Mellier}, {Abdurro'uf}, {Acevedo~Barroso}, {et~al.}}]{Mellier_2024}
{Euclid Collaboration: Mellier}, Y., {Abdurro'uf}, {Acevedo~Barroso}, J., {et~al.} 2024, \aap, accepted, arXiv:2405.13491

\bibitem[{{Euclid Collaboration: Scaramella} {et~al.}(2022){Euclid Collaboration: Scaramella}, {Amiaux}, {Mellier}, {Burigana}, {Carvalho}, {Cuillandre}, {Da Silva}, {Derosa}, {Dinis}, {Maiorano}, {Maris}, {Tereno}, {Laureijs}, {Boenke}, {Buenadicha}, {Dupac}, {Gaspar Venancio}, {G{\'o}mez-{\'A}lvarez}, {Hoar}, {Lorenzo Alvarez}, {Racca}, {Saavedra-Criado}, {Schwartz}, {Vavrek}, {Schirmer}, {Aussel}, {Azzollini}, {Cardone}, {Cropper}, {Ealet}, {Garilli}, {Gillard}, {Granett}, {Guzzo}, {Hoekstra}, {Jahnke}, {Kitching}, {Maciaszek}, {Meneghetti}, {Miller}, {Nakajima}, {Niemi}, {Pasian}, {Percival}, {Pottinger}, {Sauvage}, {Scodeggio}, {Wachter}, {Zacchei}, {Aghanim}, {Amara}, {Auphan}, {Auricchio}, {Awan}, {Balestra}, {Bender}, {Bodendorf}, {Bonino}, {Branchini}, {Brau-Nogue}, {Brescia}, {Candini}, {Capobianco}, {Carbone}, {Carlberg}, {Carretero}, {Casas}, {Castander}, {Castellano}, {Cavuoti}, {Cimatti}, {Cledassou}, {Congedo}, {Conselice}, {Conversi}, {Copin}, {Corcione}, {Costille}, {Courbin}, {Degaudenzi},
  {Douspis}, {Dubath}, {Duncan}, {Dusini}, {Farrens}, {Ferriol}, {Fosalba}, {Fourmanoit}, {Frailis}, {Franceschi}, {Franzetti}, {Fumana}, {Gillis}, {Giocoli}, {Grazian}, {Grupp}, {Haugan}, {Holmes}, {Hormuth}, {Hudelot}, {Kermiche}, {Kiessling}, {Kilbinger}, {Kohley}, {Kubik}, {K{\"u}mmel}, {Kunz}, {Kurki-Suonio}, {Lahav}, {Ligori}, {Lilje}, {Lloro}, {Mansutti}, {Marggraf}, {Markovic}, {Marulli}, {Massey}, {Maurogordato}, {Melchior}, {Merlin}, {Meylan}, {Mohr}, {Moresco}, {Morin}, {Moscardini}, {Munari}, {Nichol}, {Padilla}, {Paltani}, {Peacock}, {Pedersen}, {Pettorino}, {Pires}, {Poncet}, {Popa}, {Pozzetti}, {Raison}, {Rebolo}, {Rhodes}, {Rix}, {Roncarelli}, {Rossetti}, {Saglia}, {Schneider}, {Schrabback}, {Secroun}, {Seidel}, {Serrano}, {Sirignano}, {Sirri}, {Skottfelt}, {Stanco}, {Starck}, {Tallada-Cresp{\'\i}}, {Tavagnacco}, {Taylor}, {Teplitz}, {Toledo-Moreo}, {Torradeflot}, {Trifoglio}, {Valentijn}, {Valenziano}, {Verdoes Kleijn}, {Wang}, {Welikala}, {Weller}, {Wetzstein}, {Zamorani}, {Zoubian},
  {Andreon}, {Baldi}, {Bardelli}, {Boucaud}, {Camera}, {Di Ferdinando}, {Fabbian}, {Farinelli}, {Galeotta}, {Graci{\'a}-Carpio}, {Maino}, {Medinaceli}, {Mei}, {Neissner}, {Polenta}, {Renzi}, {Romelli}, {Rosset}, {Sureau}, {Tenti}, {Vassallo}, {Zucca}, {Baccigalupi}, {Balaguera-Antol{\'\i}nez}, {Battaglia}, {Biviano}, {Borgani}, {Bozzo}, {Cabanac}, {Cappi}, {Casas}, {Castignani}, {Colodro-Conde}, {Coupon}, {Courtois}, {Cuby}, {de la Torre}, {Desai}, {Dole}, {Fabricius}, {Farina}, {Ferreira}, {Finelli}, {Flose-Reimberg}, {Fotopoulou}, {Ganga}, {Gozaliasl}, {Hook}, {Keihanen}, {Kirkpatrick}, {Liebing}, {Lindholm}, {Mainetti}, {Martinelli}, {Martinet}, {Maturi}, {McCracken}, {Metcalf}, {Morgante}, {Nightingale}, {Nucita}, {Patrizii}, {Potter}, {Riccio}, {S{\'a}nchez}, {Sapone}, {Schewtschenko}, {Schultheis}, {Scottez}, {Teyssier}, {Tutusaus}, {Valiviita}, {Viel}, {Vriend}, \& {Whittaker}}]{Scaramella_2022}
{Euclid Collaboration: Scaramella}, R., {Amiaux}, J., {Mellier}, Y., {et~al.} 2022, \aap, 662, A112

\bibitem[{{Euclid Early Release Observations}(2025)}]{EROcite}
{Euclid Early Release Observations}. 2025, \url{https://doi.org/10.57780/esa-qmocze3}

\bibitem[{{Fossati} {et~al.}(2013){Fossati}, {Gavazzi}, {Savorgnan}, {Fumagalli}, {Boselli}, {Guti{\'e}rrez}, {Hern{\'a}ndez Toledo}, {Giovanelli}, \& {Haynes}}]{Fossati_2013}
{Fossati}, M., {Gavazzi}, G., {Savorgnan}, G., {et~al.} 2013, \aap, 553, A91

\bibitem[{{Fritz} {et~al.}(2017){Fritz}, {Moretti}, {Gullieuszik}, {Poggianti}, {Bruzual}, {Vulcani}, {Nicastro}, {Jaff{\'e}}, {Cervantes Sodi}, {Bettoni}, {Biviano}, {Fasano}, {Charlot}, {Bellhouse}, \& {Hau}}]{Fritz_2017}
{Fritz}, J., {Moretti}, A., {Gullieuszik}, M., {et~al.} 2017, \apj, 848, 132

\bibitem[{{Fumagalli} {et~al.}(2014){Fumagalli}, {Fossati}, {Hau}, {Gavazzi}, {Bower}, {Sun}, \& {Boselli}}]{Fumagalli_2014}
{Fumagalli}, M., {Fossati}, M., {Hau}, G. K.~T., {et~al.} 2014, \mnras, 445, 4335

\bibitem[{{Gannon} {et~al.}(2022){Gannon}, {Forbes}, {Romanowsky}, {Ferr{\'e}-Mateu}, {Couch}, {Brodie}, {Huang}, {Janssens}, \& {Okabe}}]{Gannon_2022}
{Gannon}, J.~S., {Forbes}, D.~A., {Romanowsky}, A.~J., {et~al.} 2022, \mnras, 510, 946

\bibitem[{{Gavazzi} {et~al.}(2001){Gavazzi}, {Boselli}, {Mayer}, {Iglesias-Paramo}, {V{\'\i}lchez}, \& {Carrasco}}]{Gavazzi_2001}
{Gavazzi}, G., {Boselli}, A., {Mayer}, L., {et~al.} 2001, \apjl, 563, L23

\bibitem[{{George} {et~al.}(2018){George}, {Poggianti}, {Gullieuszik}, {Fasano}, {Bellhouse}, {Postma}, {Moretti}, {Jaff{\'e}}, {Vulcani}, {Bettoni}, {Fritz}, {C{\^o}t{\'e}}, {Ghosh}, {Hutchings}, {Mohan}, {Sreekumar}, {Stalin}, {Subramaniam}, \& {Tandon}}]{George_2018}
{George}, K., {Poggianti}, B.~M., {Gullieuszik}, M., {et~al.} 2018, \mnras, 479, 4126

\bibitem[{{George} {et~al.}(2024){George}, {Poggianti}, {Omizzolo}, {Vulcani}, {C{\^o}t{\'e}}, {Postma}, {Smith}, {Jaffe}, {Gullieuszik}, {Moretti}, {Subramaniam}, {Sreekumar}, {Ghosh}, {Tandon}, \& {Hutchings}}]{George_2024}
{George}, K., {Poggianti}, B.~M., {Omizzolo}, A., {et~al.} 2024, \aap, 690, A337

\bibitem[{{George} {et~al.}(2023){George}, {Poggianti}, {Tomi{\v{c}}i{\'c}}, {Postma}, {C{\^o}t{\'e}}, {Fritz}, {Ghosh}, {Gullieuszik}, {Hutchings}, {Moretti}, {Omizzolo}, {Radovich}, {Sreekumar}, {Subramaniam}, {Tandon}, \& {Vulcani}}]{George_2023}
{George}, K., {Poggianti}, B.~M., {Tomi{\v{c}}i{\'c}}, N., {et~al.} 2023, \mnras, 519, 2426

\bibitem[{{Giunchi} {et~al.}(2023{\natexlab{a}}){Giunchi}, {Gullieuszik}, {Poggianti}, {Moretti}, {Werle}, {Scarlata}, {Zanella}, {Vulcani}, \& {Calzetti}}]{Giunchi_2023a}
{Giunchi}, E., {Gullieuszik}, M., {Poggianti}, B.~M., {et~al.} 2023{\natexlab{a}}, \apj, 949, 72

\bibitem[{{Giunchi} {et~al.}(2023{\natexlab{b}}){Giunchi}, {Poggianti}, {Gullieuszik}, {Moretti}, {Werle}, {Zanella}, {Vulcani}, {Tonnesen}, {Calzetti}, {Bellhouse}, {Scarlata}, \& {Bacchini}}]{Giunchi_2023b}
{Giunchi}, E., {Poggianti}, B.~M., {Gullieuszik}, M., {et~al.} 2023{\natexlab{b}}, \apj, 958, 73

\bibitem[{{Giunchi} {et~al.}(2025){Giunchi}, {Scarlata}, {Werle}, {Poggianti}, {Moretti}, {Gullieuszik}, {Vulcani}, {Ignesti}, {Marasco}, {Zanella}, \& {Wolter}}]{Giunchi_2025}
{Giunchi}, E., {Scarlata}, C., {Werle}, A., {et~al.} 2025, arXiv e-prints, arXiv:2502.15554

\bibitem[{{Gullieuszik} {et~al.}(2023){Gullieuszik}, {Giunchi}, {Poggianti}, {Moretti}, {Scarlata}, {Calzetti}, {Werle}, {Zanella}, {Radovich}, {Bellhouse}, {Bettoni}, {Franchetto}, {Fritz}, {Jaff{\'e}}, {McGee}, {Mingozzi}, {Omizzolo}, {Tonnesen}, {Verheijen}, \& {Vulcani}}]{Gullieuszik_2023}
{Gullieuszik}, M., {Giunchi}, E., {Poggianti}, B.~M., {et~al.} 2023, \apj, 945, 54

\bibitem[{{Gullieuszik} {et~al.}(2020){Gullieuszik}, {Poggianti}, {McGee}, {Moretti}, {Vulcani}, {Tonnesen}, {Roediger}, {Jaff{\'e}}, {Fritz}, {Franchetto}, {Omizzolo}, {Bettoni}, {Radovich}, \& {Wolter}}]{Gullieuszik_2020}
{Gullieuszik}, M., {Poggianti}, B.~M., {McGee}, S.~L., {et~al.} 2020, \apj, 899, 13

\bibitem[{{Gullieuszik} {et~al.}(2017){Gullieuszik}, {Poggianti}, {Moretti}, {Fritz}, {Jaff{\'e}}, {Hau}, {Bischko}, {Bellhouse}, {Bettoni}, {Fasano}, {Vulcani}, {D'Onofrio}, \& {Biviano}}]{Gullieuszik_2017}
{Gullieuszik}, M., {Poggianti}, B.~M., {Moretti}, A., {et~al.} 2017, \apj, 846, 27

\bibitem[{{Gunn} \& {Gott}(1972)}]{Gunn_1972}
{Gunn}, J.~E. \& {Gott}, J.~Richard, I. 1972, \apj, 176, 1

\bibitem[{{Hardcastle} \& {Croston}(2020)}]{Hardcastle_2020}
{Hardcastle}, M.~J. \& {Croston}, J.~H. 2020, \nar, 88, 101539

\bibitem[{{Henriksen} \& {Byrd}(1996)}]{Henriksen_1996}
{Henriksen}, M. \& {Byrd}, G. 1996, \apj, 459, 82

\bibitem[{{Hester} {et~al.}(2010){Hester}, {Seibert}, {Neill}, {Wyder}, {Gil de Paz}, {Madore}, {Martin}, {Schiminovich}, \& {Rich}}]{Hester_2010}
{Hester}, J.~A., {Seibert}, M., {Neill}, J.~D., {et~al.} 2010, \apjl, 716, L14

\bibitem[{{Hunt} \& {Hirashita}(2009)}]{Hunt_2009}
{Hunt}, L.~K. \& {Hirashita}, H. 2009, \aap, 507, 1327

\bibitem[{{Ignesti} {et~al.}(2022){Ignesti}, {Vulcani}, {Poggianti}, {Paladino}, {Shimwell}, {Healy}, {Gitti}, {Bacchini}, {Moretti}, {Radovich}, {van Weeren}, {Roberts}, {Botteon}, {M{\"u}ller}, {McGee}, {Fritz}, {Tomi{\v{c}}i{\'c}}, {Werle}, {Mingozzi}, {Gullieuszik}, \& {Verheijen}}]{Ignesti_2022}
{Ignesti}, A., {Vulcani}, B., {Poggianti}, B.~M., {et~al.} 2022, \apj, 924, 64

\bibitem[{{J{\'a}chym} {et~al.}(2014){J{\'a}chym}, {Combes}, {Cortese}, {Sun}, \& {Kenney}}]{Jachym_2014}
{J{\'a}chym}, P., {Combes}, F., {Cortese}, L., {Sun}, M., \& {Kenney}, J. D.~P. 2014, \apj, 792, 11

\bibitem[{{J{\'a}chym} {et~al.}(2019){J{\'a}chym}, {Kenney}, {Sun}, {Combes}, {Cortese}, {Scott}, {Sivanandam}, {Brinks}, {Roediger}, {Palou{\v{s}}}, \& {Fumagalli}}]{Jachym_2019}
{J{\'a}chym}, P., {Kenney}, J. D.~P., {Sun}, M., {et~al.} 2019, \apj, 883, 145

\bibitem[{{J{\'a}chym} {et~al.}(2017){J{\'a}chym}, {Sun}, {Kenney}, {Cortese}, {Combes}, {Yagi}, {Yoshida}, {Palou{\v{s}}}, \& {Roediger}}]{Jachym_2017}
{J{\'a}chym}, P., {Sun}, M., {Kenney}, J. D.~P., {et~al.} 2017, \apj, 839, 114

\bibitem[{{Joseph} {et~al.}(2025){Joseph}, {Tandon}, {Ghosh}, \& {Stalin}}]{Joseph_2025}
{Joseph}, P., {Tandon}, S.~N., {Ghosh}, S.~K., \& {Stalin}, C.~S. 2025, arXiv e-prints, arXiv:2504.00982

\bibitem[{{Kang} {et~al.}(2024){Kang}, {Hwang}, {Song}, {Park}, {Hwang}, \& {Park}}]{Kang_2024}
{Kang}, W., {Hwang}, H.~S., {Song}, H., {et~al.} 2024, \apjs, 272, 22

\bibitem[{{Kenney} {et~al.}(2015){Kenney}, {Abramson}, \& {Bravo-Alfaro}}]{Kenney_2015}
{Kenney}, J. D.~P., {Abramson}, A., \& {Bravo-Alfaro}, H. 2015, \aj, 150, 59

\bibitem[{{Kenney} {et~al.}(2014){Kenney}, {Geha}, {J{\'a}chym}, {Crowl}, {Dague}, {Chung}, {van Gorkom}, \& {Vollmer}}]{Kenney_2014}
{Kenney}, J. D.~P., {Geha}, M., {J{\'a}chym}, P., {et~al.} 2014, \apj, 780, 119

\bibitem[{{Kenney} \& {Koopmann}(1999)}]{Kenney_1999}
{Kenney}, J. D.~P. \& {Koopmann}, R.~A. 1999, \aj, 117, 181

\bibitem[{{Kenney} {et~al.}(2004){Kenney}, {van Gorkom}, \& {Vollmer}}]{Kenney_2004}
{Kenney}, J. D.~P., {van Gorkom}, J.~H., \& {Vollmer}, B. 2004, \aj, 127, 3361

\bibitem[{{Kennicutt}(1998)}]{Kennicutt_1998}
{Kennicutt}, Robert~C., J. 1998, \araa, 36, 189

\bibitem[{{Kennicutt} \& {Evans}(2012)}]{Kennicutt_2012}
{Kennicutt}, R.~C. \& {Evans}, N.~J. 2012, \araa, 50, 531

\bibitem[{{Kent} \& {Sargent}(1983)}]{Kent_1983}
{Kent}, S.~M. \& {Sargent}, W.~L.~W. 1983, \aj, 88, 697

\bibitem[{{Koopmann} {et~al.}(2006){Koopmann}, {Haynes}, \& {Catinella}}]{Koopmann_2006}
{Koopmann}, R.~A., {Haynes}, M.~P., \& {Catinella}, B. 2006, \aj, 131, 716

\bibitem[{{Koopmann} \& {Kenney}(2004{\natexlab{a}})}]{Koopmann_2004b}
{Koopmann}, R.~A. \& {Kenney}, J. D.~P. 2004{\natexlab{a}}, \apj, 613, 866

\bibitem[{{Koopmann} \& {Kenney}(2004{\natexlab{b}})}]{Koopmann_2004a}
{Koopmann}, R.~A. \& {Kenney}, J. D.~P. 2004{\natexlab{b}}, \apj, 613, 851

\bibitem[{{Kronberger} {et~al.}(2008){Kronberger}, {Kapferer}, {Unterguggenberger}, {Schindler}, \& {Ziegler}}]{Kronberger_2008}
{Kronberger}, T., {Kapferer}, W., {Unterguggenberger}, S., {Schindler}, S., \& {Ziegler}, B.~L. 2008, \aap, 483, 783

\bibitem[{{Kroupa}(2001)}]{Kroupa_2001}
{Kroupa}, P. 2001, \mnras, 322, 231

\bibitem[{{K{\"u}mmel} {et~al.}(2022){K{\"u}mmel}, {{\'A}lvarez-Ayll{\'o}n}, {Bertin}, {Dubath}, {Gavazzi}, {Hartley}, \& {Schefer}}]{Kummel_2022}
{K{\"u}mmel}, M., {{\'A}lvarez-Ayll{\'o}n}, A., {Bertin}, E., {et~al.} 2022, arXiv e-prints, arXiv:2212.02428

\bibitem[{{Larson} {et~al.}(1980){Larson}, {Tinsley}, \& {Caldwell}}]{Larson_1980}
{Larson}, R.~B., {Tinsley}, B.~M., \& {Caldwell}, C.~N. 1980, \apj, 237, 692

\bibitem[{{Laudari} {et~al.}(2022){Laudari}, {J{\'a}chym}, {Sun}, {Waldron}, {Chatzikos}, {Kenney}, {Luo}, {Nulsen}, {Sarazin}, {Combes}, {Edge}, {Voit}, {Donahue}, \& {Cortese}}]{Laudari_2022}
{Laudari}, S., {J{\'a}chym}, P., {Sun}, M., {et~al.} 2022, \mnras, 509, 3938

\bibitem[{{Longobardi} {et~al.}(2020){Longobardi}, {Boselli}, {Fossati}, {Villa-V{\'e}lez}, {Bianchi}, {Casasola}, {Sarpa}, {Combes}, {Hensler}, {Burgarella}, {Schimd}, {Nanni}, {C{\^o}t{\'e}}, {Buat}, {Amram}, {Ferrarese}, {Braine}, {Trinchieri}, {Boissier}, {Boquien}, {Andreani}, {Gwyn}, \& {Cuillandre}}]{Longobardi_2020}
{Longobardi}, A., {Boselli}, A., {Fossati}, M., {et~al.} 2020, \aap, 644, A161

\bibitem[{{Maciaszek} {et~al.}(2016){Maciaszek}, {Ealet}, {Jahnke}, {Prieto}, {Barbier}, {Mellier}, {Beaumont}, {Bon}, {Bonnefoi}, {Carle}, {Caillat}, {Costille}, {Dormoy}, {Ducret}, {Fabron}, {Febvre}, {Foulon}, {Garcia}, {Gimenez}, {Grassi}, {Laurent}, {Le Mignant}, {Martin}, {Rossin}, {Pamplona}, {Sanchez}, {Vives}, {Cl{\'e}mens}, {Gillard}, {Niclas}, {Secroun}, {Serra}, {Kubik}, {Ferriol}, {Amiaux}, {Barri{\`e}re}, {Berthe}, {Rosset}, {Macias-Perez}, {Auricchio}, {De Rosa}, {Franceschi}, {Guizzo}, {Morgante}, {Sortino}, {Trifoglio}, {Valenziano}, {Patrizii}, {Chiarusi}, {Fornari}, {Giacomini}, {Margiotta}, {Mauri}, {Pasqualini}, {Sirri}, {Spurio}, {Tenti}, {Travaglini}, {Dusini}, {Dal Corso}, {Laudisio}, {Sirignano}, {Stanco}, {Ventura}, {Borsato}, {Bonoli}, {Bortoletto}, {Balestra}, {D'Alessandro}, {Medinaceli}, {Farinelli}, {Corcione}, {Ligori}, {Grupp}, {Wimmer}, {Hormuth}, {Seidel}, {Wachter}, {Padilla}, {Lamensans}, {Casas}, {Lloro}, {Toledo-Moreo}, {Gomez}, {Colodro-Conde}, {Liz{\'a}n}, {Diaz},
  {Lilje}, {Toulouse-Aastrup}, {Andersen}, {S{\o}rensen}, {Jakobsen}, {Hornstrup}, {Jessen}, {Thizy}, {Holmes}, {Israelsson}, {Seiffert}, {Waczynski}, {Laureijs}, {Racca}, {Salvignol}, {Boenke}, \& {Strada}}]{Maciaszek_2016}
{Maciaszek}, T., {Ealet}, A., {Jahnke}, K., {et~al.} 2016, in Society of Photo-Optical Instrumentation Engineers (SPIE) Conference Series, Vol. 9904, Space Telescopes and Instrumentation 2016: Optical, Infrared, and Millimeter Wave, ed. H.~A. {MacEwen}, G.~G. {Fazio}, M.~{Lystrup}, N.~{Batalha}, N.~{Siegler}, \& E.~C. {Tong}, 99040T

\bibitem[{{Maciaszek} {et~al.}(2014){Maciaszek}, {Ealet}, {Jahnke}, {Prieto}, {Barbier}, {Mellier}, {Costille}, {Ducret}, {Fabron}, {Gimenez}, {Grange}, {Martin}, {Rossin}, {Pamplona}, {Vola}, {Cl{\'e}mens}, {Smadja}, {Amiaux}, {Barri{\`e}re}, {Berthe}, {De Rosa}, {Franceschi}, {Morgante}, {Trifoglio}, {Valenziano}, {Bonoli}, {Bortoletto}, {D'Alessandro}, {Corcione}, {Ligori}, {Garilli}, {Riva}, {Grupp}, {Vogel}, {Hormuth}, {Seidel}, {Wachter}, {Diaz}, {Gra{\~n}ena}, {Padilla}, {Toledo}, {Lilje}, {Solheim}, {Toulouse-Aastrup}, {Andersen}, {Holmes}, {Israelsson}, {Seiffert}, {Weber}, {Waczynski}, {Laureijs}, {Racca}, {Salvignol}, \& {Strada}}]{Maciaszek_2014}
{Maciaszek}, T., {Ealet}, A., {Jahnke}, K., {et~al.} 2014, in Society of Photo-Optical Instrumentation Engineers (SPIE) Conference Series, Vol. 9143, Space Telescopes and Instrumentation 2014: Optical, Infrared, and Millimeter Wave, ed. J.~M. {Oschmann}, Jr., M.~{Clampin}, G.~G. {Fazio}, \& H.~A. {MacEwen}, 91430K

\bibitem[{{Mancillas} {et~al.}(2019){Mancillas}, {Duc}, {Combes}, {Bournaud}, {Emsellem}, {Martig}, \& {Michel-Dansac}}]{Mancillas_2019}
{Mancillas}, B., {Duc}, P.-A., {Combes}, F., {et~al.} 2019, \aap, 632, A122

\bibitem[{{Mastropietro} {et~al.}(2005){Mastropietro}, {Moore}, {Mayer}, {Debattista}, {Piffaretti}, \& {Stadel}}]{Mastropietro_2005}
{Mastropietro}, C., {Moore}, B., {Mayer}, L., {et~al.} 2005, \mnras, 364, 607

\bibitem[{{Merluzzi} {et~al.}(2013){Merluzzi}, {Busarello}, {Dopita}, {Haines}, {Steinhauser}, {Mercurio}, {Rifatto}, {Smith}, \& {Schindler}}]{Merluzzi_2013}
{Merluzzi}, P., {Busarello}, G., {Dopita}, M.~A., {et~al.} 2013, \mnras, 429, 1747

\bibitem[{{Meusinger} {et~al.}(2020){Meusinger}, {Rudolf}, {Stecklum}, {Hoeft}, {Mauersberger}, \& {Apai}}]{Meusinger_2020}
{Meusinger}, H., {Rudolf}, C., {Stecklum}, B., {et~al.} 2020, \aap, 640, A30

\bibitem[{{Moore} {et~al.}(1996){Moore}, {Katz}, {Lake}, {Dressler}, \& {Oemler}}]{Moore_1996}
{Moore}, B., {Katz}, N., {Lake}, G., {Dressler}, A., \& {Oemler}, A. 1996, \nat, 379, 613

\bibitem[{{Moretti} {et~al.}(2018){Moretti}, {Paladino}, {Poggianti}, {D'Onofrio}, {Bettoni}, {Gullieuszik}, {Jaff{\'e}}, {Vulcani}, {Fasano}, {Fritz}, \& {Torstensson}}]{Moretti_2018}
{Moretti}, A., {Paladino}, R., {Poggianti}, B.~M., {et~al.} 2018, \mnras, 480, 2508

\bibitem[{{Navarro} {et~al.}(1997){Navarro}, {Frenk}, \& {White}}]{Navarro_1997}
{Navarro}, J.~F., {Frenk}, C.~S., \& {White}, S. D.~M. 1997, \apj, 490, 493

\bibitem[{{Noeske} {et~al.}(2007){Noeske}, {Weiner}, {Faber}, {Papovich}, {Koo}, {Somerville}, {Bundy}, {Conselice}, {Newman}, {Schiminovich}, {Le Floc'h}, {Coil}, {Rieke}, {Lotz}, {Primack}, {Barmby}, {Cooper}, {Davis}, {Ellis}, {Fazio}, {Guhathakurta}, {Huang}, {Kassin}, {Martin}, {Phillips}, {Rich}, {Small}, {Willmer}, \& {Wilson}}]{Noeske_2007}
{Noeske}, K.~G., {Weiner}, B.~J., {Faber}, S.~M., {et~al.} 2007, \apjl, 660, L43

\bibitem[{{Owen} {et~al.}(2006){Owen}, {Keel}, {Wang}, {Ledlow}, \& {Morrison}}]{Owen_2006}
{Owen}, F.~N., {Keel}, W.~C., {Wang}, Q.~D., {Ledlow}, M.~J., \& {Morrison}, G.~E. 2006, \aj, 131, 1974

\bibitem[{{Owers} {et~al.}(2012){Owers}, {Couch}, {Nulsen}, \& {Randall}}]{Owers_2012}
{Owers}, M.~S., {Couch}, W.~J., {Nulsen}, P. E.~J., \& {Randall}, S.~W. 2012, \apjl, 750, L23

\bibitem[{{Planck Collaboration} {et~al.}(2020){Planck Collaboration}, {Aghanim}, {Akrami}, {Ashdown}, {Aumont}, {Baccigalupi}, {Ballardini}, {Banday}, {Barreiro}, {Bartolo}, {Basak}, {Battye}, {Benabed}, {Bernard}, {Bersanelli}, {Bielewicz}, {Bock}, {Bond}, {Borrill}, {Bouchet}, {Boulanger}, {Bucher}, {Burigana}, {Butler}, {Calabrese}, {Cardoso}, {Carron}, {Challinor}, {Chiang}, {Chluba}, {Colombo}, {Combet}, {Contreras}, {Crill}, {Cuttaia}, {de Bernardis}, {de Zotti}, {Delabrouille}, {Delouis}, {Di Valentino}, {Diego}, {Dor{\'e}}, {Douspis}, {Ducout}, {Dupac}, {Dusini}, {Efstathiou}, {Elsner}, {En{\ss}lin}, {Eriksen}, {Fantaye}, {Farhang}, {Fergusson}, {Fernandez-Cobos}, {Finelli}, {Forastieri}, {Frailis}, {Fraisse}, {Franceschi}, {Frolov}, {Galeotta}, {Galli}, {Ganga}, {G{\'e}nova-Santos}, {Gerbino}, {Ghosh}, {Gonz{\'a}lez-Nuevo}, {G{\'o}rski}, {Gratton}, {Gruppuso}, {Gudmundsson}, {Hamann}, {Handley}, {Hansen}, {Herranz}, {Hildebrandt}, {Hivon}, {Huang}, {Jaffe}, {Jones}, {Karakci}, {Keih{\"a}nen},
  {Keskitalo}, {Kiiveri}, {Kim}, {Kisner}, {Knox}, {Krachmalnicoff}, {Kunz}, {Kurki-Suonio}, {Lagache}, {Lamarre}, {Lasenby}, {Lattanzi}, {Lawrence}, {Le Jeune}, {Lemos}, {Lesgourgues}, {Levrier}, {Lewis}, {Liguori}, {Lilje}, {Lilley}, {Lindholm}, {L{\'o}pez-Caniego}, {Lubin}, {Ma}, {Mac{\'\i}as-P{\'e}rez}, {Maggio}, {Maino}, {Mandolesi}, {Mangilli}, {Marcos-Caballero}, {Maris}, {Martin}, {Martinelli}, {Mart{\'\i}nez-Gonz{\'a}lez}, {Matarrese}, {Mauri}, {McEwen}, {Meinhold}, {Melchiorri}, {Mennella}, {Migliaccio}, {Millea}, {Mitra}, {Miville-Desch{\^e}nes}, {Molinari}, {Montier}, {Morgante}, {Moss}, {Natoli}, {N{\o}rgaard-Nielsen}, {Pagano}, {Paoletti}, {Partridge}, {Patanchon}, {Peiris}, {Perrotta}, {Pettorino}, {Piacentini}, {Polastri}, {Polenta}, {Puget}, {Rachen}, {Reinecke}, {Remazeilles}, {Renzi}, {Rocha}, {Rosset}, {Roudier}, {Rubi{\~n}o-Mart{\'\i}n}, {Ruiz-Granados}, {Salvati}, {Sandri}, {Savelainen}, {Scott}, {Shellard}, {Sirignano}, {Sirri}, {Spencer}, {Sunyaev}, {Suur-Uski}, {Tauber}, {Tavagnacco},
  {Tenti}, {Toffolatti}, {Tomasi}, {Trombetti}, {Valenziano}, {Valiviita}, {Van Tent}, {Vibert}, {Vielva}, {Villa}, {Vittorio}, {Wandelt}, {Wehus}, {White}, {White}, {Zacchei}, \& {Zonca}}]{Aghanim_2020}
{Planck Collaboration}, {Aghanim}, N., {Akrami}, Y., {et~al.} 2020, \aap, 641, A6

\bibitem[{{Poggianti} {et~al.}(2016){Poggianti}, {Fasano}, {Omizzolo}, {Gullieuszik}, {Bettoni}, {Moretti}, {Paccagnella}, {Jaff{\'e}}, {Vulcani}, {Fritz}, {Couch}, \& {D'Onofrio}}]{Poggianti_2016}
{Poggianti}, B.~M., {Fasano}, G., {Omizzolo}, A., {et~al.} 2016, \aj, 151, 78

\bibitem[{{Poggianti} {et~al.}(2019){Poggianti}, {Gullieuszik}, {Tonnesen}, {Moretti}, {Vulcani}, {Radovich}, {Jaff{\'e}}, {Fritz}, {Bettoni}, {Franchetto}, {Fasano}, {Bellhouse}, \& {Omizzolo}}]{Poggianti_2019}
{Poggianti}, B.~M., {Gullieuszik}, M., {Tonnesen}, S., {et~al.} 2019, \mnras, 482, 4466

\bibitem[{{Popesso} {et~al.}(2023){Popesso}, {Concas}, {Cresci}, {Belli}, {Rodighiero}, {Inami}, {Dickinson}, {Ilbert}, {Pannella}, \& {Elbaz}}]{Popesso_2023}
{Popesso}, P., {Concas}, A., {Cresci}, G., {et~al.} 2023, \mnras, 519, 1526

\bibitem[{{Portegies Zwart} {et~al.}(2010){Portegies Zwart}, {McMillan}, \& {Gieles}}]{Portegies_2010}
{Portegies Zwart}, S.~F., {McMillan}, S. L.~W., \& {Gieles}, M. 2010, \araa, 48, 431

\bibitem[{{Rawle} {et~al.}(2014){Rawle}, {Altieri}, {Egami}, {P{\'e}rez-Gonz{\'a}lez}, {Richard}, {Santos}, {Valtchanov}, {Walth}, {Bouy}, {Haines}, \& {Okabe}}]{Rawle_2014}
{Rawle}, T.~D., {Altieri}, B., {Egami}, E., {et~al.} 2014, \mnras, 442, 196

\bibitem[{{Roberts} {et~al.}(2021{\natexlab{a}}){Roberts}, {van Weeren}, {McGee}, {Botteon}, {Drabent}, {Ignesti}, {Rottgering}, {Shimwell}, \& {Tasse}}]{Roberts_2021a}
{Roberts}, I.~D., {van Weeren}, R.~J., {McGee}, S.~L., {et~al.} 2021{\natexlab{a}}, \aap, 650, A111

\bibitem[{{Roberts} {et~al.}(2021{\natexlab{b}}){Roberts}, {van Weeren}, {McGee}, {Botteon}, {Ignesti}, \& {Rottgering}}]{Roberts_2021b}
{Roberts}, I.~D., {van Weeren}, R.~J., {McGee}, S.~L., {et~al.} 2021{\natexlab{b}}, \aap, 652, A153

\bibitem[{{Roberts} {et~al.}(2022){Roberts}, {van Weeren}, {Timmerman}, {Botteon}, {Gendron-Marsolais}, {Ignesti}, \& {Rottgering}}]{Roberts_2022}
{Roberts}, I.~D., {van Weeren}, R.~J., {Timmerman}, R., {et~al.} 2022, \aap, 658, A44

\bibitem[{{Roediger} {et~al.}(2014){Roediger}, {Bruggen}, {Owers}, {Ebeling}, \& {Sun}}]{Roediger_2014}
{Roediger}, E., {Bruggen}, M., {Owers}, M.~S., {Ebeling}, H., \& {Sun}, M. 2014, \mnras, 443, L114

\bibitem[{{Salim} {et~al.}(2007){Salim}, {Rich}, {Charlot}, {Brinchmann}, {Johnson}, {Schiminovich}, {Seibert}, {Mallery}, {Heckman}, {Forster}, {Friedman}, {Martin}, {Morrissey}, {Neff}, {Small}, {Wyder}, {Bianchi}, {Donas}, {Lee}, {Madore}, {Milliard}, {Szalay}, {Welsh}, \& {Yi}}]{Salim_2007}
{Salim}, S., {Rich}, R.~M., {Charlot}, S., {et~al.} 2007, \apjs, 173, 267

\bibitem[{{Salinas} {et~al.}(2024){Salinas}, {Jaff{\'e}}, {Smith}, {Shinn}, {Crossett}, {Gullieuszik}, {Gonz{\'a}lez-Tor{\`a}}, {Piraino-Cerda}, {Poggianti}, {Vulcani}, {Biviano}, {Louren{\c{c}}o}, {Bilton}, {Kelkar}, \& {Calder{\'o}n-Castillo}}]{Salinas_2024}
{Salinas}, V., {Jaff{\'e}}, Y.~L., {Smith}, R., {et~al.} 2024, \mnras, 533, 341

\bibitem[{{Sarazin}(1986)}]{Sarazin_1986}
{Sarazin}, C.~L. 1986, Reviews of Modern Physics, 58, 1

\bibitem[{{Schulz} \& {Struck}(2001)}]{Schulz_2001}
{Schulz}, S. \& {Struck}, C. 2001, \mnras, 328, 185

\bibitem[{{Serra} {et~al.}(2023){Serra}, {Maccagni}, {Kleiner}, {Moln{\'a}r}, {Ramatsoku}, {Loni}, {Loi}, {de Blok}, {Bryan}, {Dettmar}, {Frank}, {van Gorkom}, {Govoni}, {Iodice}, {J{\'o}zsa}, {Kamphuis}, {Kraan-Korteweg}, {Loubser}, {Murgia}, {Oosterloo}, {Peletier}, {Pisano}, {Smith}, {Trager}, \& {Verheijen}}]{Serra_2023}
{Serra}, P., {Maccagni}, F.~M., {Kleiner}, D., {et~al.} 2023, \aap, 673, A146

\bibitem[{{Shimwell} {et~al.}(2017){Shimwell}, {R{\"o}ttgering}, {Best}, {Williams}, {Dijkema}, {de Gasperin}, {Hardcastle}, {Heald}, {Hoang}, {Horneffer}, {Intema}, {Mahony}, {Mandal}, {Mechev}, {Morabito}, {Oonk}, {Rafferty}, {Retana-Montenegro}, {Sabater}, {Tasse}, {van Weeren}, {Br{\"u}ggen}, {Brunetti}, {Chy{\.z}y}, {Conway}, {Haverkorn}, {Jackson}, {Jarvis}, {McKean}, {Miley}, {Morganti}, {White}, {Wise}, {van Bemmel}, {Beck}, {Brienza}, {Bonafede}, {Calistro Rivera}, {Cassano}, {Clarke}, {Cseh}, {Deller}, {Drabent}, {van Driel}, {Engels}, {Falcke}, {Ferrari}, {Fr{\"o}hlich}, {Garrett}, {Harwood}, {Heesen}, {Hoeft}, {Horellou}, {Israel}, {Kapi{\'n}ska}, {Kunert-Bajraszewska}, {McKay}, {Mohan}, {Orr{\'u}}, {Pizzo}, {Prandoni}, {Schwarz}, {Shulevski}, {Sipior}, {Smith}, {Sridhar}, {Steinmetz}, {Stroe}, {Varenius}, {van der Werf}, {Zensus}, \& {Zwart}}]{Shimwell_2017}
{Shimwell}, T.~W., {R{\"o}ttgering}, H.~J.~A., {Best}, P.~N., {et~al.} 2017, \aap, 598, A104

\bibitem[{{Shimwell} {et~al.}(2019){Shimwell}, {Tasse}, {Hardcastle}, {Mechev}, {Williams}, {Best}, {R{\"o}ttgering}, {Callingham}, {Dijkema}, {de Gasperin}, {Hoang}, {Hugo}, {Mirmont}, {Oonk}, {Prandoni}, {Rafferty}, {Sabater}, {Smirnov}, {van Weeren}, {White}, {Atemkeng}, {Bester}, {Bonnassieux}, {Br{\"u}ggen}, {Brunetti}, {Chy{\.z}y}, {Cochrane}, {Conway}, {Croston}, {Danezi}, {Duncan}, {Haverkorn}, {Heald}, {Iacobelli}, {Intema}, {Jackson}, {Jamrozy}, {Jarvis}, {Lakhoo}, {Mevius}, {Miley}, {Morabito}, {Morganti}, {Nisbet}, {Orr{\'u}}, {Perkins}, {Pizzo}, {Schrijvers}, {Smith}, {Vermeulen}, {Wise}, {Alegre}, {Bacon}, {van Bemmel}, {Beswick}, {Bonafede}, {Botteon}, {Bourke}, {Brienza}, {Calistro Rivera}, {Cassano}, {Clarke}, {Conselice}, {Dettmar}, {Drabent}, {Dumba}, {Emig}, {En{\ss}lin}, {Ferrari}, {Garrett}, {G{\'e}nova-Santos}, {Goyal}, {G{\"u}rkan}, {Hale}, {Harwood}, {Heesen}, {Hoeft}, {Horellou}, {Jackson}, {Kokotanekov}, {Kondapally}, {Kunert-Bajraszewska}, {Mahatma}, {Mahony}, {Mandal}, {McKean},
  {Merloni}, {Mingo}, {Miskolczi}, {Mooney}, {Nikiel-Wroczy{\'n}ski}, {O'Sullivan}, {Quinn}, {Reich}, {Roskowi{\'n}ski}, {Rowlinson}, {Savini}, {Saxena}, {Schwarz}, {Shulevski}, {Sridhar}, {Stacey}, {Urquhart}, {van der Wiel}, {Varenius}, {Webster}, \& {Wilber}}]{Shimwell_2019}
{Shimwell}, T.~W., {Tasse}, C., {Hardcastle}, M.~J., {et~al.} 2019, \aap, 622, A1

\bibitem[{{Smith} {et~al.}(2022){Smith}, {Shinn}, {Tonnesen}, {Calder{\'o}n-Castillo}, {Crossett}, {Jaffe}, {Roberts}, {McGee}, {George}, {Vulcani}, {Gullieuszik}, {Moretti}, {Poggianti}, \& {Shin}}]{Smith_2022}
{Smith}, R., {Shinn}, J.-H., {Tonnesen}, S., {et~al.} 2022, \apj, 934, 86

\bibitem[{{Smith} {et~al.}(2010){Smith}, {Lucey}, {Hammer}, {Hornschemeier}, {Carter}, {Hudson}, {Marzke}, {Mouhcine}, {Eftekharzadeh}, {James}, {Khosroshahi}, {Kourkchi}, \& {Karick}}]{Smith_2010}
{Smith}, R.~J., {Lucey}, J.~R., {Hammer}, D., {et~al.} 2010, \mnras, 408, 1417

\bibitem[{{Sola} {et~al.}(2022){Sola}, {Duc}, {Richards}, {Paiement}, {Urbano}, {Klehammer}, {B{\'\i}lek}, {Cuillandre}, {Gwyn}, \& {McConnachie}}]{Sola_2022}
{Sola}, E., {Duc}, P.-A., {Richards}, F., {et~al.} 2022, \aap, 662, A124

\bibitem[{{Steinhauser} {et~al.}(2016){Steinhauser}, {Schindler}, \& {Springel}}]{Steinhauser_2016}
{Steinhauser}, D., {Schindler}, S., \& {Springel}, V. 2016, \aap, 591, A51

\bibitem[{{Steyrleithner} {et~al.}(2020){Steyrleithner}, {Hensler}, \& {Boselli}}]{Steyrleithner_2020}
{Steyrleithner}, P., {Hensler}, G., \& {Boselli}, A. 2020, \mnras, 494, 1114

\bibitem[{{Sun} {et~al.}(2006){Sun}, {Jones}, {Forman}, {Nulsen}, {Donahue}, \& {Voit}}]{Sun_2006}
{Sun}, M., {Jones}, C., {Forman}, W., {et~al.} 2006, \apjl, 637, L81

\bibitem[{{Tandon} {et~al.}(2020){Tandon}, {Postma}, {Joseph}, {Devaraj}, {Subramaniam}, {Barve}, {George}, {Ghosh}, {Girish}, {Hutchings}, {Kamath}, {Kathiravan}, {Kumar}, {Lancelot}, {Leahy}, {Mahesh}, {Mohan}, {Nagabhushana}, {Pati}, {Rao}, {Sankarasubramanian}, {Sriram}, \& {Stalin}}]{Tandon_2020}
{Tandon}, S.~N., {Postma}, J., {Joseph}, P., {et~al.} 2020, \aj, 159, 158

\bibitem[{{Tandon} {et~al.}(2017){Tandon}, {Subramaniam}, {Girish}, {Postma}, {Sankarasubramanian}, {Sriram}, {Stalin}, {Mondal}, {Sahu}, {Joseph}, {Hutchings}, {Ghosh}, {Barve}, {George}, {Kamath}, {Kathiravan}, {Kumar}, {Lancelot}, {Leahy}, {Mahesh}, {Mohan}, {Nagabhushana}, {Pati}, {Kameswara Rao}, {Sreedhar}, \& {Sreekumar}}]{Tandon_2017}
{Tandon}, S.~N., {Subramaniam}, A., {Girish}, V., {et~al.} 2017, \aj, 154, 128

\bibitem[{{Taylor} \& {Webster}(2005)}]{Taylor_2005}
{Taylor}, E.~N. \& {Webster}, R.~L. 2005, \apj, 634, 1067

\bibitem[{{Tonnesen} \& {Bryan}(2012)}]{Tonnesen_2012}
{Tonnesen}, S. \& {Bryan}, G.~L. 2012, \mnras, 422, 1609

\bibitem[{{Valluri}(1993)}]{Valluri_1993}
{Valluri}, M. 1993, \apj, 408, 57

\bibitem[{{van Weeren} {et~al.}(2019){van Weeren}, {de Gasperin}, {Akamatsu}, {Br{\"u}ggen}, {Feretti}, {Kang}, {Stroe}, \& {Zandanel}}]{vanWeeren_2019}
{van Weeren}, R.~J., {de Gasperin}, F., {Akamatsu}, H., {et~al.} 2019, \ssr, 215, 16

\bibitem[{{van Weeren} {et~al.}(2024){van Weeren}, {Timmerman}, {Vaidya}, {Gendron-Marsolais}, {Botteon}, {Roberts}, {Hlavacek-Larrondo}, {Bonafede}, {Br{\"u}ggen}, {Brunetti}, {Cassano}, {Cuciti}, {Edge}, {Gastaldello}, {Groeneveld}, \& {Shimwell}}]{vanWeeren_2024}
{van Weeren}, R.~J., {Timmerman}, R., {Vaidya}, V., {et~al.} 2024, arXiv e-prints, arXiv:2410.02863

\bibitem[{{Vollmer} {et~al.}(2004){Vollmer}, {Beck}, {Kenney}, \& {van Gorkom}}]{Vollmer_2004}
{Vollmer}, B., {Beck}, R., {Kenney}, J. D.~P., \& {van Gorkom}, J.~H. 2004, \aj, 127, 3375

\bibitem[{{Vollmer} {et~al.}(2012){Vollmer}, {Wong}, {Braine}, {Chung}, \& {Kenney}}]{Vollmer_2012}
{Vollmer}, B., {Wong}, O.~I., {Braine}, J., {Chung}, A., \& {Kenney}, J.~D.~P. 2012, \aap, 543, A33

\bibitem[{{Vulcani} {et~al.}(2020){Vulcani}, {Fritz}, {Poggianti}, {Bettoni}, {Franchetto}, {Moretti}, {Gullieuszik}, {Jaff{\'e}}, {Biviano}, {Radovich}, \& {Mingozzi}}]{Vulcani_2020}
{Vulcani}, B., {Fritz}, J., {Poggianti}, B.~M., {et~al.} 2020, \apj, 892, 146

\bibitem[{{Vulcani} {et~al.}(2022){Vulcani}, {Poggianti}, {Smith}, {Moretti}, {Jaff{\'e}}, {Gullieuszik}, {Fritz}, \& {Bellhouse}}]{Vulcani_2022}
{Vulcani}, B., {Poggianti}, B.~M., {Smith}, R., {et~al.} 2022, \apj, 927, 91

\bibitem[{{Waldron} {et~al.}(2023){Waldron}, {Sun}, {Luo}, {Laudari}, {Chatzikos}, {Sivanandam}, {Kenney}, {J{\'a}chym}, {Voit}, {Donahue}, \& {Fossati}}]{Waldron_2023}
{Waldron}, W., {Sun}, M., {Luo}, R., {et~al.} 2023, \mnras, 522, 173

\bibitem[{{Yagi} {et~al.}(2007){Yagi}, {Komiyama}, {Yoshida}, {Furusawa}, {Kashikawa}, {Koyama}, \& {Okamura}}]{Yagi_2007}
{Yagi}, M., {Komiyama}, Y., {Yoshida}, M., {et~al.} 2007, \apj, 660, 1209

\bibitem[{{Yagi} {et~al.}(2017){Yagi}, {Yoshida}, {Gavazzi}, {Komiyama}, {Kashikawa}, \& {Okamura}}]{Yagi_2017}
{Yagi}, M., {Yoshida}, M., {Gavazzi}, G., {et~al.} 2017, \apj, 839, 65

\bibitem[{{Yagi} {et~al.}(2010){Yagi}, {Yoshida}, {Komiyama}, {Kashikawa}, {Furusawa}, {Okamura}, {Graham}, {Miller}, {Carter}, {Mobasher}, \& {Jogee}}]{Yagi_2010}
{Yagi}, M., {Yoshida}, M., {Komiyama}, Y., {et~al.} 2010, \aj, 140, 1814

\bibitem[{{Yoshida} {et~al.}(2012){Yoshida}, {Yagi}, {Komiyama}, {Furusawa}, {Kashikawa}, {Hattori}, \& {Okamura}}]{Yoshida_2012}
{Yoshida}, M., {Yagi}, M., {Komiyama}, Y., {et~al.} 2012, \apj, 749, 43

\bibitem[{{Yoshida} {et~al.}(2008){Yoshida}, {Yagi}, {Komiyama}, {Furusawa}, {Kashikawa}, {Koyama}, {Yamanoi}, {Hattori}, \& {Okamura}}]{Yoshida_2008}
{Yoshida}, M., {Yagi}, M., {Komiyama}, Y., {et~al.} 2008, \apj, 688, 918

\end{thebibliography}
